\documentclass{article}

\usepackage{PRIMEarxiv}

\usepackage[english]{babel}
\usepackage[utf8]{inputenc} 
\usepackage[T1]{fontenc}    
\usepackage{hyperref}       
\usepackage{url}            
\usepackage{booktabs}       
\usepackage{amsmath}
\usepackage{amsfonts}       
\usepackage{nicefrac}       
\usepackage{microtype}      
\usepackage{lipsum}
\usepackage{fancyhdr}       
\usepackage{graphicx}       
\usepackage{quantikz}
\usepackage{braket}
\usepackage{tikz}
\usepackage{orcidlink}
\usepackage[numbers,sort&compress]{natbib}
\numberwithin{equation}{section}

\usepackage{xcolor}
\newtheorem{theorem}{Theorem}
\title{Explicit Solution Equation for Every Combinatorial Problem via Tensor Networks: MeLoCoToN}

\author{
  Alejandro Mata Ali 
\orcidlink{https://orcid.org/0009-0006-7289-8827}\\
  Instituto Tecnológico de Castilla y León \\
  Burgos, Spain\\
  \texttt{alejandro.mata@itcl.es} \\
  }

\begin{document}
\maketitle

\begin{abstract}
In this paper we show that every combinatorial problem has an exact explicit equation that returns its solution. We present a method to obtain an equation that solves exactly any combinatorial problem, both inversion, constraint satisfaction and optimization, by obtaining its equivalent tensor network. This formulation only requires a basic knowledge of classical logical operators, at a first year level of any computer science degree. These equations are not necessarily computable in a reasonable time, nor do they allow to surpass the state of the art in computational complexity, but they allow to have a new perspective for the mathematical analysis of these problems. These equations computation can be approximated by different methods such as Matrix Product State compression. We also present the equations for numerous combinatorial problems. This work proves that, if there is a physical system capable of contracting in polynomial time the tensor networks presented, every NP-Hard problem can be solved in polynomial time.
\end{abstract}


\tableofcontents

\newpage
\section{Introduction}
A combinatorial problem is a problem consisting in finding a combination of elements with certain characteristics. There are basically three types of combinatorial problems: \textit{inversion problems}, which consist in obtaining the input that results in a known output through a function, \textit{constraint satisfaction problems}, which consist in finding a solution that satisfies a set of restrictions, and \textit{optimization problems}, which consist in obtaining the combination that has the lowest or highest associated value of a certain function.

Combinatorial inversion problems (do not confuse with inverse combinatorial optimization~\cite{Inverse_comb}) have application in cryptography, because if there is a function that obtains the public key of a protocol from the private key, being able to perform the reverse process would compromise its security. An example is the RSA protocol~\cite{RSA}, in which we have two prime numbers $p$ and $q$ as private information. The public information is obtained by multiplying them, obtaining a number $N=pq$. This process is very simple, but the reverse process of obtaining $p$ or $q$ from $N$ is highly expensive.

Constraint satisfaction problems~\cite{CSP} are those in which we search for a solution that satisfies a given set of constraints. This type of problems are especially interesting for real world applications such as resource relocation. Some examples of academic cases are N queens~\cite{N_Queens} or binary sudoku. However, they are usually highly expensive problems to solve, which leads to their resolution with heuristic methods~\cite{Heuristics}.

Combinatorial optimization problems~\cite{Combinatorial} are those in which we search for the vector $\vec{x}$ of integers which minimizes or maximizes a certain function $C(\cdot)$ called \textit{cost function}, given some constraints. That is, we look for the combination with the lowest or highest possible cost that satisfies the constraints. On many occasions, it is possible to include the constraints within the cost function itself as a term that greatly raises or decreases the cost of any incompatible combination. Combinatorial optimization problems are very useful for various applied use cases and academic cases. Some applied examples are route problems as the shortest path problem~\cite{A-star,Dijkstra1959} or the traveling salesman problem~\cite{TSP_overview,TSP,TSP2}; task scheduling as the job shop scheduling problem~\cite{JSSP_General,Flexible_JSSP} or flow shop scheduling problem~\cite{FlowShop} and constrained optimization as knapsack problem~\cite{Knapsack_original}, minimal spanning tree~\cite{minimaltree} or bin packing problem~\cite{Bin_Packing}.

One of the limitations in tackling all these problems is that many of them are NP-Hard, which implies that they are unassumingly expensive to solve exactly, either in space or time. There are several ways to approach this type of problems: heuristics~\cite{Heuristics}, genetic algorithms~\cite{Genetic}, approximate methods~\cite{Aproximated}, etc.

One of the technologies that has gained the most popularity in trying to address these problems is quantum computing~\cite{Bench_Quantum_Optim,Quantum_Block_Optim,Variational_Quantum_Optim,Generative_Quantum_Combinat}. Due to phenomena such as superposition and entanglement, a different type of computation can be performed. For example, in algorithms such as the Quantum Approximate Optimization Algorithm (QAOA) \cite{QAOA}, the system starts from a superposition of all the solutions and is operated in such a way that the correct solution is obtained at the end of the process with maximum probability. In many quantum computing techniques, the Quadratic Unconstrained Binary Optimization (QUBO) formulation \cite{QUBO} is used, consisting in modeling the problems as relations between pairs of unconstrained binary variables.

Due to the current limitations of quantum hardware~\cite{Quantum_Limitations}, other alternative quantum-inspired technologies have emerged. One of them is tensor networks~\cite{Tensor_networks}, which consist in the use of tensor operations to represent data and interactions.  In this way, quantum circuits can be represented as tensor networks that perform the same matrix operations~\cite{Simulation}. In addition, unknown data can be generated from knowing the rules or structure that it follows, applying them to the tensors that will generate them. It is important to realize that a tensor network is an equation. If a tensor network returns the solution to a problem, this will imply that this problem has an equation that solves it. Tensor networks have been used in recent years in various ways to address combinatorial optimization problems~\cite{TTOpt,TN_Constraint,TN_Optim_Quick,TN_Generative,TNGEO,QAOA_TN,HOBO_TN,HOBOTAN,TSP_TN,Task_TN,QUBO_Tridiagonal}, obtaining interesting results.

A very interesting work is~\cite{TN_Constraint}, in which they presented the possibility of using a tensor network simulating a quantum state. An imaginary time evolution with respect to its cost hamiltonian is applied to it. After that, tensors based on logics that eliminate the states incompatible with the restrictions are applied. Finally, the expected value of the $Z$ operator for one of the qubits is measured. In this way the correct value of the variable represented by that qubit can be determined. Although the idea presented is brilliant, this work is limited to the simulation of a quantum system, so it needs to unnecessarily increase memory and execution times by needing to contract a network tensor twice as large. The presented tensor network can be halved by performing a direct summation of the amplitudes as they are all positive, which is only a rescaling of its damping parameter $\tau$. Moreover, its way of dealing with degenerate states presents problems that are solved by projections in the iterations following the moment when one of the combinations is chosen. On the other hand, it is limited to the case of binary variables, but it is easily generalizable to positive integer variables by switching to the qudits formalism, and replacing the $Z$ operator by leaving that index free and checking
the position of the maximum. With these slight changes, the fundamental idea they present can be used to explore and design algorithms for solving not only combinatorial optimization problems, but any combinatorial problem.

In this paper we will present a general method to obtain a tensor network that solves a combinatorial problem and some techniques to improve its computation. After that, we will present the tensor networks that solve several examples of combinatorial problems and how to obtain them. It is important to emphasize that the aim of this paper is to present the method as an alternative mathematical analysis of the problems, rather than as a technique with an advantage over the state-of-the-art. For this reason, we will omit the analysis of the computational complexities, since in most of the cases that we will present, this method is exponentially more expensive than the known solutions, or even brute force. This does not imply that there are no cases in which this advantage may exist. Finally, we will present a new notation to simplify the presentation of tensor network papers.

Some parts of the paper are not available because they refer to unpublished papers. Future versions will include these papers after they have been published.

\newpage
\section{Definition of the general method MeLoCoToN}
The core of this work is that every combinatorial problem has an explicit equation that returns its exact solution. In this section we will demonstrate how such an equation is obtained based on the use of tensor networks. This method, which we call \textit{Modulated Logical Combinatorial Tensor Networks} (MLCTN or MeLoCoToN), will consist of four steps:
\begin{enumerate}
    \item Definition of the problem variables and rewriting of the functions.
    \item Creation of the associated classical logical circuit.
    \item Creation of the associated logical tensor network.
    \item Iteration on the tensor network and contraction.
\end{enumerate}

\begin{figure}[h]
    \centering
    \includegraphics[width=0.7\linewidth]{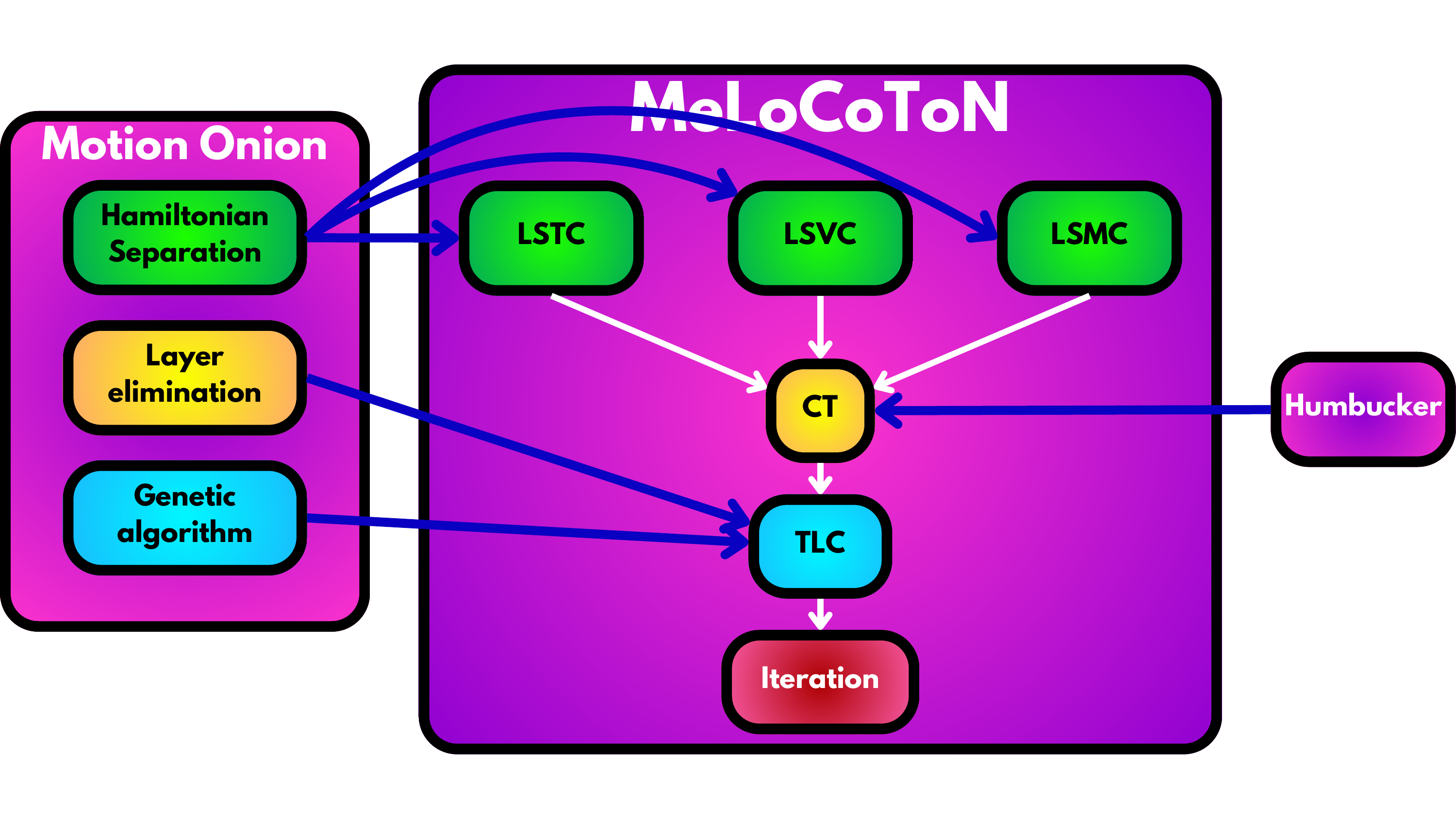}
    \caption{General scheme of ideas presented.}
    \label{fig: General Scheme}
\end{figure}

The ideas presented can be sintetized in Fig.~\ref{fig: General Scheme}.

There are three general types of combinatorial problems that we can define. The first are \textit{inversion problems}, which consist in having a function $\gamma$ which associates one output combination to each input combination, and given a known output $\vec{Y}$, we search for the input $\vec{X}$ that generates it, $\vec{Y}=\gamma(\vec{X})$. An example would be the factorization of prime numbers. The second are the \textit{constraint satisfaction problems}, which consist in obtaining a solution that satisfies a set of constraints. An example is the N queens. The third are the \textit{optimization problems}, which consist in having a function which associates a cost to each input, these are not transformed, and obtaining the input with lower cost. An example is the knapsack problem. Both types can be considered combinatorial problems for the purposes of what we will discuss later.

\subsection{Quantum computing explanation of the method}
Before explaining the details of the tensor method, let us explain the quantum motivation for it, for those more familiar with quantum computing. Our method will consist in the simultaneous evaluation of all possible solutions, so that, when measuring the system, it will return with maximum probability the basis state that encodes the optimal solution. In the cases of inversion and constraint satisfaction problems, when measuring the system, only the correct solution can be obtained, while in the cases of optimization problems it can be obtained with higher probability.

For inversion problems, we have two registers. The measurement register, with the qudits we will measure, and the post-selection register, with the qudits we will operate and post-select. 
The system starts with uniform superposition on the qudits from both registers, making a Bell state between each qudit in each register.

\begin{equation}
    \ket{\psi_0} = \bigotimes_{k=0}^{N-1} \left(\sum_{x_k=0}^{d_k-1} \ket{x_k}_{m,k}\ket{x_k}_{p,k}\right)=\sum_{\vec{x}}\left(\ket{\vec{x}}_{m}\ket{\vec{x}}_{p}\right),
\end{equation}
where $\ket{x_k}_{m,k}$ is the state of the $k$-th measurement qudit and $\ket{x_k}_{p,k}$ is the state of the $k$-th post-selection qudit.

After that, we apply an oracle $\mathcal{T}$ on the post-selection register, so that for each base state its state becomes the output associated to the input of that state through the function $\gamma$ to be inverted.
\begin{equation}
    \ket{\psi_1} = \sum_{\vec{x}}\left( \ket{\vec{x}}_{m}\mathcal{T}\ket{\vec{x}}_{p}\right)=\sum_{\vec{x}} \left( \ket{\vec{x}}_{m}\ket{\gamma(\vec{x})}_{p}\right).
\end{equation}
Now, we post-select the state of the qudits in the second register, so that they are only in the state of the known output $\vec{Y}$ of the function we want to invert. This is a non-unitary operation, so we cannot perform it in a quantum system. In this way, in the measurement register, the only state $\ket{\vec{X}}$ that will remain is the one generated by the output that we have post-selected.
\begin{equation}
    \ket{\psi_2} = \left(\mathbb{I}\otimes \ket{\vec{Y}}\bra{\vec{Y}}\right)  \sum_{\vec{x}} \left( \ket{\vec{x}}_{m}\ket{\gamma(\vec{x})}_{p}\right) = \sum_{\vec{x}} \left( \ket{\vec{x}}_{m}\ket{\vec{Y}}\braket{\vec{Y}|\gamma(\vec{x})}_{p}\right) = \ket{\vec{X}}_{m}\ket{\vec{Y}}_p.
\end{equation}
Now, if we measure the first register, we can only get the correct solution.

For optimization and constraint satisfaction problems we only use one register, which starts in uniform superposition
\begin{equation}
    \ket{\psi_0} = \bigotimes_{k=0}^{N-1} \left(\sum_{x_k=0}^{d_k-1} \ket{x_k}\right)=\sum_{\vec{x}}\ket{\vec{x}}.
\end{equation}
After that, we apply an operator that performs an imaginary time evolution~\cite{ITE}, having as Hamiltonian the cost function for each combination
\begin{equation}
    \ket{\psi_1} = \sum_{\vec{x}}e^{-\tau C(\vec{x})}\ket{\vec{x}}.
\end{equation}
Finally, we apply an $\mathcal{R}$ operator that applies the constraints of the problem, cancelling the amplitude of the states that do not satisfy them
\begin{equation}
    \ket{\psi_2} = \sum_{\vec{x}}\mathcal{R}e^{-\tau C(\vec{x})}\ket{\vec{x}} = \sum_{\vec{x}\in R}e^{-\tau C(\vec{x})}\ket{\vec{x}},
\end{equation}
being $R$ the subspace of combinations satisfying the constraints. Again, these are non-unitary operations not directly implementable in a quantum system. After this, the basis state with less associated cost is the most probable state to measure.

\subsection{Definition of the problem variables and rewriting of the functions}

The first step is choosing which variables are going to be optimized to solve the problem, and rewrite it according to these variables. Many problems may have different sets of variables such that solving the problem in one set returns the same solution as solving it in another ones. For example, in the traveling salesman problem we can choose as variables to optimize the vector $\vec{y}$ whose component $y_k$ indicates the time step in which we are at node $k$, or the vector $\vec{x}$ whose component $x_t$ indicates in which node we are at time step $t$. Both formulations are equivalent, but the former allows us to express the cost function, variable transformations and dependencies in the problem in a much simpler way.

Once the variables $\vec{x}$ to optimize have been chosen, we must rewrite the problem in function to these variables. In case of inversion problems, it will be necessary to determine which operations are performed on the input variables to obtain the output variables values. For example, if the problem is to determine two numbers of a set that added result in a certain value, it is possible to take as variables the bits of each number and do the binary addition process until obtaining the output number, also in binary variables. Something similar happens in constraint satisfaction problems, but this time we need to determine when a combination is unfeasible.

In the optimization problems the cost function must be written as a function that receives the values $\vec{x}$ and returns a number. This can be expressed simply as
\begin{equation}
    C(\vec{x}) = C_{x_0,x_1,\dots,x_{N-1}},
\end{equation}
being $C(\cdot)$ the cost function and $C$ its associated cost tensor.

However, to simplify the posterior implementation of the tensor network, the cost function should be expressed as the least-number of variables dependent cost operation. An example is the cost given as a QUBO
\begin{equation}\label{eq: QUBO}
    C(\vec{x}) = \sum_{i,j}Q_{i,j}x_i x_j.
\end{equation}

Another example is the Tensorial Quadratic Unconstrained D-ary Optimization (T-QUDO) formulation, which consists in expressing the cost as a sum of tensor elements of two indexes, these being the values of the variables,
\begin{equation}\label{eq: T QUDO cost}
    C(\vec{x}) = \sum_{k} C_{k,x_{a_k},x_{b_k}},
\end{equation}
where $a_k$ and $b_k$ are the identifiers of the first and second variable involved in the $k$-th term and $\vec{x}$ is a vector of natural values.

In the case of the traveling salesman problem~\cite{TSP_General}, if the variables $x_t$ are defined as the node where we are at time $t$, it can be expressed as
\begin{equation}
    C(\vec{x}) = \sum_{t} C_{x_t,x_{t+1}}.
\end{equation}

Defined the cost function, it is necessary to define the constraints. There are many ways to define the constraints, but a convenient one is to use auxiliary variables that indicate the activation or not of a certain condition. For example, in the traveling salesman problem the constraints are to end at the same node where we start and not to repeat any node. This is
\begin{equation}
\begin{gathered}
    x_0=x_{N},\\
    x_t \neq x_{t'}\ \forall t\neq t'.
\end{gathered}
\end{equation}
Another way of expressing it with auxiliary variables is as follows
\begin{itemize}
    \item $y_r = x_0 \Rightarrow x_N=y_r$ for ending at the start.
    \item $\forall i,t$ for each node $i$ and time step $t$.
    \begin{itemize}
        \item if $i\in \{x_0, x_1, \dots, x_t\} \Rightarrow y_{i,t} = 1$
        \item else $y_{i,t} = 0$
        \item if $y_{i,t} =1 \Rightarrow x_{t'} \neq i\ \forall t'>t$
        \item else $\exists t'>t\ |\ x_{t'}=i$
    \end{itemize}
\end{itemize}
where $y_{i,t}$ takes into account if the node $i$ has being visited in some step up to the step $t$.

\subsection{Creation of the associated classical logical circuit}
Once we have determined the variables that we will use to solve the problem, we have to build a classical 
logical circuit of the problem. The type of circuit to build depends on the type of problem to solve. These circuits make use of what we call \textit{internal signals}, internal information of the circuit, which is not part of the output, comes from some operators and conditions the action of others who receive it. The internal signal is the problem relevant information that is sent between operators. It is the only information they need to perform their operations correctly, and depends directly or indirectly on the problem input. In addition, we can interpret the inputs and outputs that connect to the outside as \textit{external signals}. For this reason, the construction method is called the \textit{Signals Method}. To understand how it works, we will start with the construction for inversion problems, then the CSP and finally the optimization problems.

\subsubsection{Inversion Problem}
For an inversion problem, we have to make a circuit that implements the function to invert, receiving the inputs $\vec{x}$ and returning the corresponding output $\vec{y} = \gamma(\vec{x})$. This can always be done, as it is a known function, making use of a classical logic circuit that transforms the information it receives. This can be implemented by means of fundamental logical operators or by means of more complex ones. We call this circuits \textit{Logical Signal Transformation Circuits} (LSTC), since each operator transforms its input signals into output signals using logical rules. These circuits also serve to solve the problem of calculating $\gamma(\vec{x})$, which we call \textit{forward problems}. Let us give a few examples to make this class of circuits easier to understand.

\paragraph{Sum of two numbers in binary}
$ $

The problem is, given a number $c$, to determine two numbers $a$ and $b$ such that $c=a+b$. That is, we want to invert the addition function. To do this, we use as variables the bits of the numbers. In this way, we will have to build the LSTC that performs the binary addition function.
\begin{figure}[h]
    \centering
    \includegraphics[width=0.7\linewidth]{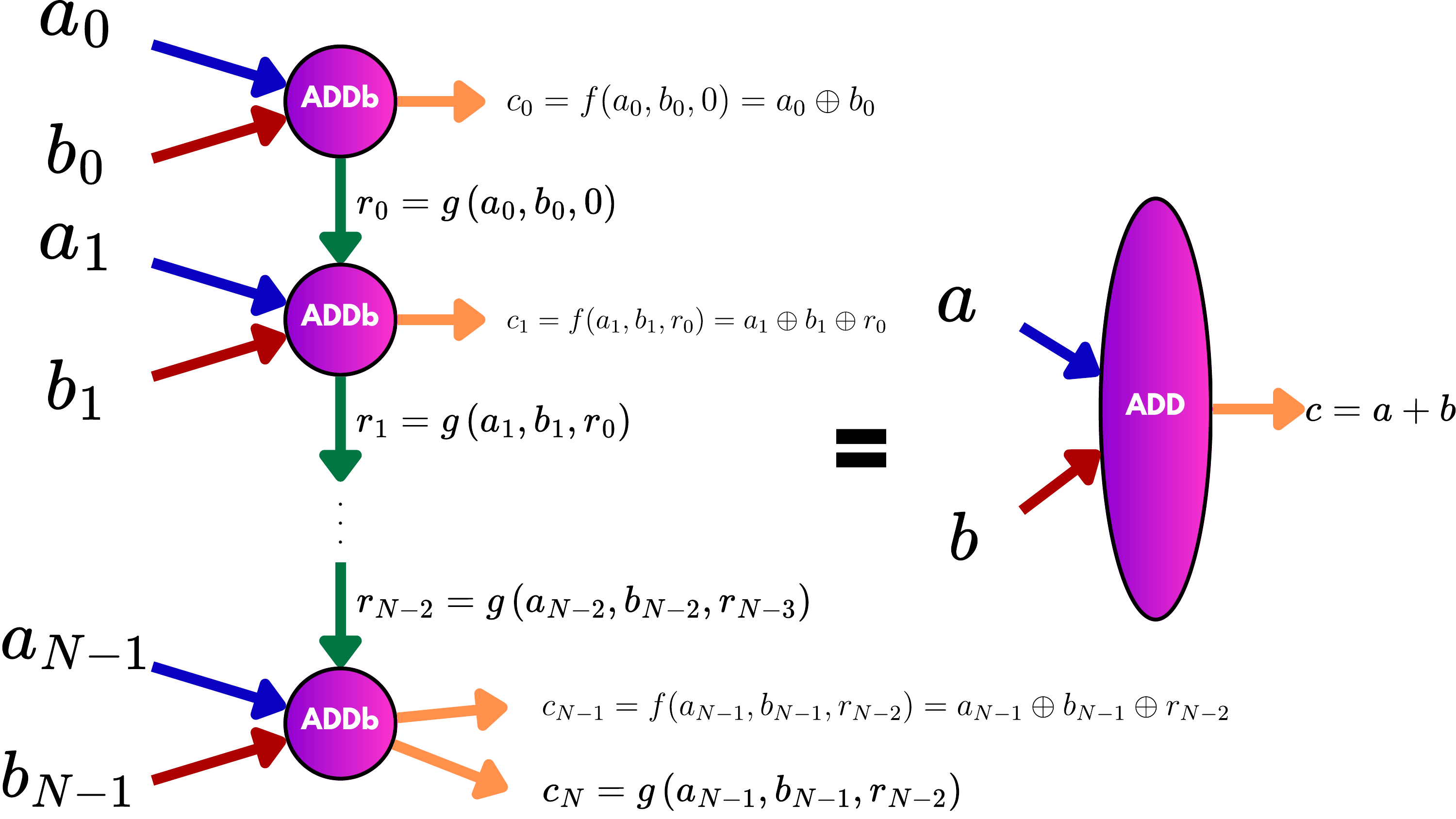}
    \caption{LSTC to add two numbers $a$ and $b$ to obtain a number $c$.}
    \label{fig: ADD circuit}
\end{figure}

The way to build this circuit is shown in Fig.~\ref{fig: ADD circuit}, where each pair of bits, one of $a$ and the other of $b$, enters in each logical operator $ADDb$, which are in charge of doing the part of the global sum corresponding to those bits. If each binary sum is performed in an $ADDb$ operator, we can make them return both the modular sum of the bits and the carry for the sum of the following bits. This carry information is sent in what we call the \textit{internal signal}.

The $ADDb$ operator has three inputs, which are the three bits to be added, and two outputs. The first output is the function $f(x,y,z) =x\oplus y \oplus z$, the modular sum of the 3 bits, marked in orange. The second output is the function $g(x,y,z)=\left\lfloor\frac{x+y+z}{2}\right\rfloor$, which outputs the carry, marked in green. The first output is part of the circuit output, but the second is internal information that is part of the input of the next $ADDb$ operator.
This circuit uses minimal amounts of information to communicate between its parts to obtain the final output, in addition to being small in size. These are two properties necessary for the resulting tensor network to be computable.

\paragraph{Multiply two numbers in binary}
$ $

In this problem we have a number $c$ and we want to obtain two numbers $a$ and $b$ such that $c=a\times b$. To do this, we have to make the LSTC that generates the multiplication of two numbers. We generate it based on conditional binary sums. For this, we have two different internal signals. The first one is the main signal, which keeps track of how much we have added up to a certain point, and that will be the circuit output at the end, and the second signal in each sum keeps track of the carry and the condition. That is, the main signal starts with the value $0$, and the circuit adds to it $b$ if $a_0$ is equal to $1$. That is, it adds $a_0b$.
\begin{figure}
    \centering
    \includegraphics[width=0.7\linewidth]{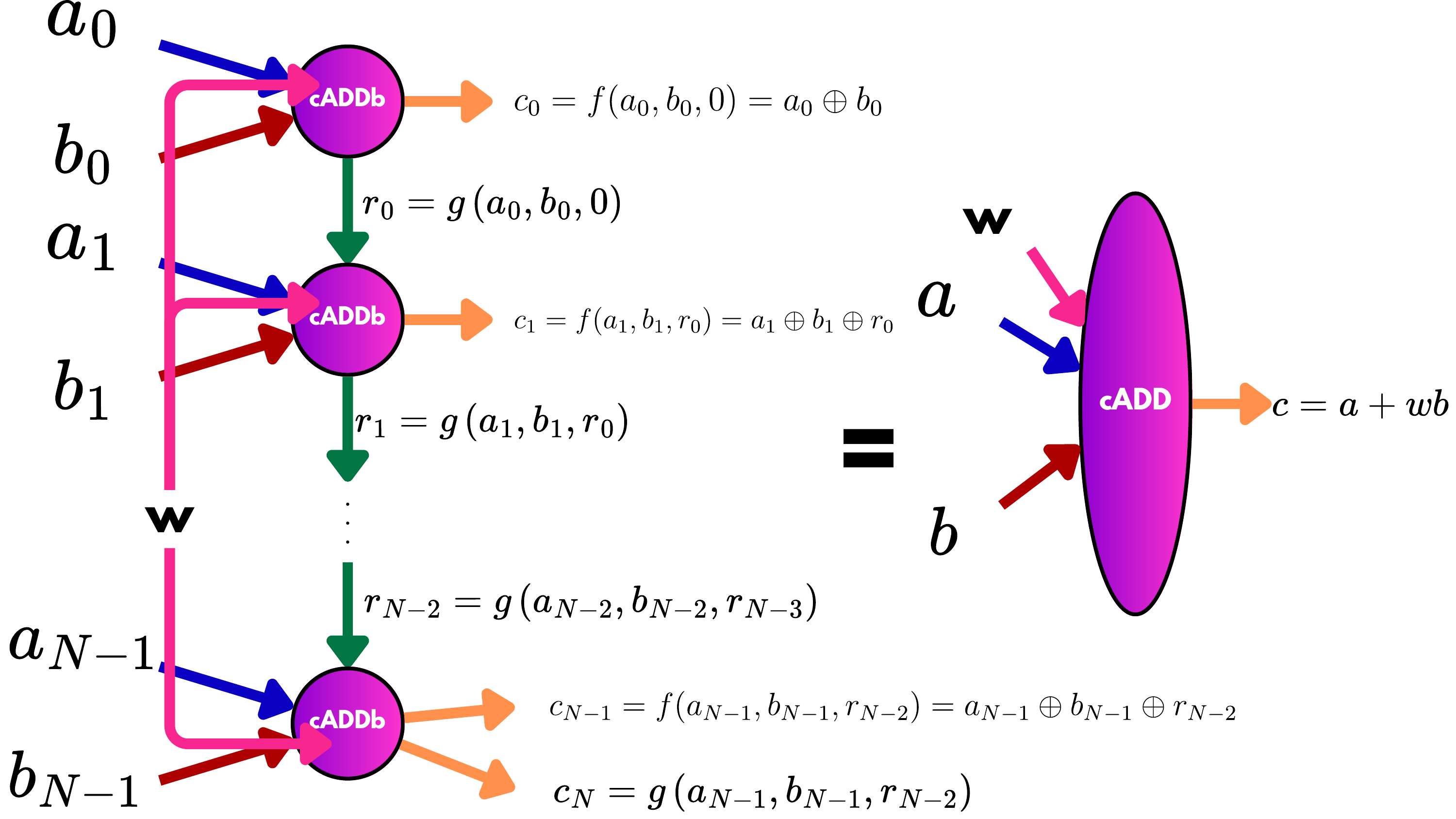}
    \caption{LSTC to add two numbers $a$ and $b$ to obtain a number $c$ if $w=1$.}
    \label{fig: cADD circuit}
\end{figure}

The $ADD$ circuit is similar as before, but now each operator has an extra input, which indicates if the addition is done or not. The $cADD$ circuit is in Fig.~\ref{fig: cADD circuit}. Thus, after doing the first conditional addition, the result is the input of another $cADD$ operator, which has to add the value $b$ multiplied by $2$, but this time if $a_1=1$. With this, the main signal is
\begin{equation}
    r_1 = a_0b + 2a_1b.
\end{equation}
Repeating this step $N$ times gives the signal
\begin{equation}
    r_{N-1} = \sum_{n=0}^{N-1} 2^n a_n b = a\times b=c.
\end{equation}

\begin{figure}
    \centering
    \includegraphics[width=0.7\linewidth]{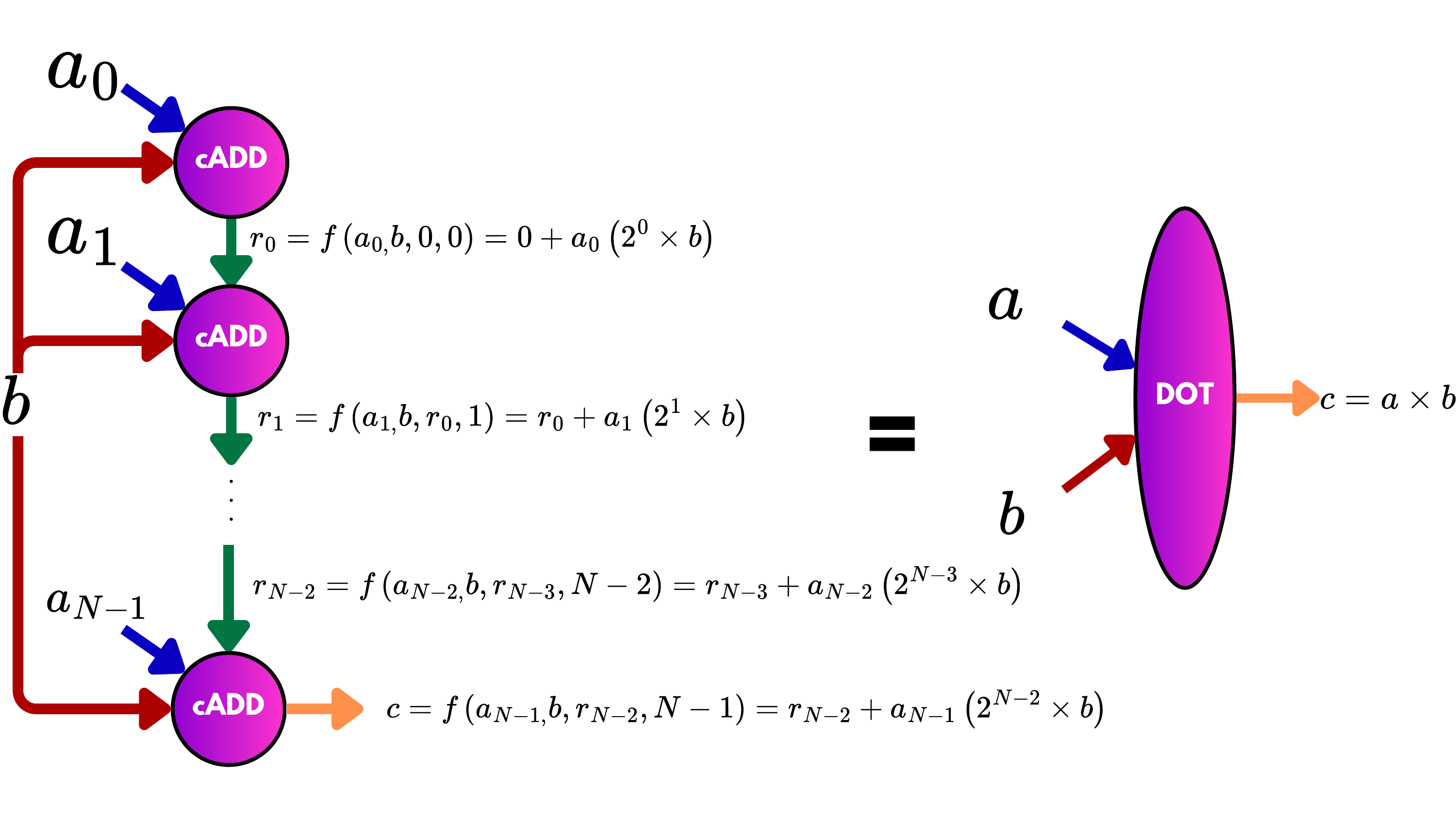}
    \caption{LSTC that performs the multiplication of two numbers $a$ and $b$.}
    \label{fig: DOT circuit}
\end{figure}
As we can see in the circuit in Fig.~\ref{fig: DOT circuit}, there is an \textit{intermediate state} that transforms until finally resulting in the final output. The intermediate state is a class of signal which is internal through the computation, and becomes external at the end of it.

\subsubsection{Constraint Satisfaction Problem}
In these problems the objective is that the circuit receives an input and returns the same value as output only if the combination satisfies the constraints. That is, if the input does not meet the constraints, we will not return any output. To do this, in addition to a set of internal and external signals, we have a value associated with the combination. This number is the \textit{amplitude} of the combination, which we will understand better in the section of optimization problems. Up to this point, it is only an internal number starting at $1$, and in case at some point it is detected that the input violates the constraints, an operator will change it to $0$.

Our circuit is composed of a set of operators that send a set of internal signals to each other, each one being in charge of analyzing a specific part of the input and detecting if any constraint is violated. We call this circuit \textit{Logical Signal Verification Circuits} (LSVC).

\paragraph{Single One Input}
$ $

This problem consists in finding a string of binary numbers such that only one of them is equal to $1$, and the rest are $0$.
\begin{figure}[h]
    \centering
    \includegraphics[width=0.7\linewidth]{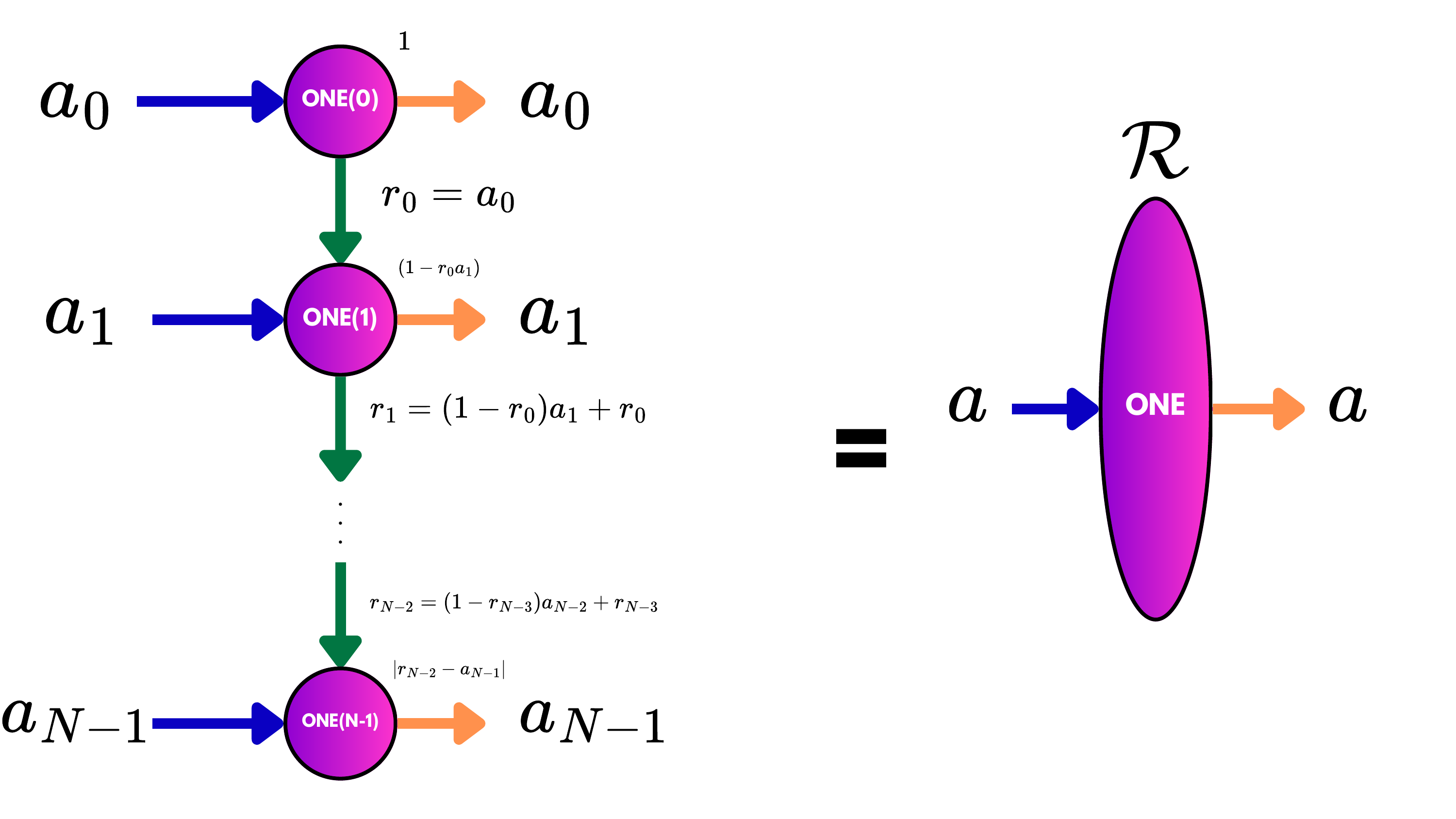}
    \caption{LSVC to determine if a number has only one bit on $1$.}
    \label{fig: One bit}
\end{figure}

To do this, the input to the circuit is the value of each bit, and each one enters in an $ONE$ operator, which determines whether its value is $1$ and how many $1$ have appeared up to that point. They have two inputs and two outputs. The first input is the value of its bit, and the second is the signal indicating how many $1$ have appeared in the combination so far. The first output is the value of its bit, and the second is the signal that tells how many $1$ have appeared in the combination so far, including the current bit. If the operator receives that $1$ has already appeared in the combination, then it will make the amplitude $0$ if the input signal of its bit is also equal to $1$, since there cannot be two $1$ in the combination. The last operator forces that, if it receives that no $1$ has appeared, the amplitude of the combination is $0$ if it does not receive a $1$ on its bit. The circuit is shown in Fig.~\ref{fig: One bit}.

\subsubsection{Optimization Problem}
For optimization problems we have to change the approach slightly. In these problems we are not looking for an output nor an input which only satisfies constraints, but rather each input has an associated cost that we can calculate, but generally do not need to know. We want the state with the lowest associated cost, which satisfies the restrictions. Therefore, we will create a logical circuit whose input and output are the same, but which has a number associated with its state. We can visualize it as an optical circuit that can receive waves at discrete frequencies with a certain amplitude. Then, the circuit, depending on the frequency of that wave, changes its amplitude. Thus, the output of the circuit is a wave of the same frequency, but with a different amplitude. For example, if the circuit receives the value $x$, the output will also be $x$, but the internal value of the state will be $f(x)$. We call this internal value \textit{amplitude}, in analogy to quantum computing. It is important to note that this circuit is NOT a quantum circuit, nor does it work on superposition. It is a classical circuit that, depending on what it receives, amplifies or reduces the amplitude of the internal state. We call these circuits \textit{Logical Signal Modulation Circuits} (LSMC), since each operator transforms the internal signals it receives to modulate the amplitudes of the inputs according to logical rules. 

Due to the properties of the tensors that we will explain later, the changes in amplitude can only be multiplicative. That is, each operator can only multiply the amplitude by a number. This may seem restrictive, but it is sufficient to tackle any problem. Due to the types of existing problems and this restriction, we choose that given an input $\vec{x}$, which is a solution combination, the circuit multiplies its amplitude (initially $1$) by $e^{-\tau C(\vec{x})}$, being $\tau$ a constant. In this way, a combination with higher cost has an associated amplitude exponentially smaller than one with lower cost. This process is called \textit{imaginary time evolution}. This also allows taking advantage of the exponential property
\begin{equation}
    \prod_{i} e^{a_i}= e^{\sum_i a_i}.
\end{equation}
In case of constrained optimization problems, the circuit will also implement the LSVC logics of the constraint satisfaction problems. We will understand it better with three examples.

\paragraph{Linear function}
$ $

This combinatorial optimization problem has a cost function
\begin{equation}
    C(\vec{x})=\sum_{i=0}^{N-1} a_i x_i
\end{equation}
for a set of $a_i\in \mathbb{R}$ values, where $\vec{x}$ is a vector of binary values. The exponential of the cost function can be expressed as
\begin{equation}
    e^{-\tau C(\vec{x})} = \prod_{i=0}^{N-1} e^{-\tau a_i x_i}.
\end{equation}

\begin{figure}
    \centering
    \includegraphics[width=0.7\linewidth]{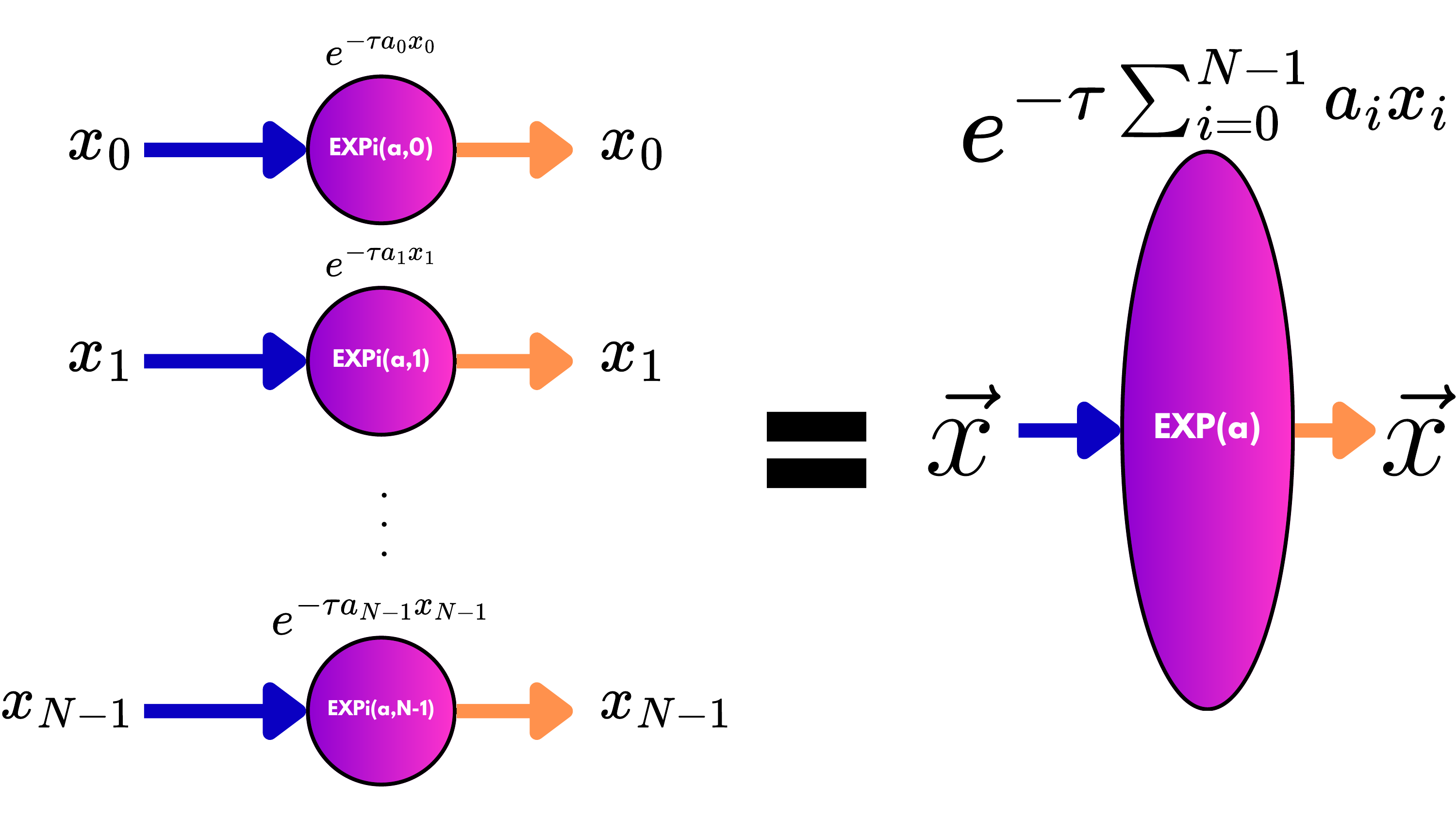}
    \caption{LSMC that multiplies the amplitude of an input $\vec{x}$ by $e^{-\tau \sum_{i=0}^{N-1} a_i x_i}$.}
    \label{fig: Linear circuit}
\end{figure}

As each product only depends on one variable, its LSMC can be expressed as in Fig.~\ref{fig: Linear circuit}. If we multiply the amplitude of each part of the input by a value, the amplitude of the global input is multiplied by the product of all these values. This property allows the circuit to make the amplitude contain information from all the input without the need to transmit all the information at the same time. In this case we have not needed signals between operators as in the case of the addition, but we are going to see a more complicated case.

\paragraph{Quadratic function with a single neighbor in a linear chain}
$ $

This combinatorial optimization problem has a cost function
\begin{equation}
    C(\vec{x})=\sum_{i=0}^{N-1} (Q_{i,i} x_i^2 + Q_{i,i+1} x_ix_{i+1}).
\end{equation}
As before, the exponential can be expressed as products of exponentials. In this case, each operator needs, in addition to the information of the variable that corresponds to it, the value of the previous variable in the chain. The signal that each operator gives to the next one is its variable state.
\begin{figure}
    \centering
    \includegraphics[width=0.7\linewidth]{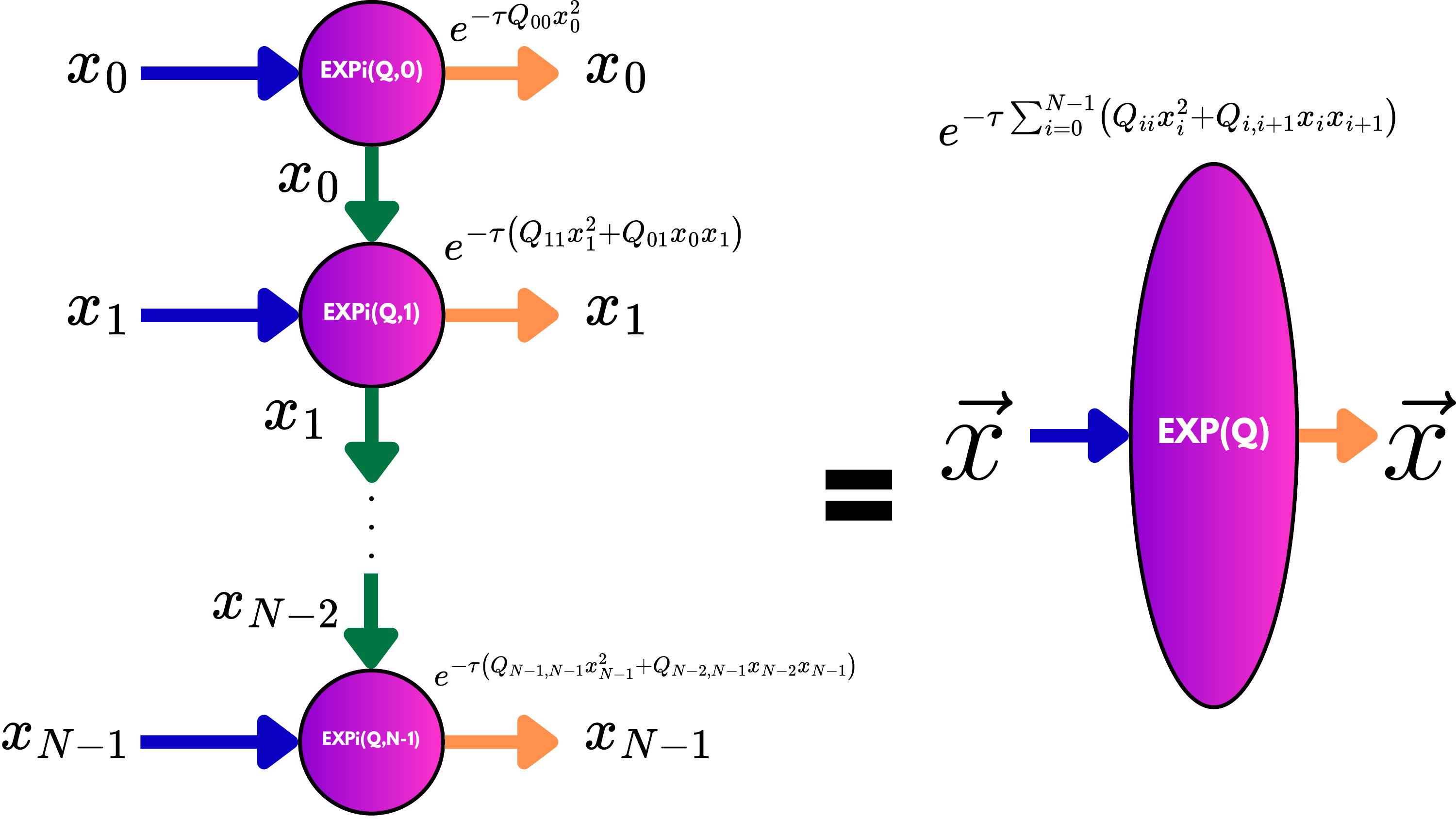}
    \caption{LSMC that multiplies the amplitude of an input $\vec{x}$ by $e^{-\tau \sum_{i=0}^{N-1} (Q_{i,i} x_i^2 + Q_{i,i+1}x_ix_{i+1})}$.}
    \label{fig: Quadratic one neighbor circuit}
\end{figure}

The LSMC is given by Fig.~\ref{fig: Quadratic one neighbor circuit}. As noted above, this circuit is simple, where each functional part depends on few elements, and has a small size.

\paragraph{Natural sum total function problem}
$ $

\begin{figure}
    \centering
    \includegraphics[width=0.7\linewidth]{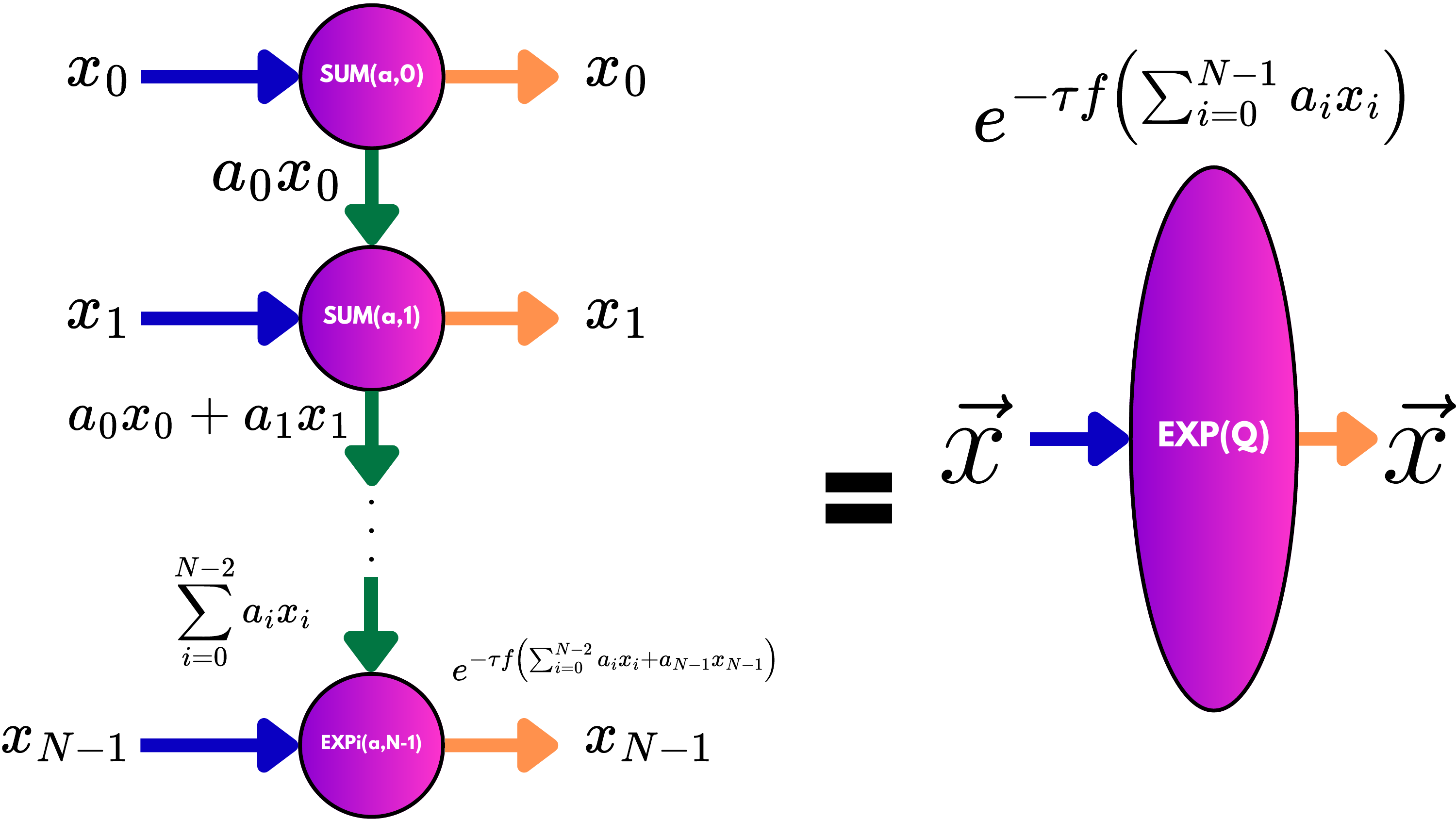}
    \caption{LSMC that multiplies the amplitude of an input $\vec{x}$ by $e^{-\tau f\left(\sum_{i=0}^{N-1} a_i x_i\right)}$.}
    \label{fig: Natural sum circuit}
\end{figure}

The combinatorial optimization problem has a cost function
\begin{equation}
    C(\vec{x})=f\left(\sum_{i=0}^{N-1} a_i x_i\right),
\end{equation}
where $a_i\in \mathbb{N}$ and $f(\cdot)$ is some known function. In this case, we can use as a signal the sum $\sum_{i=0}^{m-1} a_i x_i$ up to the $m$-th variable, so that the last operator only does the evolution on the application of $f(\cdot)$ on the signal. This makes the last one operator the only one that multiplies the amplitude. Each previous operator only adds to the signal a value $a_i$ multiplied by its input. In this way, the LSMC is given by Fig.~\ref{fig: Natural sum circuit}.

\subsection{Creation of the associated logical tensor network}
After understanding the three types of circuits, we have to build the tensor network associated with these circuits. The key of this tensor network is that, given its structure, it is possible to take all possible inputs at once, and return all possible corresponding outputs, with their associated amplitudes. This is because, when tensorizing, the inputs and outputs of the operators become the basis states of tensors. This allows to apply a superposition of all possible inputs, generating their corresponding outputs, and thus propagating the signals through the circuit by means of their entanglement.

Since we have all the possible outputs, in the inversion problems we only have to force the output to be the one we know. That is, having the circuit, we  put as `inverse input' the desired output, making the input of the circuit only the one which generates that output. This will result in a tensor in which the only non-zero elements are in the input values we are looking for. In cases of constraint satisfaction, we have a tensor network that represents a diagonal tensor. The only non-zero elements of this tensor will be those that satisfy the imposed constraints. In the case of optimization problems we have a similar phenomenon. As we have all the possible inputs with their amplitudes associated to their costs, we have a diagonal tensor where each element has the amplitude of that combination, so we only have to look for the one with the highest amplitude.

In both cases, the only thing we have to do is to change each operator of the circuit by a tensor with as many indexes as inputs and outputs the operator has. The values of the output indexes for the non-zero elements depend on the values of the input indexes, following the equations of the outputs of the associated operator. The values of the non-zero elements are the amplitudes of the corresponding operators, which in the case of inversion and contraint satisfaction are always 1. The inputs to the circuit are converted for each of the variables into a vector of ones, which will express the uniform superposition of all possible values of that variable. All tensors are connected to each other in exactly the same way as in the associated circuit, by the same indexes in the same way. We call this process \textit{Circuit Tensorization} (CT), and the resulting tensor network is the \textit{Tensor Logical Circuit} (TLC).

Translated into equations, it means that if we have an operator $U$ with 3 inputs $x,y,z$ and 2 outputs $\mu, \nu$, calculated as $\mu=f(x,y,z),\ \nu=g(x,y,z)$, which multiplies the amplitude of the state by $h(x,y,z)$, then its associated tensor $U$ has as non-zero elements those that satisfy
\begin{equation}
    \begin{gathered}
        \mu=f(x,y,z),\ \nu=g(x,y,z),\\
        U_{x,y,z,\mu,\nu}= h(x,y,z).
    \end{gathered}
\end{equation}
We call this process \textit{Input-Output Indexing} (IOI). As we can see, the relation between the inputs and outputs of each operator implies a constraint on the tensor representing it.

It is important to note that at this point, after creating the TLC of the problem, we are not yet going to contract it. This is because if we were to contract it at this point, we would have a tensor that collects the amplitude for all possible combinations, which is not our objective. The tensor network to be contracted is the one mentioned in the next subsection. We now present a few examples of CT.

\subsubsection{Inversion Problems}
In this case, each tensor transforms the position of the elements of the input $\vec{x}$ to the positions of the elements of the output $\vec{y}$, but not its amplitude. That is, we apply a function on the indexes of the tensor, and not on the values of the elements themselves. There is a function $\gamma$, which we do not need to know explicitly, represented through the LSTC, which receives the input and returns the output. That is, $\vec{y}=\gamma(\vec{x})$. Thus, the tensor network that replaces the circuit is such that, when contracted, it results in a $T$ tensor of elements $T_{x_0,x_1,x_2,\dots, y_0, y_1, y_2, \dots}=1$ only if $\vec{y}=\gamma(\vec{x})$, for all possible values of $\vec{x}$, and otherwise equals $0$. 

In order to force the output to be the correct one $\vec{Y}$ to get the correct input $\vec{X}$, we have to project the tensor on the subspace on that the last indexes of it are $\vec{y}=\vec{Y}$. This is equivalent to performing the contraction operation
\begin{equation}
    T'_{x_0,x_1,x_2,\dots} = \sum_{\vec{y}} T_{x_0,x_1,x_2,\dots, y_0, y_1, y_2, \dots}\delta^{Y_0}_{y_0}\delta^{Y_1}_{y_1}\delta^{Y_2}_{y_2}\dots
\end{equation}
being $\delta^{b}$ a vector of all zero elements except one equals to $1$ at position $b$. To do this, we put in each $k$-th output line a $\delta^{Y_k}$ vector. This causes the only non-zero element of the $T'$ tensor to be $T'_{X_0,X_1,X_2,\dots}=1$, for all the inputs $\vec{X}$ that generate the output $\vec{Y}$.

\paragraph{Sum of two numbers in binary}
$ $

\begin{figure}
    \centering
    \includegraphics[width=0.7\linewidth]{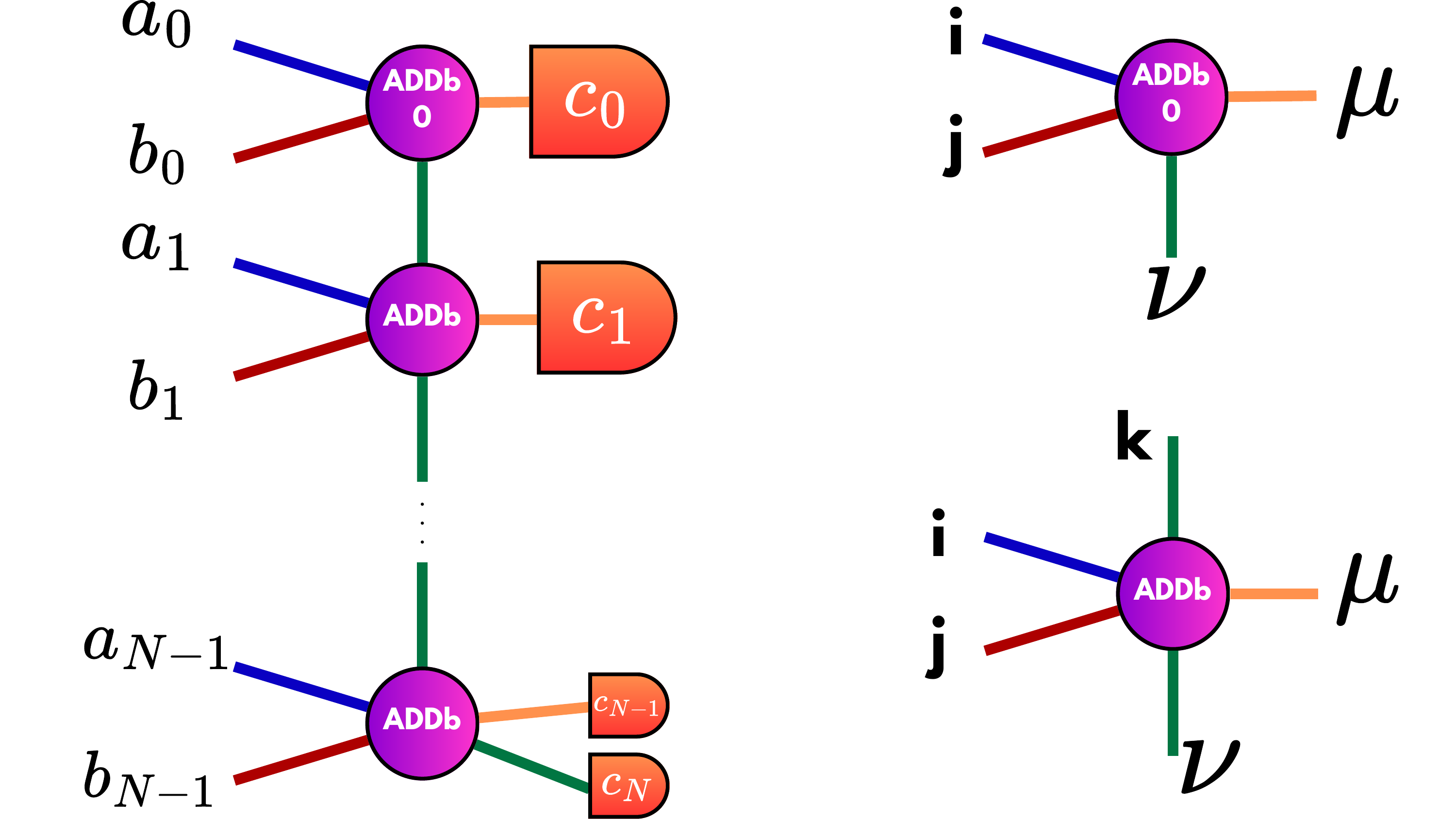}
    \caption{TLC of Fig.~\ref{fig: ADD circuit} and its index correspondence for the tensors.}
    \label{fig: TN ADD}
\end{figure}

In this case, each operator $ADDb$ must be replaced by its corresponding tensor. The TLC, and the names of the indexes are shown in Fig.~\ref{fig: TN ADD}. Note that the first tensor is different from the others, since it has no index to tell the previous carried, since it is always 0. In this case, the initial $ADDb0$ tensor is a 4-index tensor of $2\times 2\times 2\times 2$ dimensions with non-zero elements $ADDb0_{ij\mu\nu}=1$ when the following is true
\begin{equation}
\begin{gathered}
    \mu = f(i,j,0) =i\oplus j,\\
    \nu = g(i,j,0) = \left\lfloor\frac{i+j}{2}\right\rfloor.
\end{gathered}
\end{equation}

For the rest of the tensors $ADDb$, these have 5 indexes of $2\times 2\times 2\times 2 \times 2$ dimensions, whose non-zero elements $ADDb_{ijk\mu\nu}=1$ are those that satisfy
\begin{equation}
\begin{gathered}
    \mu = f(i,j,k) =i\oplus j \oplus k,\\
    \nu = g(i,j,k) = \left\lfloor\frac{i+j+k}{2}\right\rfloor.
\end{gathered}
\end{equation}

Finally, the vectors $c_b$ are those with their non-zero element at position $c_b$. 

If we contract this tensor network, the resulting tensor will have its nonzero elements in the positions in which its indexes are those corresponding to the bits of the $a$ and $b$ that generate as a result the $c$ that we have imposed.

\subsubsection{Constraint Satisfaction Problem}
In this case, the tensor network represents a diagonal tensor, in which all non-zero elements are equal to 1 when its indexes indicate a solution compatible with the constraints. Since the tensor is diagonal, we can eliminate half of the indexes, which are going to be a repetition of the other half, so we add a set of ones vectors in each output. This tensors are called the \textit{Plus Vectors} or `+' tensors. Thus, the tensor represented by the tensor network is
\begin{equation}
    T_{x_0,x_1,x_2,\dots} = 1\ \forall \vec{x} \in R.
\end{equation}

\paragraph{Single One Input}
$ $

\begin{figure}
    \centering
    \includegraphics[width=0.7\linewidth]{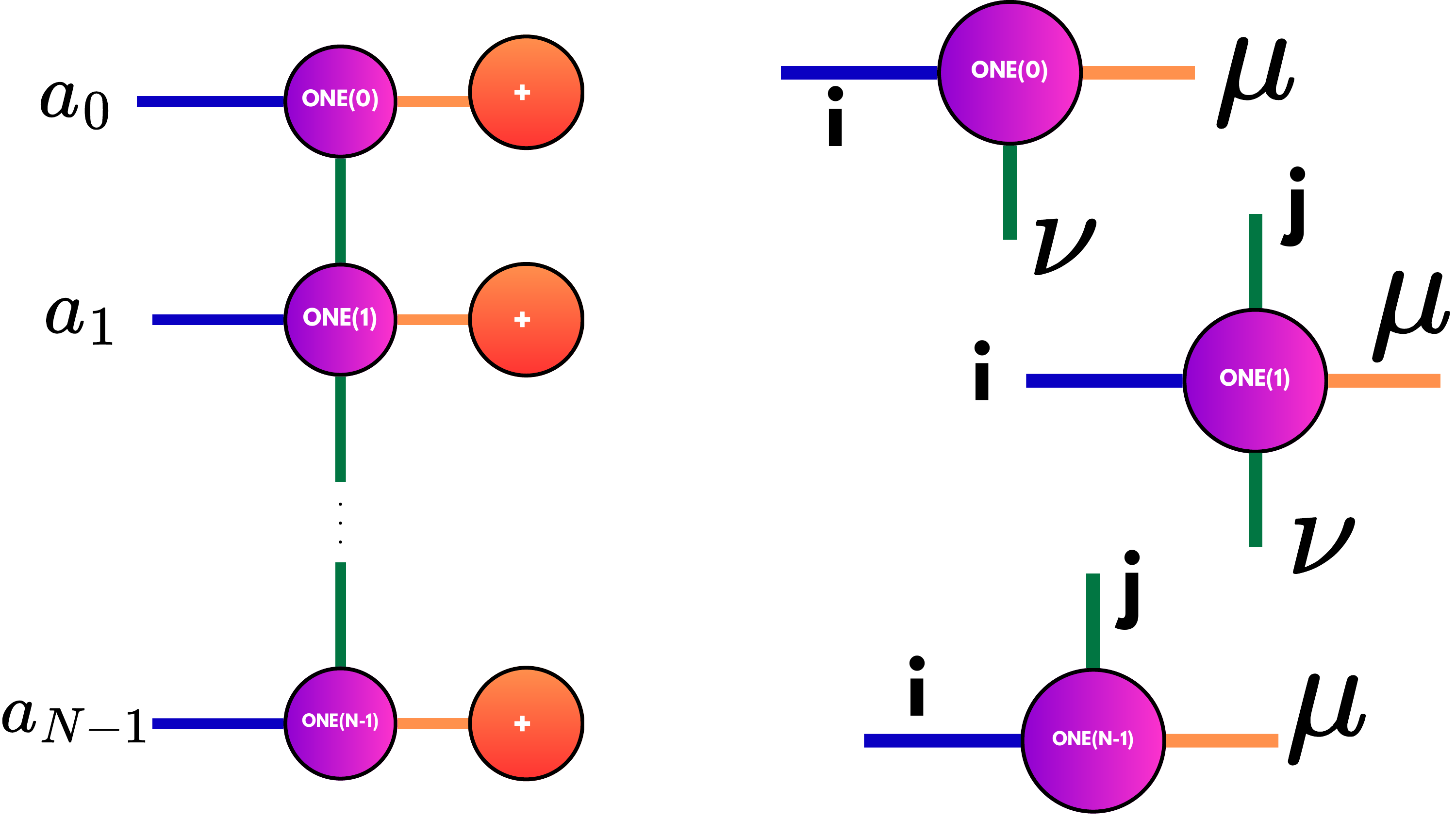}
    \caption{TLC of Fig.~\ref{fig: One bit} for the Single One Input problem.}
    \label{fig: One bit TN}
\end{figure}

In this case, each $ONE$ operator is replaced by the corresponding $ONE$ tensor to obtain the tensor network in Fig.~\ref{fig: One bit TN}. The non-zero elements of these tensors are
\begin{equation}
\begin{gathered}
    \mu = \nu =i,\\
    ONE(0)_{i\mu\nu}=1,
\end{gathered}
\end{equation}
\begin{equation}
\begin{gathered}
    \text{if } j=0\Rightarrow \nu = i,\quad\text{ else } \nu=j,\ i=0,\\
    \mu =i,\\
    ONE(k)_{ij\mu\nu}=1,
\end{gathered}
\end{equation}
\begin{equation}
\begin{gathered}
    i = j\oplus 1,\quad\mu =i,\\
    ONE(N-1)_{ij\mu}=1.
\end{gathered}
\end{equation}

\subsubsection{Optimization problem}
As before, in optimization problems we do not want to impose a specific output. Our first objective is to create a tensor whose non-zero elements are those in which the input equals the output and whose values are the exponentials of their associated costs. That is, to perform an imaginary time evolution for a diagonal operator. After that, as the input and output information is the same, resulting redundant, we add a set of Plus Vectors in each output. This is equivalent to allowing any output, so nothing change in the problem. Each tensor that replaces an operator has to perform the multiplication of the amplitude that the operator performed. To do this, the element of the tensor associated to the corresponding input and output indexes has to have the value by which the operator multiplied the amplitude. As the contractions multiply the values of the tensor elements, this do the process we imposed on the circuit.

By doing this, the tensor $T$ represented by this tensor network has elements
\begin{equation}
    T_{x_0,x_1,x_2,\dots} = e^{-\tau C(\vec{x})}.
\end{equation}
In case of having constraints in the problem, we only have to add after the optimization circuit, a constraint circuit, so that the tensor represented will be
\begin{equation}
    T_{x_0,x_1,x_2,\dots} = e^{-\tau C(\vec{x})} \ \forall \vec{x} \in R.
\end{equation}

\paragraph{Linear problem}
$ $

\begin{figure}
    \centering
    \includegraphics[width=0.7\linewidth]{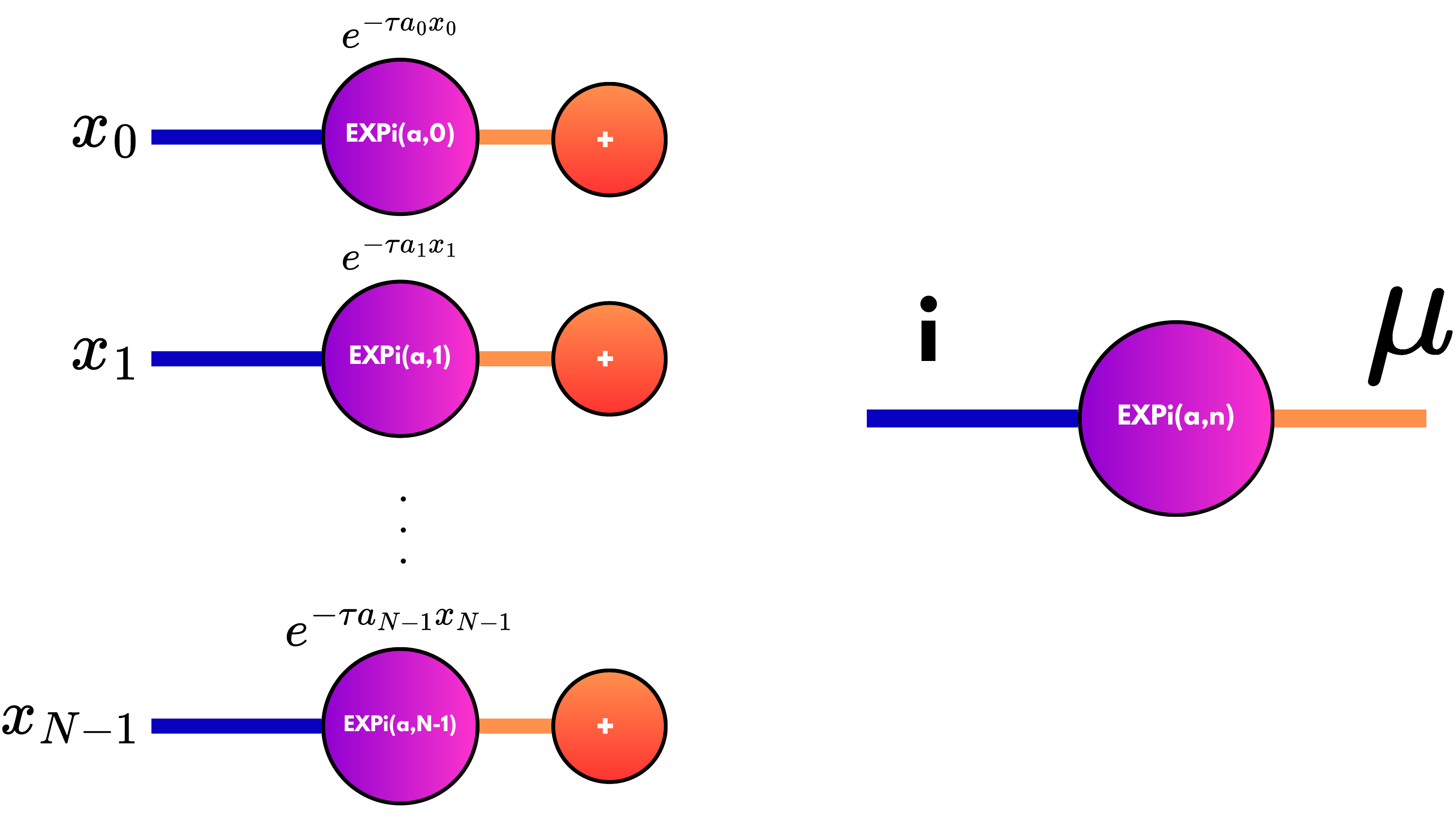}
    \caption{TLC of Fig.~\ref{fig: Linear circuit} and its index correspondence for the tensors.}
    \label{fig: Linear TN}
\end{figure}

In this case, we only have matrices of dimension $2\times 2$, represented in Fig.~\ref{fig: Linear TN}, whose non-zero elements are
\begin{equation}
\begin{gathered}
    \mu=i,\\
    EXPi(a,n)_{i,\mu} = e^{-\tau a_n i}.
\end{gathered}
\end{equation}

\paragraph{Quadratic function with a single neighbor in a linear chain}
$ $

\begin{figure}
    \centering
    \includegraphics[width=0.7\linewidth]{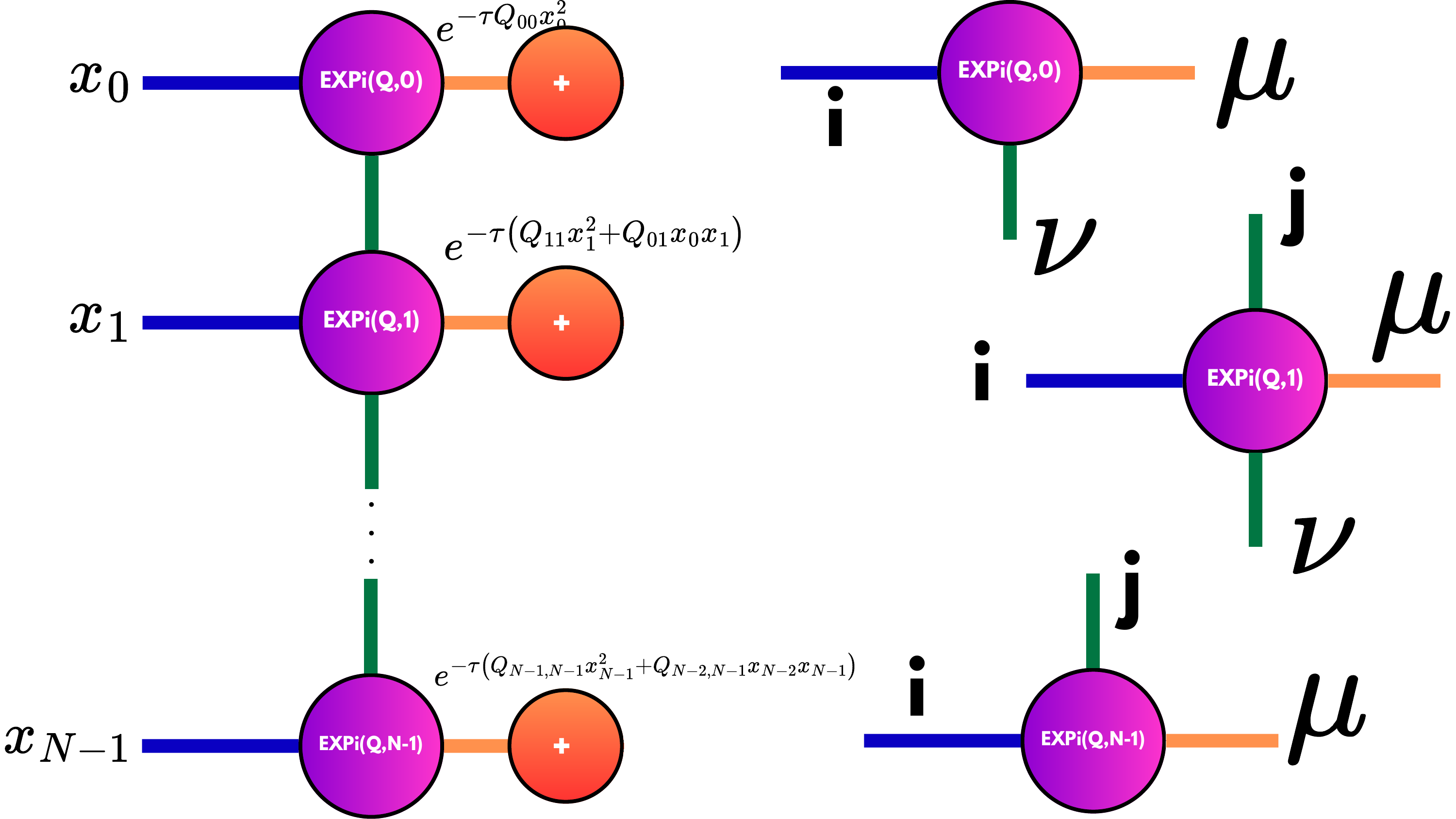}
    \caption{TLC of Fig.~\ref{fig: Quadratic one neighbor circuit} and its index correspondence for the tensors.}
    \label{fig: Quadratic TN}
\end{figure}

Unlike the previous case, now we have tensors of 3 and 4 indexes, divided into 3 groups: the initial tensor, the intermediate tensor and the final tensor. The tensor network is represented in Fig.~\ref{fig: Quadratic TN}. All have dimension $2$ in their indexes, with non-zero elements
\begin{equation}
    \begin{gathered}
        \mu=\nu=i,\\
        EXPi(Q,0)_{i\mu\nu}=e^{-\tau Q_{00}i},\\
        EXPi(Q,n)_{ij\mu\nu}=e^{-\tau (Q_{n,n}i^2+Q_{n-1,n}ij)},
    \end{gathered}
\end{equation}
\begin{equation}
    \begin{gathered}
        \mu=i,\\
        EXPi(Q,N-1)_{ij\mu}=e^{-\tau (Q_{N-1,N-1}i^2+Q_{N-2,N-1}ij)}.
    \end{gathered}
\end{equation}

\paragraph{Natural sum total function problem}
$ $

\begin{figure}
    \centering
    \includegraphics[width=0.7\linewidth]{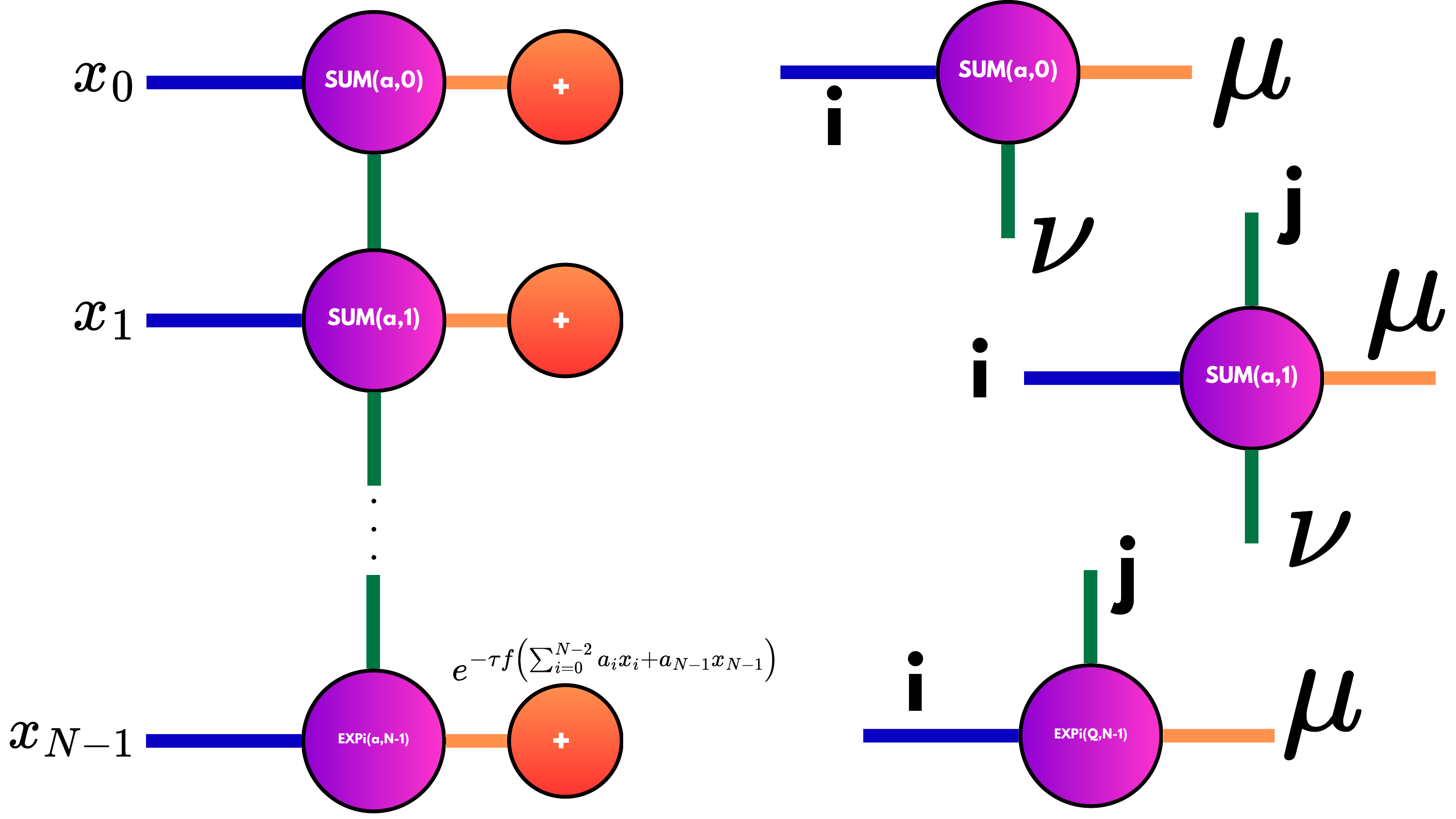}
    \caption{TLC of Fig.~\ref{fig: Natural sum circuit} and its index correspondence for the tensors.}
    \label{fig: Natural sum TN}
\end{figure}

This case is similar to the previous one, but now the tensors have for their upper and lower indexes a dimension that allows sending any of the possible partial sums. Therefore, the $n$-th tensor has for its upper index a dimension of $\sum_{i=0}^{n-1}a_i$, while for the lower one a dimension of $\sum_{i=0}^{n}a_i$, and for the side ones a dimension of $2$. In the tensor network expressed in Fig.~\ref{fig: Natural sum TN} we have 3 types of tensors, whose non-zero elements are
\begin{equation}
    \begin{gathered}
        \mu=i,\quad \nu=a_0 i,\\
        EXPi(Q,0)_{i\mu\nu}=1,
    \end{gathered}
\end{equation}
\begin{equation}
    \begin{gathered}
        \mu=i,\quad \nu=j+a_n i,\\
        EXPi(Q,n)_{ij\mu\nu}=1,
    \end{gathered}
\end{equation}
\begin{equation}
    \begin{gathered}
        \mu=i,\\
        EXPi(Q,N-1)_{ij\mu}=e^{-\tau f(j+a_{N-1}i)}.
    \end{gathered}
\end{equation}

\subsection{Iteration and contraction of the tensor network}
Once we have the tensor network that gives us the result tensor we search, we need to extract the relevant information from it without having to store it in memory. That is, we want to somehow be able to look at a reduced version of it that gives us the information we really need to get the solution. To do this, we are going to determine the correct value of each variable iteratively. Therefore, in each iteration we only want to know what is the value of the $n$-th variable in the optimal combination. To do this we  perform an integral over all the other variables, so that we only have to find the maximum in a vector of exponentially smaller dimension, and that information is included in the next iteration. We call this process \textit{Half Partial Trace}.

This can be visualized as a process similar to that performed when measuring a quantum state. We measure the qubits in order, so that the state of all the qubits that remain to be measured conditions the probability of obtaining a result on the one we are measuring, and the result of the already measured fix and alter the following probabilities in a fixed way. In our case, we do not use the amplitudes in the same way as in quantum mechanics, but the amplitudes are our `probabilities'. Instead of using the density matrix $\rho_{AB}$ of the $|\psi_{AB}\rangle$ state and performing an operation $P_{A,i} = Tr_{B}(\rho_{AB} M_i)$, we apply directly $P_{A,i} = \langle i,+^{\otimes N-1}|\psi_{AB}\rangle$. In~\cite{Escanez_Notation} simplified notation, $P_{A,i} = {}^{0}_{i}\langle +|\psi_{AB}\rangle$.

With this in mind, we will see it more clearly first in the inversion case and then in the optimization case.

\subsubsection{Inversion problem}
For simplicity, we begin by addressing the case in which there is only one solution to the problem. In this case, there is only one input $\vec{X}$ that results in that output $\vec{Y}$. This implies that there is only one non-zero element in the tensor represented by the TLC we have constructed, whose indexes give us the solution. Therefore, we can create a vector of dimension equal to that of the first index of the tensor, that is, of the number of possible values of the first variable. In the $k$-th component of this vector we store the sum of all the elements of the tensor whose first index has the value $k$. Although this operation seems computationally very expensive, it can be performed by putting a Plus Vector in all the indexes except the first one. Since there is only one correct solution, there is only one non-zero element in the tensor, so only one of these sums has a non-zero summand. This makes the position of the non-zero element of the vector match the value of the first index for the non-zero element of the global tensor.

Therefore, to determine the first variable we create the TLC and we connect Plus Vectors in all the indexes except the first one, and the value of the correct variable is the position with a 1 in the vector. To determine the second variable we can do exactly the same, but this time putting Plus Vectors in all the indexes except the second one. We can repeat this process for each variable and we will obtain the solution to the problem.

\begin{figure}
    \centering
    \includegraphics[width=\linewidth]{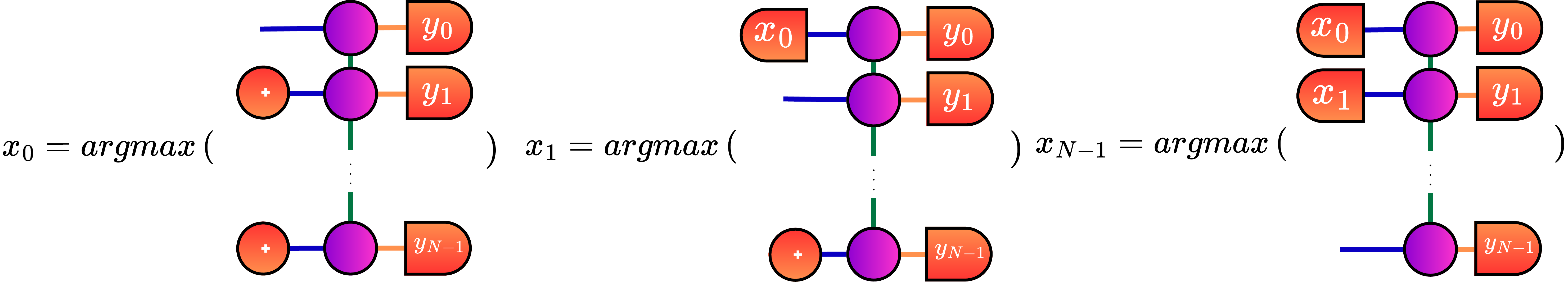}
    \caption{Iterative method for the determination of the solution variables in an inversion problem, for a chain-type tensor network.}
    \label{fig: Inversion iteration}
\end{figure}

To generalize to problems in which there are several solutions, we have to take into account that the tensor may have more than one non-zero value. However, our reasoning is the same. In case we want one of the solutions, we only have to take in each iteration as the correct value of the variable the position of any of the non-zero components of the vector, since it indicates that there is a solution that has that value for that variable. One possibility is to choose the position with the largest element. However, since there are several solutions, performing the process exactly the same can lead to mixing several degenerated solutions, resulting in an incorrect result. To avoid this problem, we only need to introduce in each iteration the information of all previous results. That is, in iteration $n$ we have already determined the previous $n-1$ variables, so we can project the tensor to the solutions with the first $n-1$ indexes equal to the values already determined. We can do this just as in the projection onto the output, with vectors having only one non-zero element at the position corresponding to the value we want to impose. We call this \textit{Projection Vectors}. In this way we do not mix solutions. An example of the operation of this method is shown in Fig.~\ref{fig: Inversion iteration}.

In case the variables are binary, instead of evaluating a vector we can evaluate a scalar. If we have only one solution, there are only two possible vectors for a variable: $(1,0)$ if the value has to be $0$ or $(0,1)$ if the value has to be $1$. Therefore, we can connect to the variable we are going to measure a vector $(-1,1)$, which will subtract to the amplitude associated to the value $1$ the amplitude associated to the value $0$. We call this vector \textit{Minus Vector}. If the correct value is $1$, the resulting scalar will be $1$, while if it is $0$, the scalar will be $-1$. In case of having several solutions, we choose $1$ if the scalar is positive or $0$ and $0$ if it is negative.

In the case of not being binary variables, we can perform the same process if we binarize the variable to be determined at the beginning of the circuit. That is, if we do a splitting of the free index so that we have $\log_2(d)$ binary indexes, we can determine the correct value of each of them as if they were different variables. We can also do the equivalent by placing a vector that performs the corresponding additions and subtractions between the groups.
Therefore, the resolution can always be expressed as an equation of the type
\begin{equation}
    x_i = H(\Omega_i),
\end{equation}
being $\Omega_i$ the scalar resulting from contracting the tensor network of the $i$-th variable, and $H$ the Heaviside step function. This way, if $\Omega_i<0$, $x_i=0$, else $x_i=1$. As the value to be introduced in the next iteration of the tensor network will be dependent on the previous one, through the projection tensor to the obtained result, the equation that determines the solution of the combinatorial problem (both the inversion problem and any other) is given by a nesting of Heaviside step functions within tensor networks. The constraint satisfaction problems follow the same mechanism.

\subsubsection{Optimization problems}
\begin{figure}
    \centering
    \includegraphics[width=\linewidth]{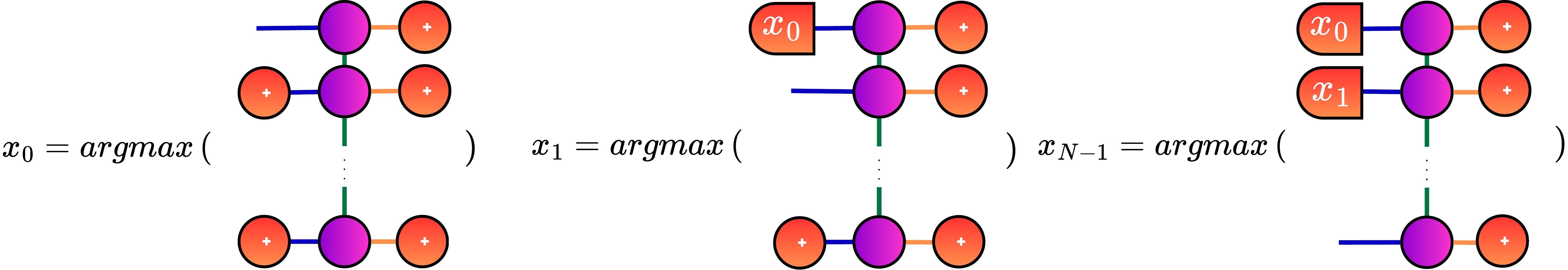}
    \caption{Iterative method for the determination of the solution variables in an constraint satisfaction and optimization problem, for a chain-type tensor network.}
    \label{fig: Diagonal iteration}
\end{figure}

For the optimization cases the process is the same as described above, but with a completely different motivation. To begin with, we have to visualize the amplitude map for all possible combinations. Since we have chosen as the amplitude for each combination the negative exponential of its cost multiplied by a constant, the combination with the lowest cost will have the largest amplitude. If we increase the value of $\tau$, the amplitudes of the suboptimal combinations will decrease exponentially faster than the amplitude of the optimal combination. Thus, in the limit $\tau\rightarrow\infty$, if we renormalize the tensor by dividing each element by the sum of all the elements of the tensor, only the amplitude of the optimal combination $\vec{X}$ will remain, exactly as in the inversion case. The limit is
\begin{equation}
    \lim_{\tau\rightarrow\infty} \sum_{\vec{x}\in R}\frac{e^{-\tau C(\vec{x})}}{\sum_{\vec{x}\in R}e^{-\tau C(\vec{x})}}\ket{\vec{x}}=
    \lim_{\tau\rightarrow\infty} \sum_{\vec{x}\in R}\frac{e^{-\tau C(\vec{x})}}{e^{-\tau C(\vec{X})}}\ket{\vec{x}}=
    \lim_{\tau\rightarrow\infty}\sum_{\vec{x}\in R}e^{-\tau (C(\vec{x})-C(\vec{X}))}\ket{\vec{x}} = \ket{\vec{X}}.
\end{equation}
In case of having degeneracy, we will have the case of several solutions and we solve it as presented for inversion problems. The method is presented in Fig.~\ref{fig: Diagonal iteration}.

However, it is not necessary to go to the infinite limit to extract information. For a sufficiently large finite value of $\tau$, the peak amplitude in the optimal combination will be large enough so that, when summing over the other variables values to obtain the vector of amplitudes of the variable we want to determine, this amplitude will be greater than the sum of all the suboptimals. This can be seen analogously to the case of measuring a quantum system in which there is a basis state with a probability much higher than the other basis states. Since the probability of measuring that state is higher, when measuring the first qubit of the system it will be more likely to measure the first bit of the state of maximum probability, and so on with all the qubits. Therefore, with a sufficiently large value of $\tau$, this same procedure is valid if we choose in each iteration as correct the position with the largest element. However, in case of not taking a sufficiently large $\tau$ value, the suboptimal states for an incorrect value of the variable to be determined may have a sum of amplitudes greater than that of the correct value.

To dampen this error, we can replace the sum by a complex sum, changing the Plus Vectors to vectors of complex numbers of unit modulus. We call them the \textit{Phase Vectors}. In this way, the `noise' generated by the amplitudes of the suboptimal ones will not sum in the same direction. Since the amplitude of the optimal combination is always the highest, it is the dominant value in their sum. If we distribute the phases of the numbers evenly, we decrease the probability that all the suboptimal ones are in the opposite phase to the optimum, so they will cancel each other. We call this method \textit{Humbucker}, and it was presented for the first time in~\cite{QUBO_Tridiagonal}.

\subsection{General equation}
In the previous subsections we have presented the method and demonstration of how to create the tensor network that solves the problem, and extracting the solution from it. Now, we can formulate the exact equation that solves any combinatorial problem.

\begin{theorem}
    Given a combinatorial problem, be it an inversion problem, a constraint satisfaction problem or an optimization problem, there is an exact explicit equation for its solution (or solutions).
\end{theorem}

\begin{theorem}
    Given a combinatorial problem, be it an inversion problem, a constraint satisfaction problem or an optimization problem, the exact explicit equation that solves it can be obtained in a polynomial time with respect to the time needed to formulate it.
\end{theorem}

\begin{theorem}
    Due to the symmetries of tensor networks, there are infinitely many equations that solve a combinatorial problem, all equivalent to each other.
\end{theorem}

This equation is obtained by following the steps listed in this section. The steps are
\begin{enumerate}
    \item Choose a set of variables $\vec{x}$ for the problem, which will encode the solution, input and/or output. For simplicity, and without loss of generality, the variables are considered to be binary, although it is general for natural variables.

    \item Build the logical circuit corresponding to the problem. If it is an inversion problem, the LSTC, if it is a constraint satisfaction problem, the LSVC, and if it is an optimization problem, the LSMC. 

    \item Tensorize the logical circuit to obtain the TLC of the problem.

    \item Perform the half partial trace of the TLC, adding the corresponding Minus Vector for the first variable, which we will call $x_0$. This tensor network is equal to the scalar value ${\color{blue} \Omega_0}$. Thus, the correct value of the first variable is
    \begin{equation}\label{eq: first variable}
        x_0={\color{blue} H(\Omega_0)}.
    \end{equation}

    \item To determine the next variable, the half partial trace of the TLC is made by adding now the corresponding Minus Vector for the second variable. This tensor network is a function of the value already determined for $x_0$, since the input index for that variable now has a projection vector at that value. Therefore, the tensor network has a $\delta^{x_0}$ tensor of $\delta^{x_0}_{i}$ components, so all components are zero except for the $x_0$-th component. If we substitute the expression~\ref{eq: first variable}, the tensor is $\delta^{{\color{blue} H(\Omega_0)}}$. The traced TLC in this case is equal to the value ${\color{teal}\Omega_1}$, which, depending on the value of $x_0$, can be expressed as ${\color{teal}\Omega_1(}x_0{\color{teal} )}$, which by substituting the expression~\ref{eq: first variable} becomes ${\color{teal}\Omega_1(}{\color{blue} H(\Omega_0)}{\color{teal} )}$. Thus, the value of the second variable is
    \begin{equation}
        x_1 = {\color{teal} H(\Omega_1(}{\color{blue} H(\Omega_0)}{\color{teal} ))}.
    \end{equation}
    
    \item The third TLC depends on the value of the two previous ones for the same reason as in the previous step, so it is a function
    \begin{equation}
    \Omega_2(x_0,x_1)=\Omega_2({\color{blue} H(\Omega_0)},{\color{teal} H(\Omega_1(}{\color{blue} H(\Omega_0)}{\color{teal} ))}).
    \end{equation}
    This way,
    \begin{equation}
        x_2 = H(\Omega_2({\color{blue} H(\Omega_0)},{\color{teal} H(\Omega_1(}{\color{blue} H(\Omega_0)}{\color{teal} ))})).
    \end{equation}

    \item The correct value of the $n$-th variable is
    \begin{equation}
        x_n = {\color{red}H( \Omega_n(}{\color{blue} H(\Omega_0)},{\color{teal} H(\Omega_1(}{\color{blue} H(\Omega_0)}{\color{teal} ))}{\color{red},\dots,}{\color{violet} H(\Omega_{n-1}}({\color{blue} H(\Omega_0)},{\color{teal} H(\Omega_1(}{\color{blue} H(\Omega_0)}{\color{teal} ))}{\color{violet},\dots ))}{\color{red} ))}.
    \end{equation}
    Since each variable ultimately depends on the initial tensor network, we can say that its value is actually given by a function $\Xi_n$, so that $x_n=\Xi_n({\color{blue}\Omega_0})$.
    \item The solution of the problem always can be expressed as
    \begin{equation}
        \vec{x}=(\Xi_0({\color{blue}\Omega_0}), \Xi_1({\color{blue}\Omega_0}), \Xi_2({\color{blue}\Omega_0}),\dots, \Xi_{N-1}({\color{blue}\Omega_0})).
    \end{equation}
\end{enumerate}

Since the construction of the tensor network is as fast as the construction of the logic circuit, and this can be done in a polynomial time with respect to the formulation of the problem, the equation can be obtained in a polynomial time with respect to the formulation.

Moreover, since between two tensors one can always place an $A$ matrix and its inverse and have the same tensor represented, and the number of possible invertible $A$ matrices is infinite, then there are infinitely many possible TLC, and therefore infinitely many equations for each problem.

Since we already understand how the general method works to obtain the tensor network, and therefore, the equation that solves any combinatorial problem, either inversion, constraint satisfaction or optimization one, we will present a wide range of examples of the application of this method to problems. Due to the large number of examples, many of them with shared concepts, we will explain their logics in a more or less superficial way, providing the form of the tensors and tensor networks to be constructed with examples. In this way, we can be sure to explain the cases and their generalizations without overextending ourselves. In case it generates a lot of interest, concrete cases can be dealt with in greater depth in future versions. It is important to note that most of these methods have not been thoroughly investigated in an attempt to obtain either the simplest formulation or the most efficiently computable tensor network.

\newpage
\section{The Motion Onion Method}
Once the general method, the MeLoCoToN, is known, we can perform several optimizations in order to compute the calculation of this equation  in an exact or approximate way. The set of techniques to make the computation of the MeLoCoToN lighter is called the \textit{Motion Onion}. Its name comes from the fact that, dynamically, we modify the way of constructing the tensor network in each iteration, modifying also its layers. This method is specialized in optimization problems, although it could be applicable to the other types in some way.

Even so, the objective of this section is not to present how to best approximate each problem, a topic that can be left for future work. The objective is to present the fundamental ideas.

\subsection{Hamiltonian separation}
The first and most important part of the Motion Onion is the separation of the parts of the cost function (or Hamiltonian) into a tensorizable and a non-tensorizable part. That is, we will separate the part of the cost function (and constraints) into one that we can implement cheaply in our tensor network, and one that is prohibitively expensive to implement. We perform this process because we are going to delegate this second part in the iterative solving process.

In each iteration, the tensor network implements the tensorizable part of the problem, gives a result for the variable to be determined, and in the next step, the tensor network is created so that it takes into account the non-tensorizable part as part of its tensor values. The process is as follows:

\begin{enumerate}
    \item We initialize the TLC with the tensorizable part of the problem, contract and determine the first variable.
    \item We initialize the TLC with the tensorizable part of the problem and the information of the previous variable, contract and determine the second variable.
    \item Depending on the obtained variables, we change the initialization, evolution or constraint tensors, so that they add the part of the cost and constraints not easily computable due to having selected them.
    \item We contract the network tensor and determine the next variable.
    \item We repeat steps 3 and 4 until all the variables are determined.
\end{enumerate}

We can understand it better with the following example.

The problem is to find a sequence of a set of $N$ elements, which can be repeated as many times as we want. The cost associated with the sequence depends on the order in which the elements appear, which means that the cost function is
\begin{equation}
    C(\Vec{x})=\sum_{t=0}^{N-2} C_{x_t,x_{t+1}},
\end{equation}
being $x_t$ the selected element at position $t$ in the sequence.

In addition, there is the soft constraint (highly recommended, but not mandatory) that each element must appear about $N_i$ times in the sequence, and this ratio must also be satisfied in local parts of the sequence. That is, its occurrence ratio must be about $N_i/N$ along the sequence.

As can be seen, the first part of the problem, the cost function, is very easy to minimize by means of the tensor network. However, the second part, the uniformity part, is terribly expensive, since each element would need a layer of tensors that have control that the ratio is fulfilled at each step. This type of layer would be similar to the $F$ layers of the TSP that we will see in Ssec.~\ref{ssec: TSP}, which make the computational cost of the realization of the algorithm scale exponentially.

\begin{figure}
    \centering
    \includegraphics[width=0.9\linewidth]{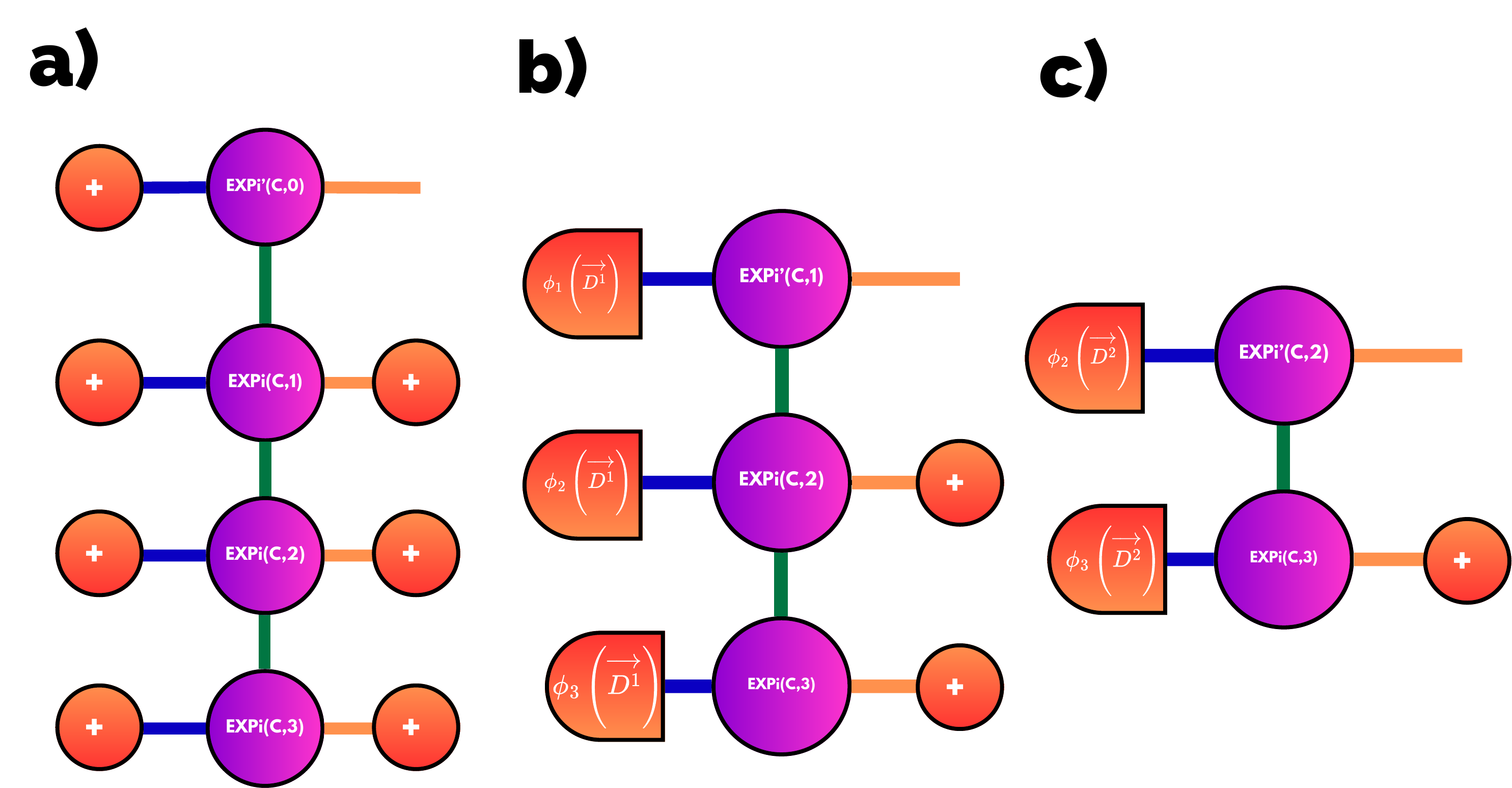}
    \caption{Tensor Network for solving the minimization problem in the a) first iteration, b) second iteration, c) third iteration.}
    \label{fig: Motion Onion TN}
\end{figure}

For this reason, the second part will be solved by the iterative method. In the first step we create a chain tensor network, as presented in Fig.~\ref{fig: Motion Onion TN} a, which implements the imaginary time evolution with the cost function. Once we have determined the first variable, in the next iteration, instead of the system starting in a uniform superposition, it will start in a superposition that favors that the same result does not appear again. That is, we make that in the following qudits there will be an extra cost associated to the value $x_0$ determined in the first iteration. As it is more important not to move further away from the recommended distribution in the first following variables to be determined, the state of each qudit will be different. The $t$-th qudit will be initialized with the vector $\phi_t(\vec{D^m})$, with $\vec{D^m}$ being how much each has been moved away the number of times we have chosen each element from the recommended distribution. This vector is calculated as
\begin{equation}
    D^m_i=\frac{N_i}{N} - \frac{n_i}{m},
\end{equation}
being $D^m_i$ the value associated to the $i$-th element when we determine the $m$-th variable and $n_i$ the number of times $i$ appears in the sequence. This way, the initialization of the $t$-th qudit at iteration $m$ is
\begin{equation}
    \phi_t(\vec{D^m})_i = e^{\tau \frac{\lambda}{t+1-m}D^{m}_i}.
\end{equation}
In this way, there is a greater probability of choosing the elements that we have chosen the least proportionally to their distribution, and we make that priority decrease as we advance in the sequence. In addition, we put in a proportionality factor $\lambda$ to consider how relevant this condition is in relation to the cost function.

In addition, to reduce the amount of calculations, we can also eliminate the tensors associated to the variables already determined, and include their information in the remaining tensors, in an exact way. For example, in this case, it would be to make the new tensors $EXPi'$, first in the chain, take into account the cost associated with the value of the previous variable determined. This is an optimization that does not add an approximation, since it is exact.

Thus, in the second iteration we have the tensor network of Fig.~\ref{fig: Motion Onion TN} b, where we have included the information of the extra non-tensorizable constraint in the initializations. This allows us to maintain a polynomial scaling of the complexity, in exchange for making the method now approximate. This technique can also be used to change the evolution and constraint layers, so that we need less of them. Moreover, it allows the combination of the method with genetic algorithms and heuristics, useful in initializations, or to improve them thanks to tensor networks.

\subsection{Approximation by elimination}
A second technique is the layer removal approach. That is, remove constraint layers, hard constraints, and implement them in the iterative method. The process is as follows:

\begin{enumerate}
    \item The first variable is determined with a tensor network with fewer constraint layers. The value of the first variable prevents certain values in other variables.
    \item In the new tensor network to determine the next variable, instead of starting with a uniform superposition, each qudit starts with a uniform superposition only of the allowed values given the results already obtained.
    \item The tensor network is contracted and the next variable is determined, which, having eliminated the possibilities that violate the constraints, satisfies them.
    \item Steps 2 and 3 are repeated until all variables are determined.
\end{enumerate}

Obviously, this method is a huge approximation, as it may result in the fact that, in some iteration, there is no combination that meets the remaining constraints, given the previous results.

An interesting example is presented in the paper~\cite{TSP_TN}, for the traveling salesman problem. In this problem, $N$ layers of constraints are needed to guarantee that no node is repeated in the route. However, to prevent the contraction complexity of the tensor network from scaling as $2^N$, one option is to use a fixed number $L<<N$ of layers in each iteration.

In the first iteration the system starts with a uniform superposition, and a few constraint layers. The first variable $x_0$ is determined. Since a node cannot be repeated in the path, for the next iteration, the state of all qudits starts with an superposition of all possible nodes except the one we have already selected. Obviously, since the constraint has already been imposed at initialization, no constraint layer is needed for it. The variable $x_1$ is determined and for the next one, all qudits are initialized on all possible nodes, except those already selected. This is repeated until the end. To avoid problems associated with choosing inadequate constraints in the tensor network, and compensating them with each other, it is convenient to change the chosen constraint layers at each iteration.

Another example is in~\cite{Task_TN}, where we have a large set of directed constraints. These constraints mean that if one set of variables has a certain value, another can only have a specific value. However, since the constraints are obtained from historical data, not all of them are relevant to the problem. Therefore, you can start solving without constraint layers, and each time you obtain a solution that violates a constraint, add its layer to the tensor network and start over. In this way, we ensure that we only need the minimum number of constraint layers to solve the problem.

\subsection{Genetic algorithm}
An interesting combination of technologies is that of MeLoCoToN with Motion Onion and genetic algorithms. There are two possible visions in this union: using genetic algorithms to leverage tensor networks or using tensor networks to leverage genetic algorithms. We will focus on the first possibility, explored in~\cite{Task_TN}.

This union consists in having a population of initializations to remove constraints. That is, each individual in the population considers that in its solution only a subset of the values of each variable are possible, the possibilities for each variable not being equal. These are its chromosomes. Each chromosome will be the possibilities for one variable, and a subset of constraints may also be included. In this way, each individual can have a smaller number of constraints or layers in general implemented, and obtain a solution. The individuals with the best results will cross with each other, creating descendents with their chromosomes crossed. This is highly interesting for certain problems with historically obtained constraints, which may not all be relevant to obtain a quality solution.

\newpage
\section{Unconstrained Optimization Problems}
The first group of problems to be addressed are unconstrained combinatorial optimization problems~\cite{Unconstrained}. In these problems we do not need operators that cancel the amplitude of incompatible solutions with some constraints, because they do not exist.
\subsection{QUBO}
The first type of problems are the most famous in the world of quantum computing. These are the QUBO problems, which have quadratic relations between pairs of binary variables. The cost function in these problems is
\begin{equation}\label{eq: cost QUBO}
    C(\vec{x})=\sum_{i}^{N-1}\sum_{j=0}^{i} Q_{ij}x_ix_j,
\end{equation}
where $Q$ is a $N\times N$ matrix that accounts for the relationships between pairs of variables and $\vec{x}$ is a binary vector to optimize. In this problem we want to minimize $C(\vec{x})$. This kind of problems are usually solved with QAOA~\cite{QAOA} or Quantum Annealing~\cite{Annealing}.

In this problem, each variable is related to all the others, so we need a circuit that relates them in pairs. To do this, each variable must have an operator of $N$ connections, which is in charge of receiving the value of that variable through its first input, transmitting it to the other operators through their outputs and receiving the values of the other variables through their other inputs. With this information, the operator carries out the partial evolution.

Since the product $x_i x_j$ is the same as the product $x_j x_i$, we only have to account for this interaction once. Each operator receives as input the signal from the operators of the variables with lower index, and sends its signal to those with higher index. Therefore, the $i$-th operator makes the evolution with
\begin{equation}
    e^{-\tau \sum_{j=0}^{i} Q_{ij}x_ix_j},
\end{equation}

\begin{figure}[h]
    \centering
    \includegraphics[width=0.9\linewidth]{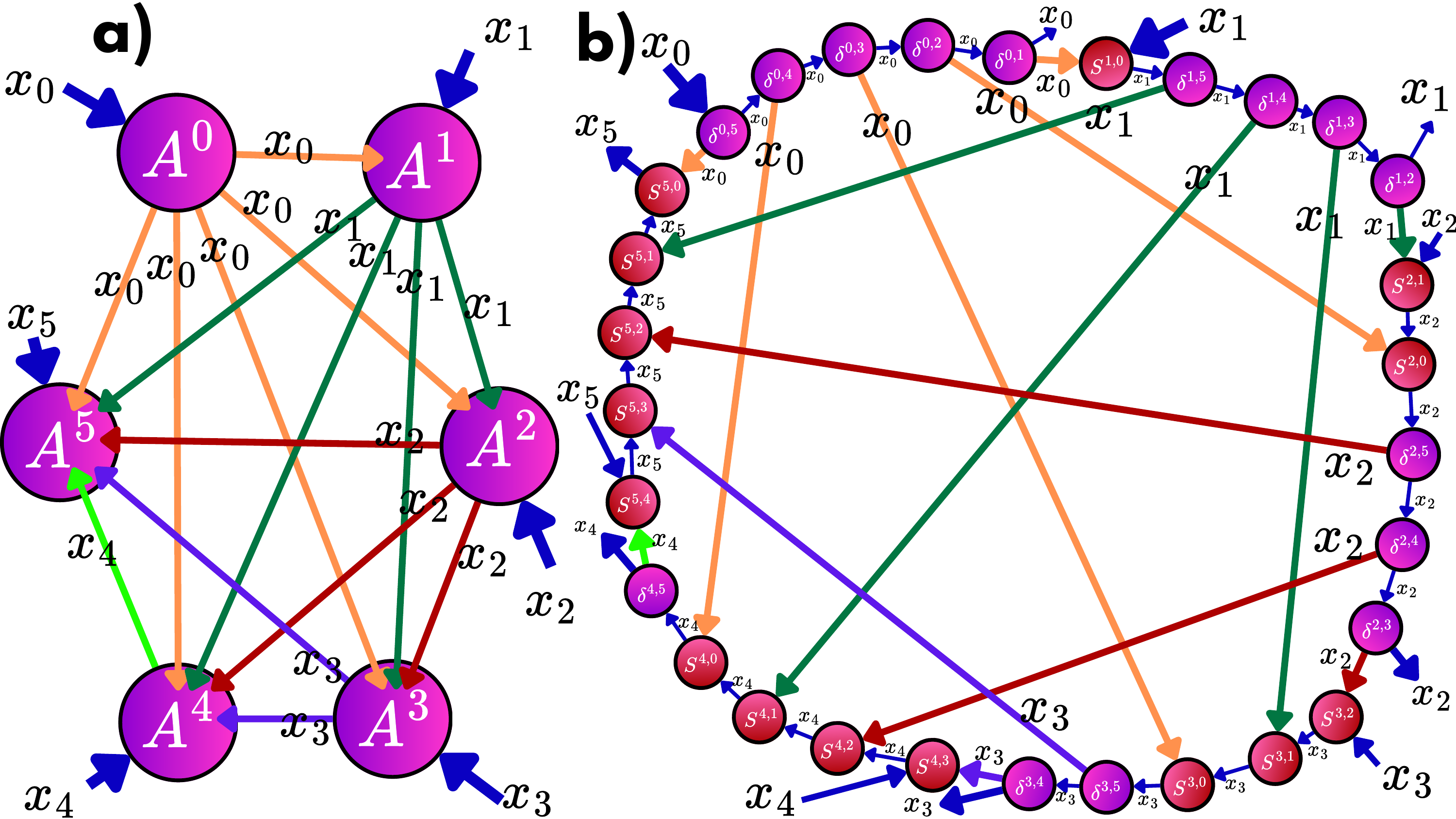}
    \caption{LSMC for the QUBO/QUDO/T-QUDO problem with 6 variables. a) One operator per variable, b) One operator per pair of variables.}
    \label{fig: QUBO Circuit}
\end{figure}

\begin{figure}
    \centering
    \includegraphics[width=0.9\linewidth]{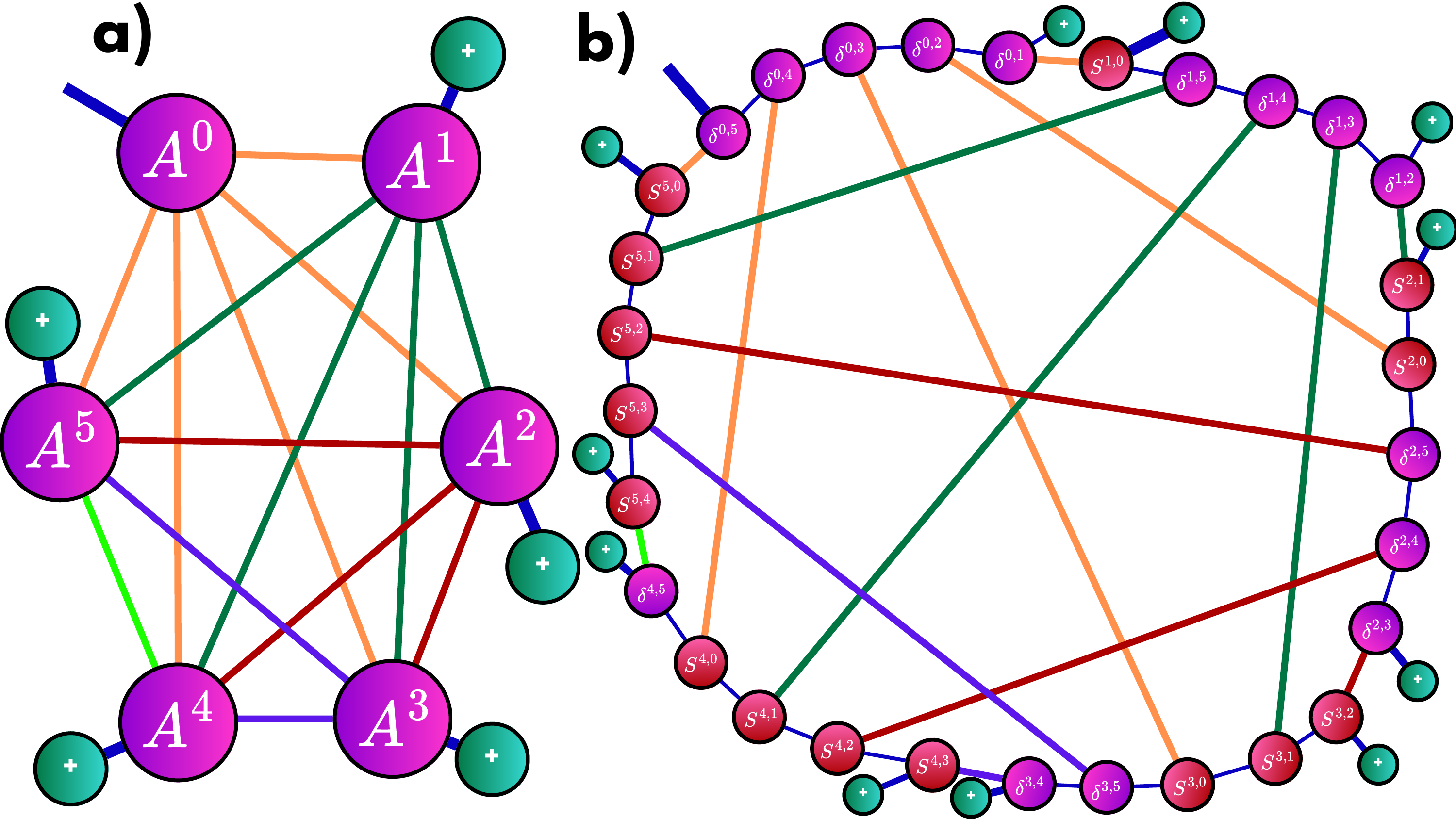}
    \caption{Tensor Network for the QUBO/QUDO/T-QUDO problem with 6 variables to determine the first variable value. a) One tensor per variable, b) One tensor per pair of variables.}
    \label{fig: QUBO TN}
\end{figure}

The original circuit is the one shown in Fig.~\ref{fig: QUBO Circuit} a. Its equivalent tensor network is shown in Fig.~\ref{fig: QUBO TN} a. We can notice that having formulated the circuit in this way, we have tensors with a huge number of indexes, although each one is highly sparse. We can also note that we could delegate all the evolution to the last tensor, since it receives all the information of the system, which is undesirable.

To avoid this, and to have a greater separation of the information, we can realize that each tensor sends by its output exactly the same signal, and that the modulations depend each one only on one of the inputs and the input of the variable. Therefore, we can formulate the circuit as in Fig.~\ref{fig: QUBO Circuit} b. Each original operator is replaced by a chain of operators (its tensor train~\cite{Tensor_Train}) that send each other the same signal, and each one receives or sends to the operators of the other variables a single signal. Thus, each operator deals with a single term of the imaginary time evolution, or sending its signal to one of the other operators. The associated tensor network is presented in Fig.~\ref{fig: QUBO TN} b. The $S^{a,b}_{2\times 2\times 2}$ operator corresponds to the $a$-th variable, and receives its value and the state of the $b$-th variable to perform the imaginary time evolution, and passes the signal of $a$-th variable to the next operator of the same variable. The operator $\delta^{a,b}_{2\times 2\times 2}$ corresponds to the variable $a$-th, and sends its state to the operator $S^{b,a}$. The non-zero elements of these tensors are 
\begin{equation}\label{eq: QUBO delta}
    \begin{gathered}
        \mu = \nu = i\\
        \delta^{a,b}_{i,\mu,\nu} = 1\\
        \delta^{a,a+1}_{i,\mu,\nu} = e^{-\tau Q_{a,a}i^2}
    \end{gathered}
\end{equation}
\begin{equation}\label{eq: QUBO S}
    \begin{gathered}
        \nu = i\\
        S^{a,b}_{i,j,\nu} = e^{-\tau Q_{a,b}ij}\\
        S^{N-1,N-2}_{i,j,\nu} = e^{-\tau (Q_{N-1,N-2}ij+Q_{N-1,N-1}i^2)}
    \end{gathered}
\end{equation}
where we have made the last $\delta$ tensor of each variable and the last $S$ tensor of the last variable take into account the diagonal cost term.

If a term $Q_{a,b}$ of the cost matrix is zero, then there is no direct relationship between the variable $x_a$ and the variable $x_b$. Because of this, we can eliminate all zero element connections. This implies that, for all $a,b$ such that $Q_{a,b}=0$ we can eliminate the tensors $\delta_{b,a}$ and $S_{a,b}$ from the circuit and the tensor network.

\begin{figure}
    \centering
    \includegraphics[width=0.9\linewidth]{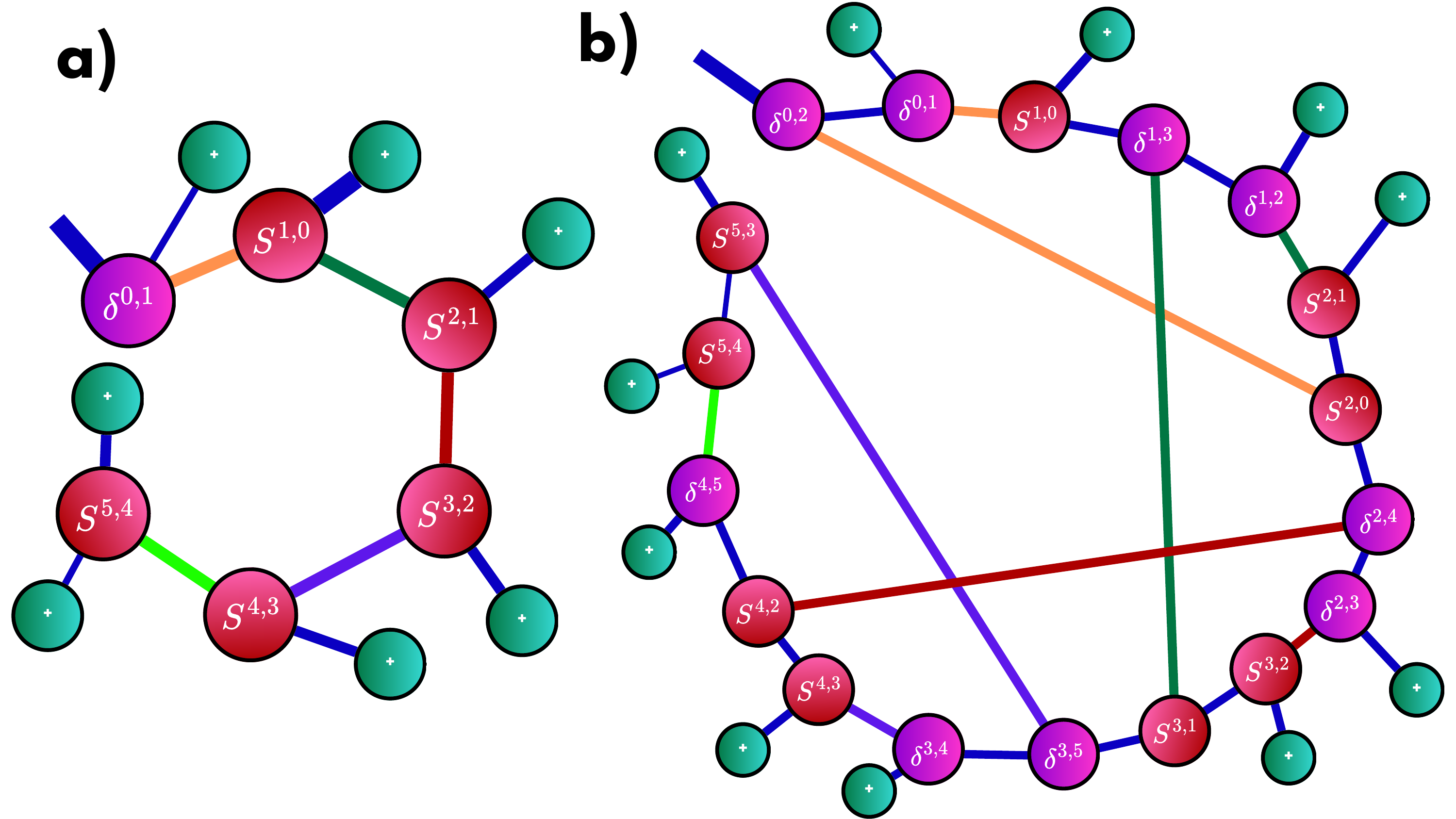}
    \caption{Tensor network for determine the first variable in the QUBO/QUDO/T-QUDO problem in a linear chain with 6 variables. a) First neighbor, b) Second neighbor.}
    \label{fig: QUBO Tridiag}
\end{figure}

A particularly interesting case is one in which each variable is only related to the next. That is, a case in which the $Q$ matrix is tridiagonal, which is going to be a case of linear chain neighbor interaction. The circuit in this case can be realized with a single operator for each variable, having the equivalent tensor network shown in Fig.~\ref{fig: QUBO Tridiag} a. As can be seen, this tensor network can be contracted in an efficient way as it is a chain. This construction can be optimized to require less memory and computation time, as studied in \cite{QUBO_Tridiagonal}.  It is also studied for $m$-neighbors in {\color{red} [pending to publish]}.

In the case of having an interaction with the two nearest neighbors, a five-diagonal $Q$ matrix, we have three tensors for each variable, as shown in Fig.~\ref{fig: QUBO Tridiag} b. As in the tridiagonal case, it is possible to contract a more optimal version of this tensor network in polynomial time, as a chain. In the case with interactions reaching up to the $m$-th neighbor, the tensor network has $m$ 3-index tensors for each variable. One can also solve the problem with a linear equivalent tensor network with indexes of dimension $2^m$ or with one in the form of a grid with the upper triangle and part of the lower one cut off.

\begin{figure}
    \centering
    \includegraphics[width=0.9\linewidth]{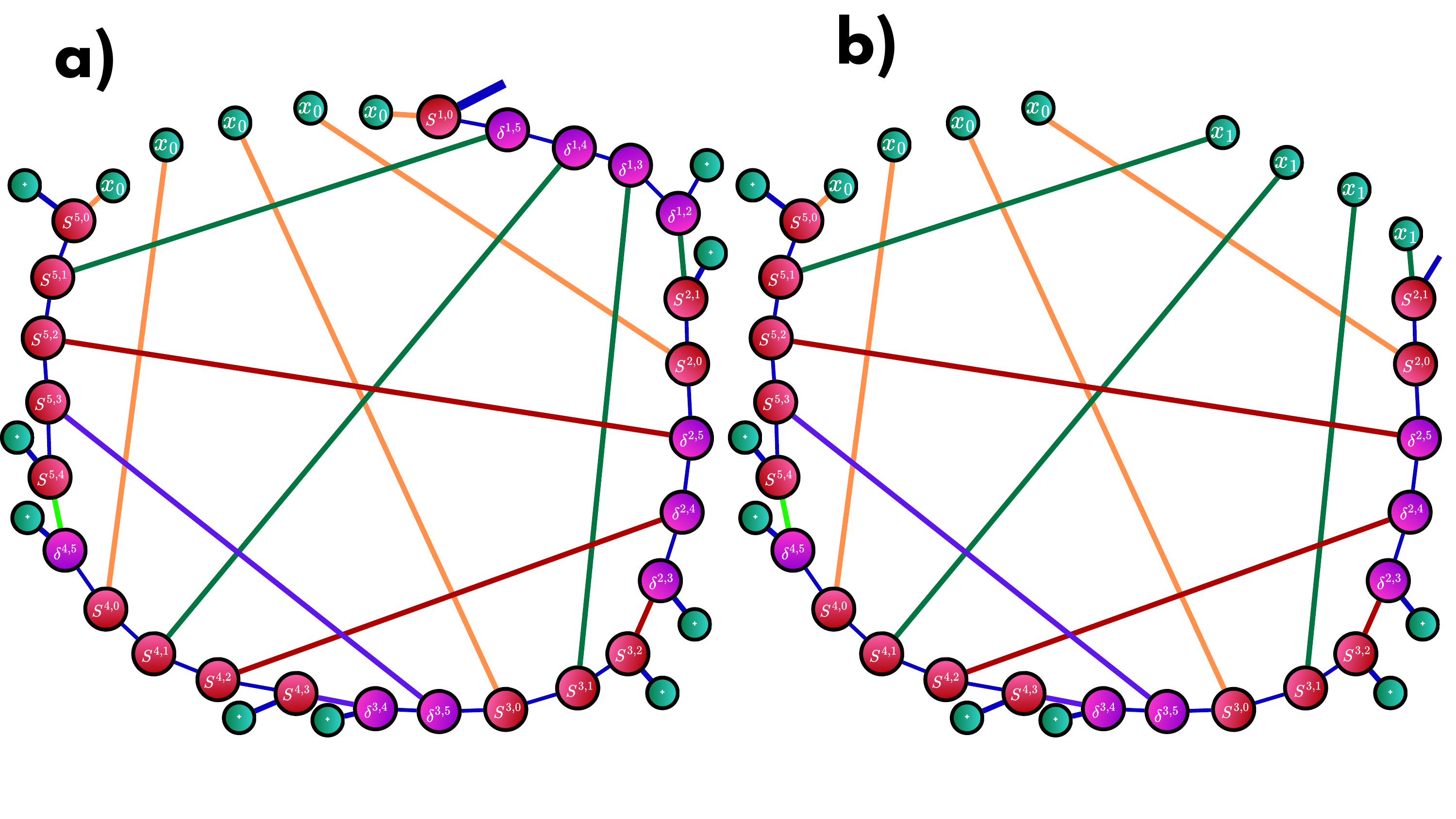}
    \caption{Tensor network for the QUBO/QUDO/T-QUDO problem with 6 variables to determine, a) the second variable, b) the third variable.}
    \label{fig: QUBO Iteration}
\end{figure}

In the iterative process of determining the variables, each time we create the network for a variable, we eliminate all the operators of the variables already determined, and in the lines that connect with them we put a vector of zeros with a 1 in the position of the value determined for that variable. The tensor network for the second and third iteration is shown in Fig.~\ref{fig: QUBO Iteration}.

\subsection{QUDO}
The natural extension of the QUBO problem is the QUDO problem, in which the variables $x_i$ are now natural numbers instead of binary. That is, the cost function is still \eqref{eq: cost QUBO}, but now $x_i\in[0,d_i-1]$. This problem can be solved with the same circuit and tensor network that we have presented for the QUBO problem, but increasing the dimensionality of the tensors, so that they can send signals with the dimensionality of the $x_i$. That is, the equations \eqref{eq: QUBO delta} and \eqref{eq: QUBO S} are still the right ones, but now the tensors are of dimension $\delta^{a,b}_{d_a\times d_a\times d_a}$ and $S^{a,b}_{d_a\times d_b\times d_a}$. All other analyses performed for the QUBO case are directly extensible to this one. This one is also optimized for a linear chain with one neighbor in~\cite{QUBO_Tridiagonal}, and for $m$-neighbors in {\color{red} [pending to publish]}.

\subsection{T-QUDO}
Another natural extension is the T-QUDO, where the cost function is expressed in \eqref{eq: T QUDO cost}. However, we can rewrite it in an alternative way that is easier to interpret
\begin{equation}
    C(\vec{x}) = \sum_{i=0}^{N-1}\sum_{j=0}^{i} C_{i,j,x_i,x_j},
\end{equation}
where the cost depends on which variables are involved and their values throught the tensor $C$. The QUDO case is a particular case of this, in which $C_{i,j,x_i,x_j} = Q_{i,j}x_ix_j$.

In this case, the construction of the TLC is exactly the same as in the QUDO case, only changing the value of the tensor elements, to take into account this form of the cost expression. The new values are
\begin{equation}
    \begin{gathered}
        \mu = \nu = i\\
        \delta^{a,b}_{i,\mu,\nu} = 1\\
        \delta^{a,a+1}_{i,\mu,\nu} = e^{-\tau C_{a,a,i,i}}
    \end{gathered}
\end{equation}
\begin{equation}
    \begin{gathered}
        \nu = i\\
        S^{a,b}_{i,j,\nu} = e^{-\tau C_{a,b,i,j}}\\
        S^{N-1,N-2}_{i,j,\nu} = e^{-\tau (C_{N-1,N-2,i,j}+C_{N-1,N-1,i,i})}.
    \end{gathered}
\end{equation}
 This one is also optimized for a linear chain with one neighbor in~\cite{QUBO_Tridiagonal}, and for $m$-neighbors in {\color{red} [pending to publish]}.

\subsection{HOBO, HODO and T-HODO}
The next extension of the problem is one in which the interactions move from couples to larger groups of $M$ variables. That is, the cost function is expressed as
\begin{equation}\label{eq: cost HOBO}
    C(\vec{x})=\sum_{i_0}^{N-1}\sum_{i_1=0}^{i_0}\cdots \sum_{i_{M-1}=0}^{i_{M-2}} Q_{i_0, i_1, \dots, i_{M-1}}\prod_{k=0}^{M-1}x_{i_k}.
\end{equation}
This is a Higher-Order Binary Optimization (HOBO) problem~\cite{VQC_HOBO,HOBO_Quantum_2}. This problem has been addressed with tensor networks before in~\cite{HOBO_TN,HOBOTAN}. This formulation is also valid for the Higher-Order D-ary Optimization (HODO) problems, where $x_i\in [0,d_i]$. As in the QUDO and QUBO cases, it can be generalized to the tensorial case T-HODO with a cost function
\begin{equation}
    C(\vec{x})=\sum_{i_0}^{N-1}\sum_{i_1=0}^{i_0}\cdots \sum_{i_{M-1}=0}^{i_{M-2}} C_{i_0, i_1, \dots, i_{M-1}, x_{i_0}, x_{i_1}, \dots, x_{i_{M-1}}}.
\end{equation}
For simplicity, the terms in which $i_k=i_{k'}$ for $k\neq k'$ will be introduced in the terms in which the index values are not repeated. That is, we eliminate the self-interactions introducing its effect in the interactions with other variables.

In this case we have two options. The first is to mimic the structure previously presented for the quadratic case, making now each tensor $S$ receive $M-1$ inputs from the previous variables to perform the imaginary time evolution. The second one is that each variable has only $\delta$ tensors, and passes its signal to some new intermediate tensors that, receiving each one the $M$ signals, perform the evolution. However, this second approach ends up requiring a larger number of tensors to perform, for practical reasons, the same calculation as the first one.

\begin{figure}
    \centering
    \includegraphics[width=0.9\linewidth]{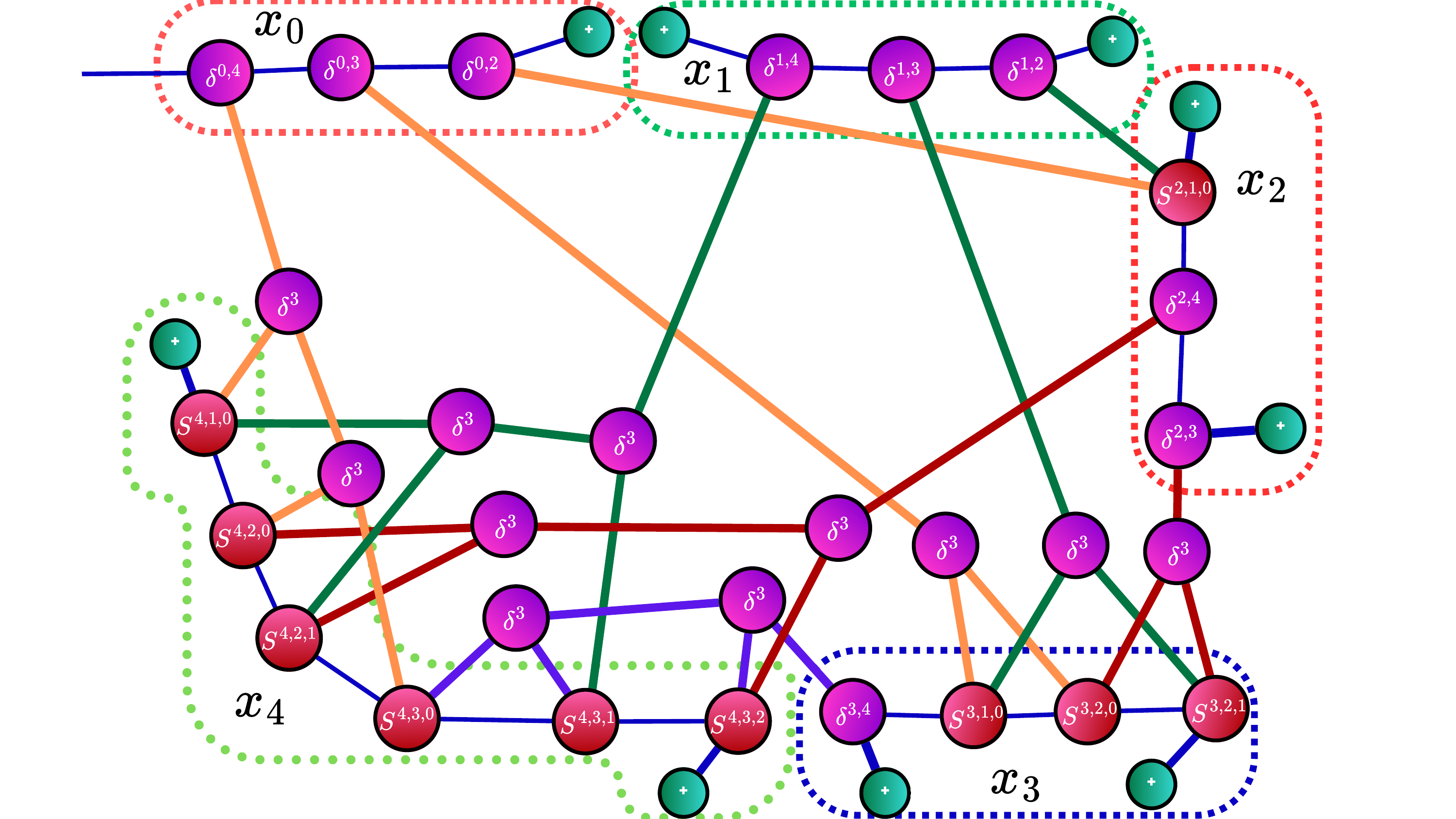}
    \caption{Tensor network for the T-HODO problem with 6 variables and $M=3$ to determine the first variable.}
    \label{fig: T-HODO 3 TN}
\end{figure}

For the first case, the tensor network with $M=3$ is the one presented in Fig.~\ref{fig: T-HODO 3 TN}. In this case we have to add $\delta^3$ tensors, which are Kronecker deltas of 3 indexes, to indicate to all the minimization tensors of each variable the value of the variables involved in its term. The non-zero values of the tensors are
\begin{equation}
    \begin{gathered}
        \mu = \nu = i,\\
        \delta^{a,b}_{i,\mu,\nu} = 1,
    \end{gathered}
\end{equation}
\begin{equation}
    \begin{gathered}
        \nu = i,\\
        S^{a,b_0,b_1,b_2,\dots}_{i,j_0,j_1,\dots,\nu} = e^{-\tau C_{a,b_0,b_1,\dots,i,j_0,j_1,\dots}}.
    \end{gathered}
\end{equation}

\subsection{Integer sum total function problem}
The cost function of this problem is
\begin{equation}
    C(\vec{x})=f\left(\sum_{i=0}^{N-1} a_i x_i\right),
\end{equation}
where $a_i\in \mathbb{Z}$ are fixed cost constants, $x_i\in [0,d_i-1]$ and $f(\cdot)$ is a non-linear function. The reason $f(\cdot)$ is a nonlinear function is that if it were linear, we would be back to the case of the linear cost function problem, which can be solved by finding the minimum of a list.

To solve this problem, the signal is the partial sum of the argument of the function. That is, the signal is $r_m=\sum_{i=0}^{m-1} a_i x_i$. Since tensors can only send information through positive integers, we perform a shift of all signals. If $c_-=-\sum_{i=0}^{N-1} \min(a_i,0) d_i$ is the minimum possible value for the argument, then we define the signal as $r_m=\sum_{i=0}^{m-1} a_i x_i + c_-$. Thus, if $c_+=\sum_{i=0}^{N-1} \max(a_i,0) d_i$ is the maximum value the argument can take, the dimension of the signal is $D=c_-+c_+$. This dimension can be optimized operator by operator taking into account what is the maximum range of values that can be taken up to the $m$-th variable, but for our analysis we will not do so.
\begin{figure}
    \centering
    \includegraphics[width=0.8\linewidth]{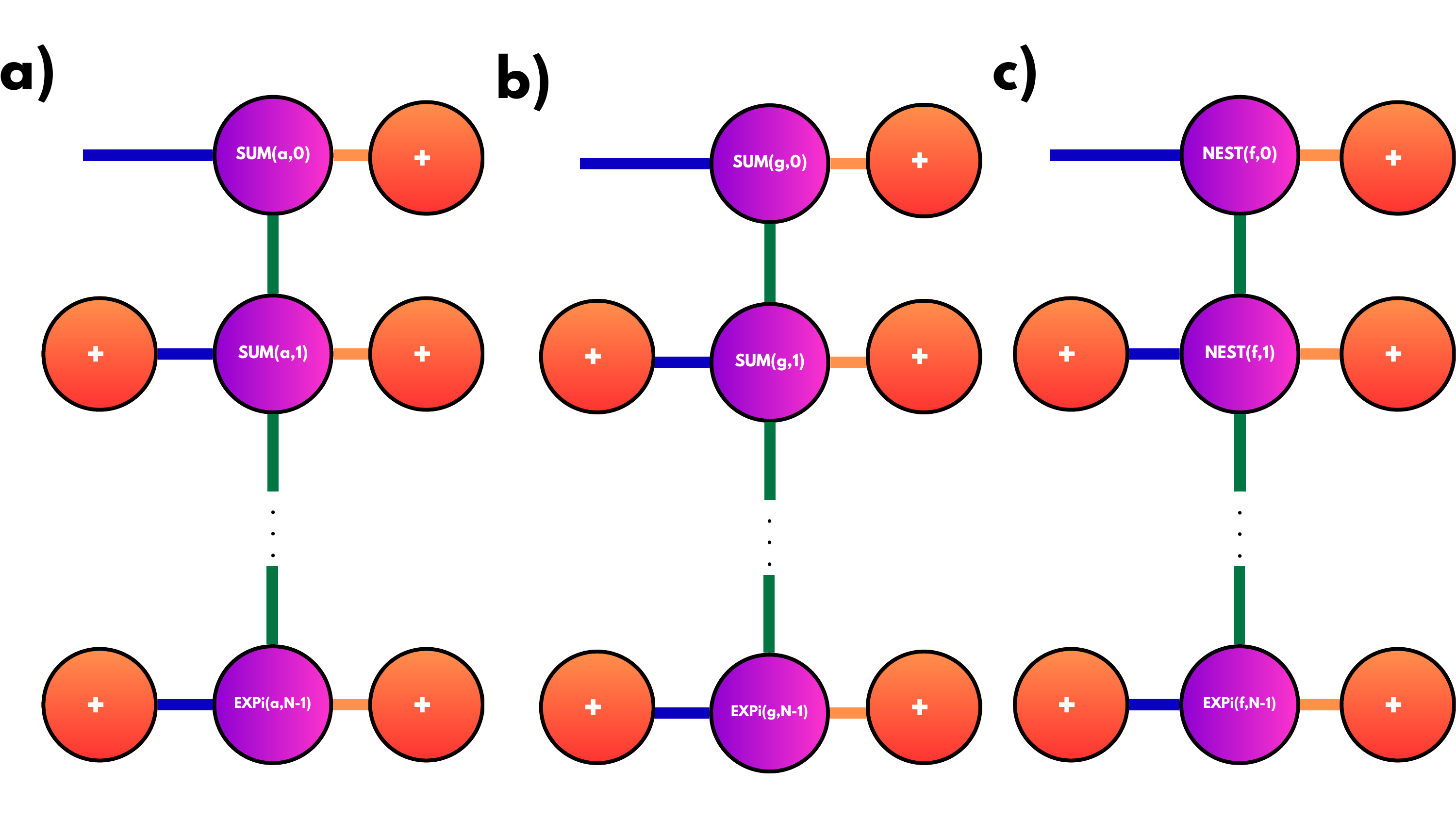}
    \caption{Tensor network for the Integer sum total function problem with a) lineal argument b) non-linear argument. c) Tensor network for the Nested cost function problem.}
    \label{fig: Integer Sum TN}
\end{figure}

The tensor network that we will have to create is similar to the one shown in Fig.~\ref{fig: Natural sum TN} above, but taking into account this displacement in order to generalize to integer values. It is shown in Fig.~\ref{fig: Integer Sum TN} a. The non zero elements of the tensors $SUM(a,0)_{d_0\times d_0 \times D}$, $SUM(a,k)_{d_k\times D \times d_k \times D}$ and $EXPi(a,N-1)_{d_{N-1}\times d_{N-1}\times D}$ are
\begin{equation}
     \begin{gathered}
         \mu=i,\quad \nu=a_0 i+c_-,\\
         SUM(a,0)_{i,\mu,\nu} = 1,
     \end{gathered}
\end{equation}

\begin{equation}
     \begin{gathered}
         \mu=i,\quad \nu=j+a_k i+c_-,\\
         SUM(a,k)_{i,j,\mu,\nu} = 1,
     \end{gathered}
\end{equation}

\begin{equation}
     \begin{gathered}
         \mu=i,\\
         EXPi(a,N-1)_{i,j,\mu} = e^{-\tau f(j+a_{N-1}i-c_-)}.
     \end{gathered}
\end{equation}

We can further generalize the problem by making each summand of the argument of the function have a tensor dependence instead of being linear with $x_i$, such that
\begin{equation}
    C(\vec{x})=f\left(\sum_{i=0}^{N-1} g_i(x_i)\right) = f\left(\sum_{i=0}^{N-1} g_{i,x_i}\right),
\end{equation}
where $g_i: \mathbb{Z}\rightarrow \mathbb{Z}$ and $g_{i,x_i}$ is its associated tensor of integer numbers. To solve this problem, we start from the same tensor network, but now taking into account the tensors that must add the tensor term instead of directly adding it, it must add the $g_i(x_i)$. Now we define  $c_-=-\sum_{i=0}^{N-1} \min(\min(g_i),0)$ and $c_+=\sum_{i=0}^{N-1} \max(\max(g_i),0)$. The tensor network is shown in Fig.~\ref{fig: Integer Sum TN} b. The tensor values now are
\begin{equation}
     \begin{gathered}
         \mu=i,\quad \nu=g_{0,i}+c_-,\\
         SUM(g,0)_{i,\mu,\nu} = 1
     \end{gathered}
\end{equation}

\begin{equation}
     \begin{gathered}
         \mu=i,\quad \nu=j+g_{k,i}\geq 0,\\
         SUM(g,k)_{i,j,\mu,\nu} = 1
     \end{gathered}
\end{equation}

\begin{equation}
     \begin{gathered}
         \mu=i,\\
         EXPi(g,N-1)_{i,j,\mu} = e^{-\tau f(j+g_{N-1,i}-c_-)}.
     \end{gathered}
\end{equation}

\subsection{Nested cost function problem}
This is a generalization of the previous case. The cost function of this problem is defined as
\begin{equation}
    \begin{gathered}
        C(\vec{x})=Q_{N-1}\\
        Q_i = f_i(x_i,Q_{i-1})=f_{i,x_i,Q_{i-1}}\ \forall i\in[1,N-1],\\
        Q_0 = f_0(x_0,0)=f_{0,x_0,0},
    \end{gathered}
\end{equation}
where $f_i: \mathbb{Z}^2\rightarrow \mathbb{Z}$ and $f_{i,x_i,Q_{i-1}}$ is its associated tensor of integer numbers. In other words, the cost function is a nested function
\begin{equation}
    C(\vec{x})=f_{N-1}(x_{N-1},f_{N-2}(x_{N-2},\dots f_{1}(x_{1},f_0(x_0,0))\dots)).
\end{equation}

To solve this problem, we start from the same tensor network, but now taking into account the tensors that must add the tensor term instead of adding it, it transforms the signal in function of its variable value. Now we define  $c_-=-\min(\min(f),0)$ and $c_+=\max(\max(f),0)$. The tensor network is shown in Fig.~\ref{fig: Integer Sum TN} c. The tensor values now are
\begin{equation}
     \begin{gathered}
         \mu=i,\quad \nu=f_{0,i,0}+c_-,\\
         NEST(f,0)_{i,\mu,\nu} = 1,
     \end{gathered}
\end{equation}

\begin{equation}
     \begin{gathered}
         \mu=i,\quad \nu=f_{k,i,j-c_-}+c_-,\\
         NEST(f,k)_{i,j,\mu,\nu} = 1,
     \end{gathered}
\end{equation}

\begin{equation}
     \begin{gathered}
         \mu=i,\\
         EXPi(f,N-1)_{i,j,\mu} = e^{-\tau f_{N-1,i,j-c_-}}.
     \end{gathered}
\end{equation}

\newpage
\section{Inversion Problems}
Now that we have solved some unconstrained combinatorial optimization problems, we are going to deal with combinatorial inversion problems. These are useful in cryptography to determine if a cryptographic protocol is secure or not, but we will study also other cases.

\subsection{Cybersecurity}
\subsubsection{Prime Factorization}
{\color{red} Subsubsection not available due to paper pending publication.}


\subsubsection{SHA-3 inversion}
{\color{red} Subsubsection not available due to paper pending publication.}

\subsection{Systems of linear equations}
The objective is to obtain the vector $\vec{x}$ that satisfies the equation $A\vec{x}=\vec{b}$ , where $A$ is an invertible square matrix and $\vec{b}$ is a vector, both known. To formulate it as a combinatorial inversion problem, we must work with a matrix $A$ of positive integers, a vector $\vec{b}$ of positive integers and that the solution $\vec{x}$ is also of positive integers. However, it can be extended to all integers using the corresponding shifts and mappings.

\begin{figure}[h]
    \centering
    \includegraphics[width=\linewidth]{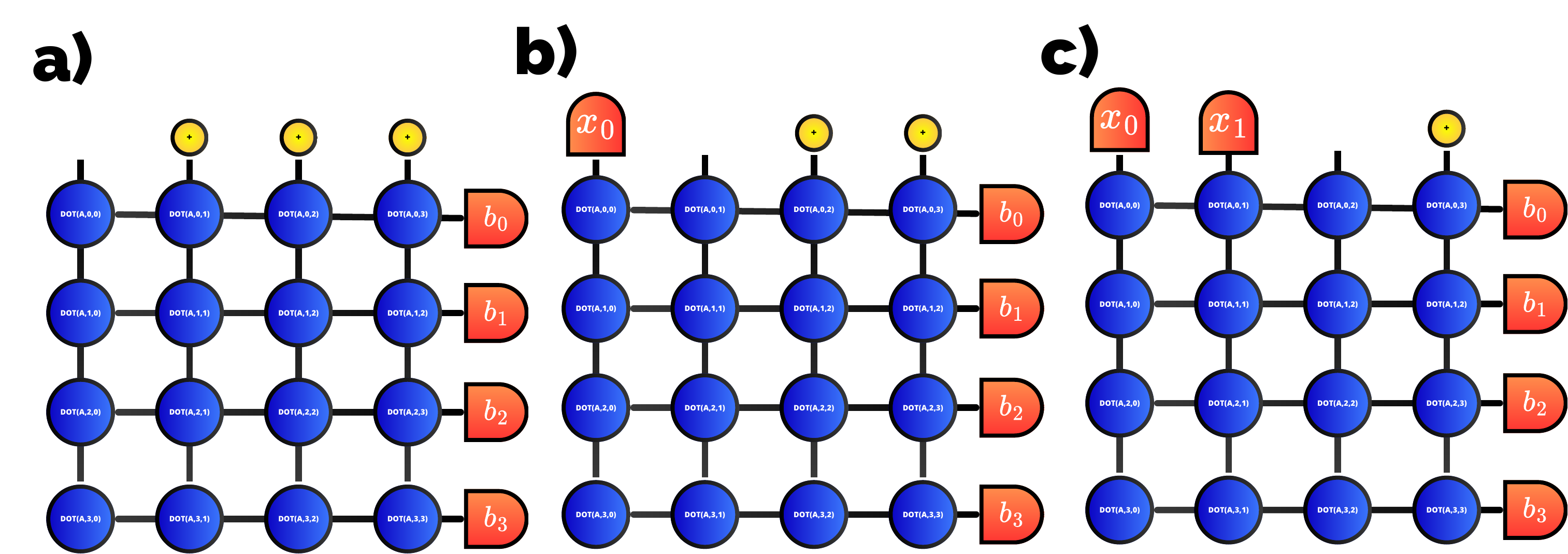}
    \caption{MeLoCoToN for solving systems of linear equations with a $4\times 4$ matrix $A$. a) Determining the $x_0$. b) Determining the $x_1$. c) Determining the $x_2$.}
    \label{fig: Lineal Solver}
\end{figure}

\begin{figure}[h]
    \centering
    \includegraphics[width=0.4\linewidth]{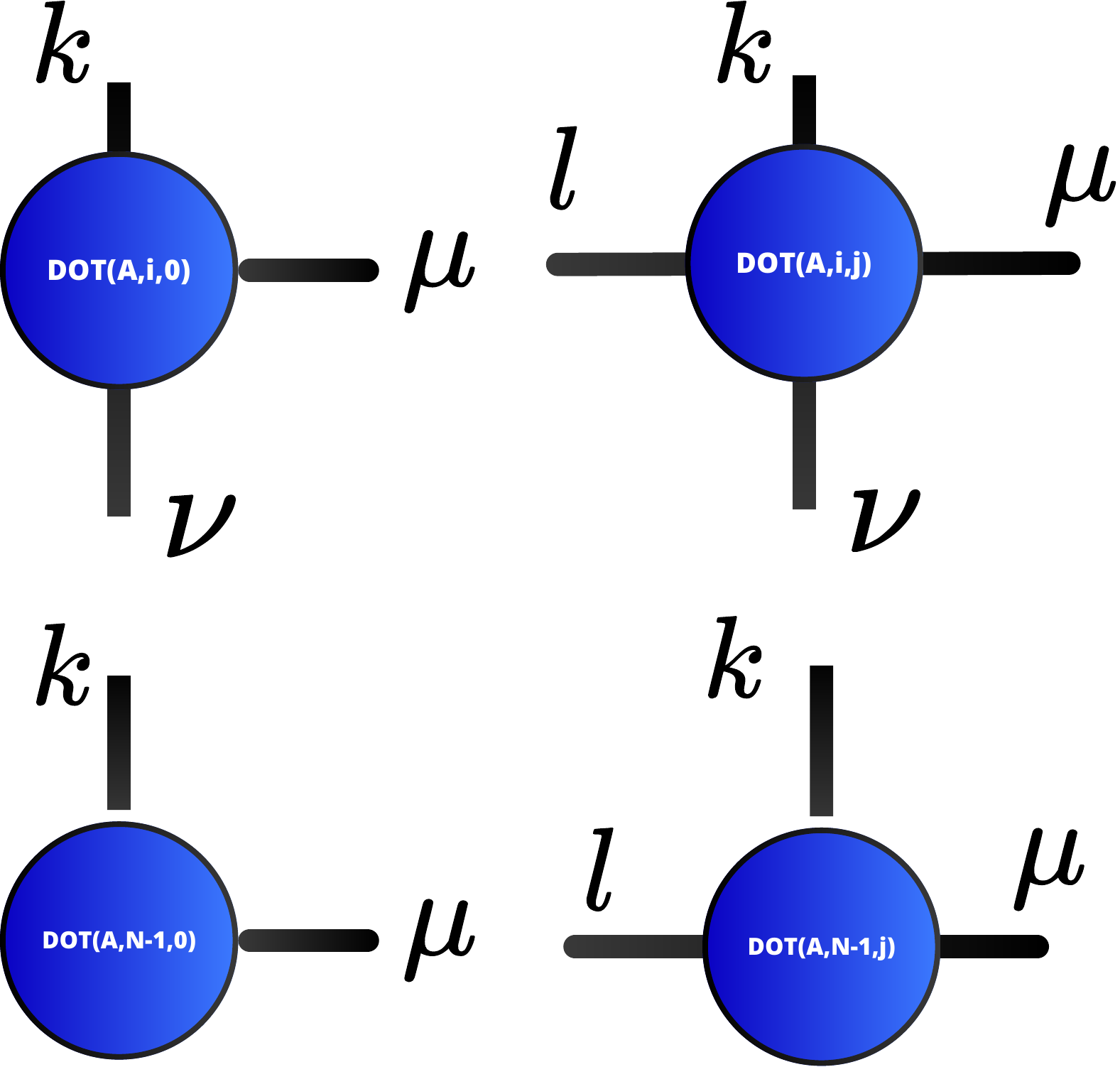}
    \caption{Indexes names for the tensor of the linear solver.}
    \label{fig: Notation_Linear}
\end{figure}

In this case, we must make a tensor network that performs the product of all possible $\vec{x}$ vectors with the known $A$ matrix, and postselects the final result to that of $\vec{b}$. Therefore, we make a square tensor network, in which each tensor in the grid performs the product and sum with its corresponding $A$ element for that row and column. The tensor network can be seen in Fig.~\ref{fig: Lineal Solver}. The operator $DOT(A,i,j)$ of row $i$ and column $j$ receives by its upper index the value of the component of $x_j$ and by its left index the value of the sum of products up to that point $\sum_{k=0}^{j-1}A_{ik}x_k$. With that information, it returns by its right index the value $\sum_{k=0}^{j}A_{ik}x_k$. It transmits downward the same signal it receives upward. The indexes names are presented in Fig.~\ref{fig: Notation_Linear}. The non-zero elements of each tensor for the $N\times N$ case are
\begin{equation}
    \begin{gathered}
        \mu = A_{i,0} k,\quad \nu=k,\\
        DOT(A,i,0)_{k\mu\nu} = 1,
    \end{gathered}
\end{equation}
\begin{equation}
    \begin{gathered}
        \mu = A_{N-1,0} k,\\
        DOT(A,N-1,0)_{k\mu} = 1,
    \end{gathered}
\end{equation}
\begin{equation}
    \begin{gathered}
        \mu = l+A_{i,j} k,\quad \nu=k,\\
        DOT(A,i,j)_{lk\mu\nu} = 1,
    \end{gathered}
\end{equation}
\begin{equation}
    \begin{gathered}
        \mu = l+A_{N-1,j} k,\\
        DOT(A,N-1,j)_{l,k,\mu} = 1.
    \end{gathered}
\end{equation}
In case some elements of matrix $A$ are equal 0, we can omit their corresponding tensors, and simply pass the signal. This way, for example, if we have a tridiagonal matrix, we only need three diagonals in the tensor network, allowing its contraction more efficiently.

\subsection{Closure finding problem}
This problem consists in finding the closure of a graph. The closure of a graph is the graph that is achieved by iteratively adding edges between non-connected vertices such that the sum of their degrees exceeds or equals the number of vertices $V$. 

\begin{figure}[h]
    \centering
    \includegraphics[width=\linewidth]{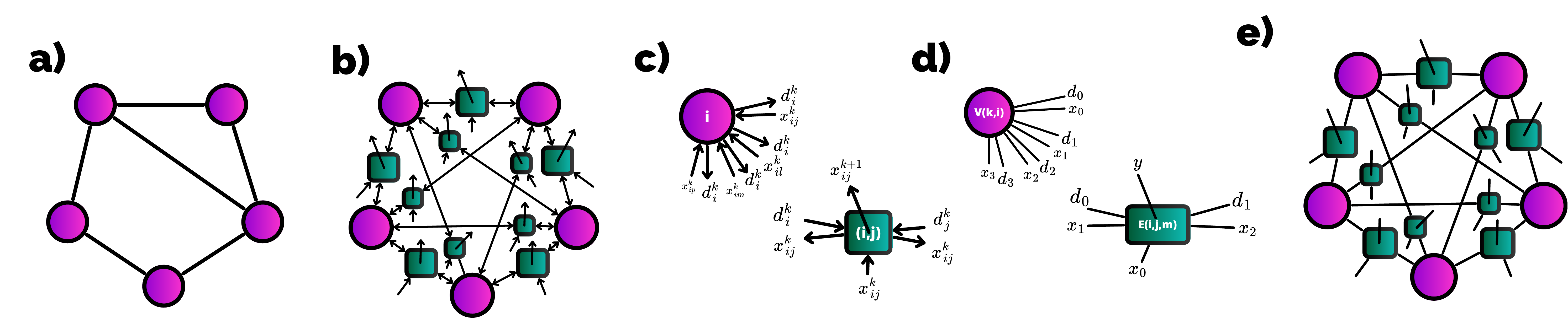}
    \caption{a) Original graph. b) LSTC of one step of the Closure finding problem. c) Operator signals. d) Tensors of the TLC. e) TLC of one step. Each index connecting nodes is two indexes.}
    \label{fig: Closure 1}
\end{figure}

To understand how the tensor network is constructed, and to be able to visualize it, we will work on two levels of tensor complexity. The first is the layer level and the second is the operator level. The tensor network to be constructed is one in which each vertex of the graph (Fig.~\ref{fig: Closure 1} a) is replaced by a tensor of as many double indexes as edges the graph would have if it were completely connected, and on each possible edge there is a tensor with five indexes (Figs.~\ref{fig: Closure 1} b and c). The vertex tensors receive for half of their indexes the signal of whether that edge is activated, and for the other half they communicate how many edges they have activated. The edge tensors send if that edge is activated by two of its indexes to the two vertex tensors that join, and receive from them how many active edges each one has through other two indexes. If that edge was not activated, and the sum of signals it receives equals or exceeds the number of vertices of the graph, it activates the edge, sending the signal through its superior index. The tensor network can be seen in Fig. ~\ref{fig: Closure 1} e and the tensors in Fig.~\ref{fig: Closure 1} d.

As this process must be repeated in several iterations, a layer is placed for each iteration, so that the edge tensors receive the signal from the previous layer, to know if they were activated in the previous iteration, and perform the process. It continues until the process is finished. Finally it is measured.

\begin{figure}[h]
    \centering
    \includegraphics[width=\linewidth]{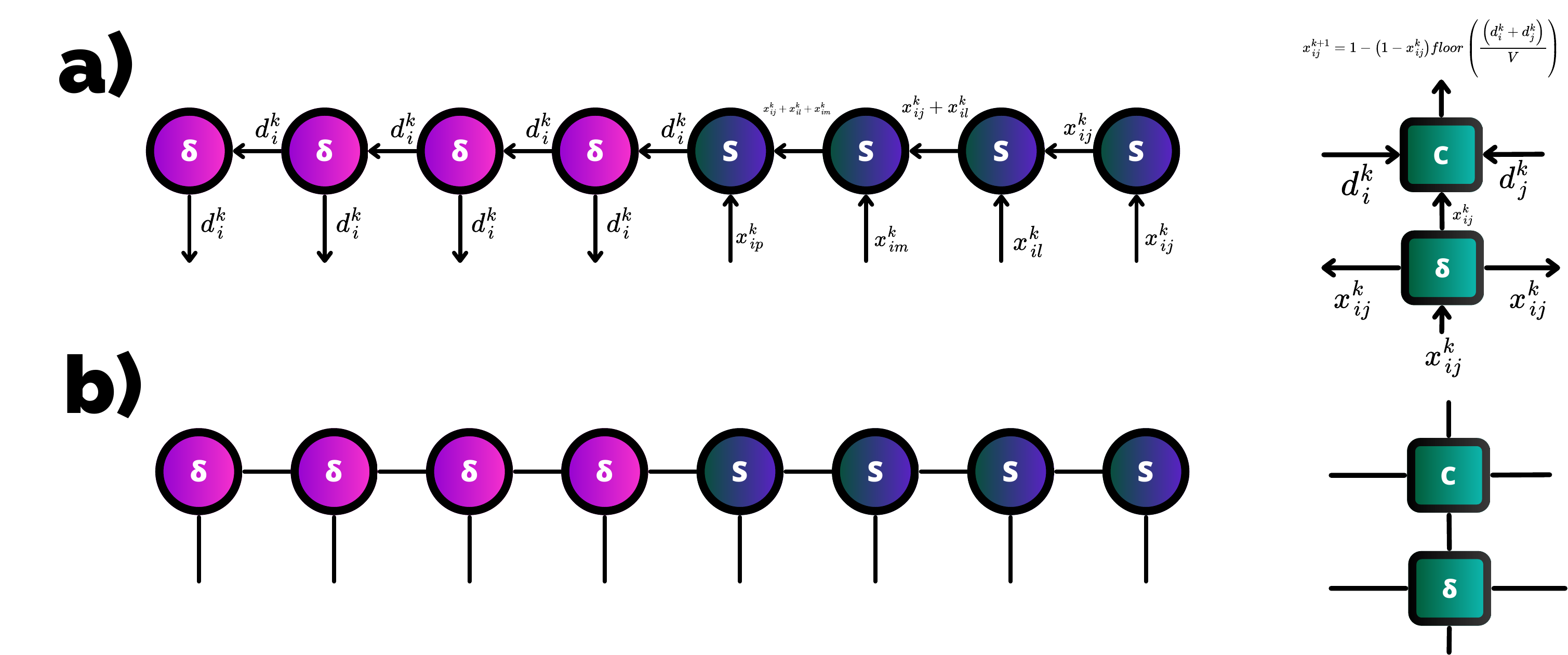}
    \caption{a) Operator decomposition for vertex and edge operators. b) Tensor decomposition for vertex and edge tensors.}
    \label{fig: Closure 1 TT}
\end{figure}

Each vertex tensor can be decomposed into a tensor train, where half of the nodes perform the sum of the active edges, and the other half communicate this sum. Each edge tensor is composed of 2 tensors, one that receives the signal of the previous state and communicates it to the vertices, and another that receives the sum signal and makes the modification. We can observe them in Fig.~\ref{fig: Closure 1 TT}. Their non-zero elements are
\begin{equation}
    \begin{gathered}
    \mu = i,\\
    \delta_{i\mu} = 1,
    \end{gathered}
\end{equation}
\begin{equation}
    \begin{gathered}
    \mu =\nu = i,\\
    \delta_{i\mu\nu} = 1,
    \end{gathered}
\end{equation}
\begin{equation}
    \begin{gathered}
    \mu =\nu =\eta= i,\\
    \delta_{i\mu\nu\eta} = 1,
    \end{gathered}
\end{equation}
\begin{equation}
    \begin{gathered}
    \mu = i,\\
    S_{i\mu} = 1,
    \end{gathered}
\end{equation}
\begin{equation}
    \begin{gathered}
    \mu = i+j,\\
    S_{ij\mu} = 1,
    \end{gathered}
\end{equation}
\begin{equation}
    \begin{gathered}
    \text{if } i=0 \text{ or } j+k<V \Rightarrow \mu = 0, \quad\text{else } \mu = 1,\\
    C_{ijk\mu} = 1.
    \end{gathered}
\end{equation}

Up to this point, the problem is a forward problem. However, we can formulate the problem of, given a known closure, obtaining the graphs that have that closure associated with it. For this, the only thing necessary is, with the same transformation layers of the tensor network, to impose by means of Projection Vectors the closure in the output of the TLC, leaving the input free to perform the Half Partial Trace.

\newpage
\section{Constraint Satisfaction Problems}
These problems consist in finding a combination that satisfies a set of constraints. They are a large family of problems, with great application and interest.

\subsection{K-colouring}
Given a graph $G$ with $V$ vertices and an adjacency matrix $E_{ij}$, the problem is to find a coloring pattern with $k$ colors of the vertices of the graph such that two vertices connected by an edge cannot have the same color. This problem has been solved with quantum algorithms~\cite{Colouring_Quantum}. This problem is a Constraint Satisfaction Problem.
\begin{figure}[h]
    \centering
    \includegraphics[width=0.9\linewidth]{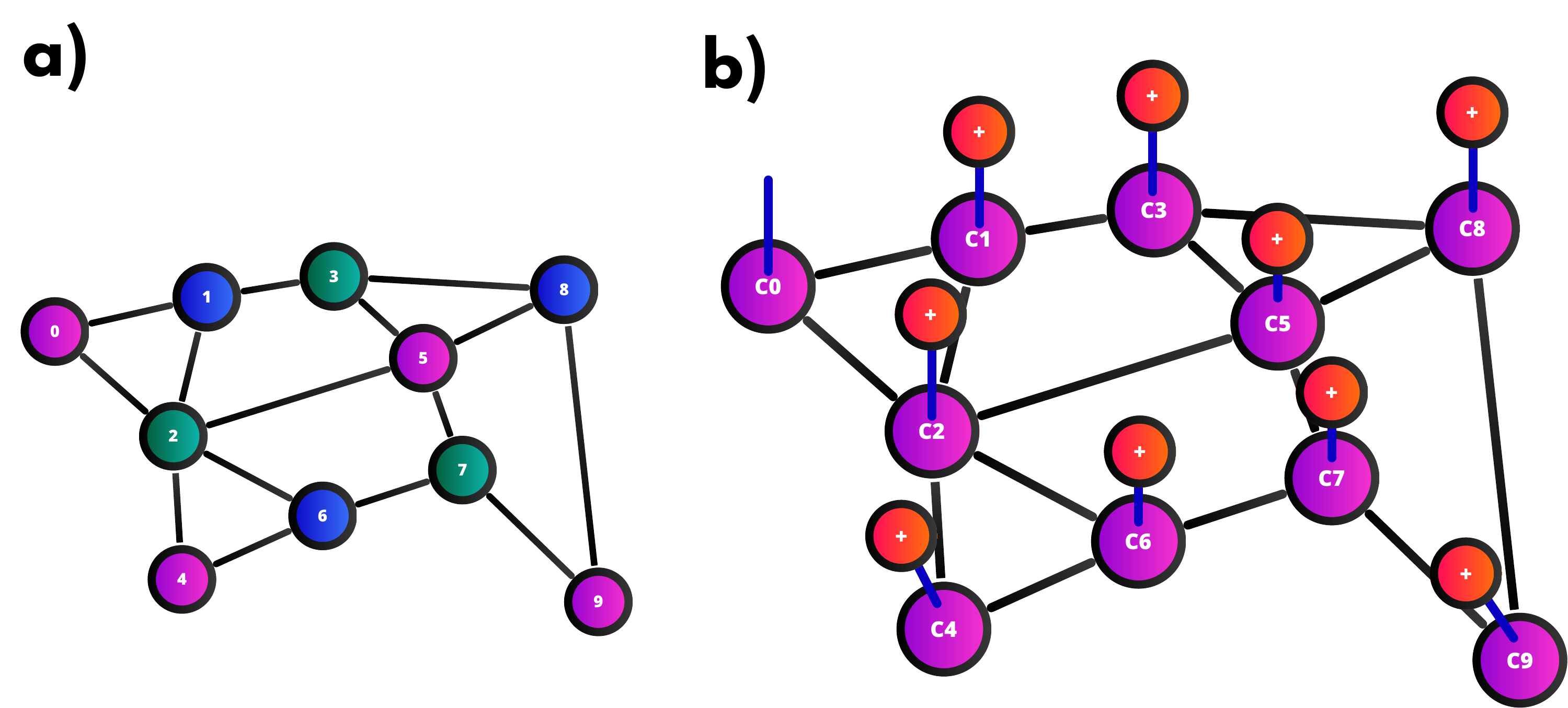}
    \caption{a) k-colouring graph with $k=3$ and 10 vertices, b) Tensor Network for the problem to determine the colour of the first vertex.}
    \label{fig: kcolouring}
\end{figure}

To solve this problem we have to make a tensor network with the same shape of the graph, so that each tensor substitutes a vertex and each index of it is an edge of the graph. Each tensor sends to those on its right the value of its color, which it sends through its upper index, and receives the value of those on its left. The tensor only has a non-zero value when the input values are all different from its output value. That is, the non-zero elements of each tensor are
\begin{equation}
    \begin{gathered}
        j_0 = j_1 = \dots= j,\quad i_n\neq j\ \forall n,\\
        C_{j,i_0,i_1,\dots, j_0,j_1, \dots} = 1
    \end{gathered}
\end{equation}

However, each tensor has as many indexes as the vertex has edges, which is not desirable, since there is an excess of information processed at the same time. For this, we obtain its tensor train for every vertex tensor.
\begin{figure}[h]
    \centering
    \includegraphics[width=0.9\linewidth]{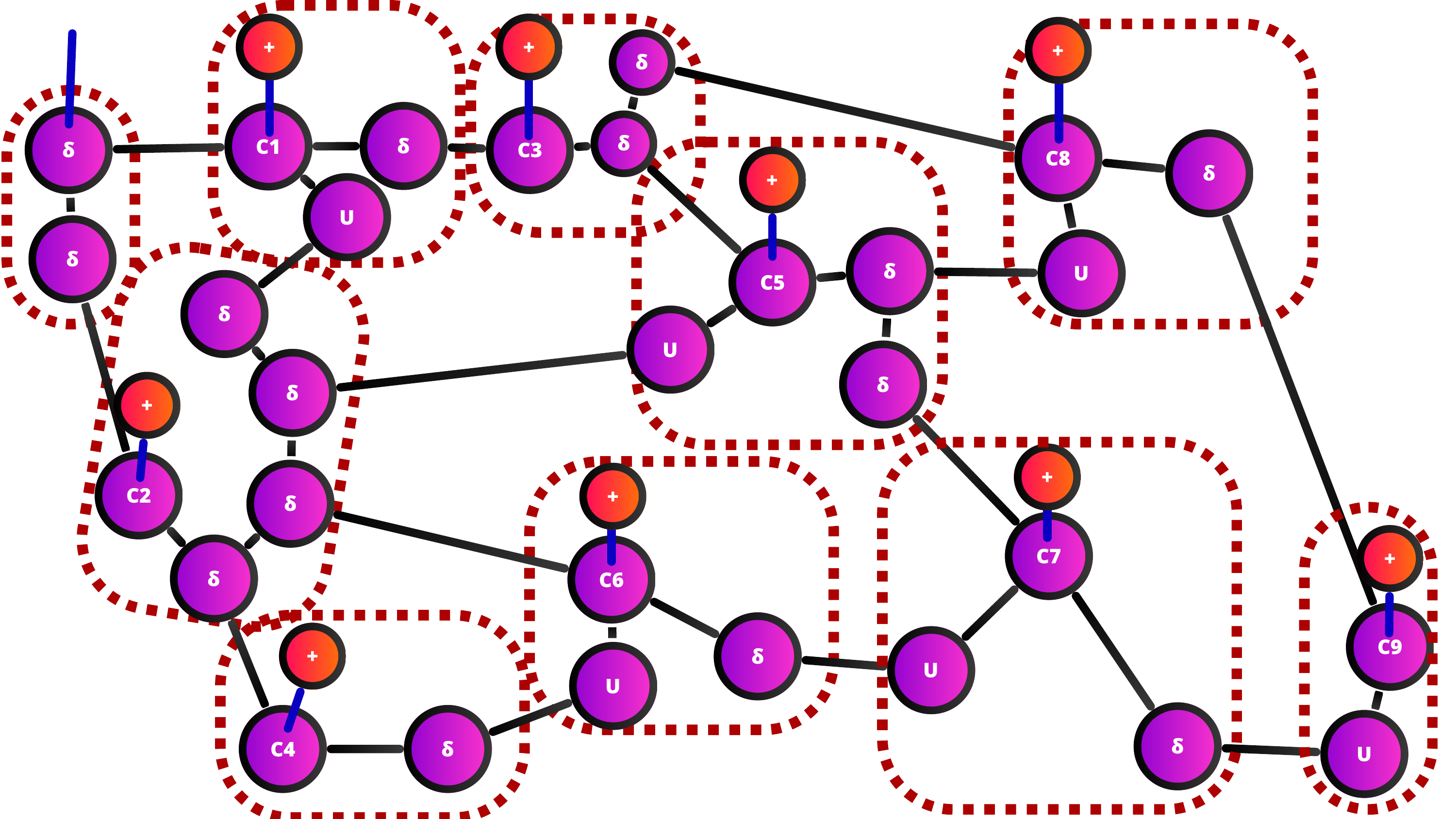}
    \caption{Simplified k-colouring tensor network.}
    \label{fig: kcolouring simple}
\end{figure}

To do this, we continue with the $C$ tensors defined previously, but this time they only have at most 4 indexes. The first is that of the color of the vertex itself, the second is one of input from a previous vertex, and the other two are for the other two types of tensors in the chain. The first new tensor type is the $\delta$, which is a Kronecker delta that only transmits the color signal to the following vertices. The second is the $U$ tensor, which receives through its top index $j_0$ the color of the same vertex from $C$, through its other index $j_1$ transmits the signal to the next $U$ tensor in the chain and from its left index $i$ receives the color of the previous vertex. The new tensor network is shown in Fig.~\ref{fig: kcolouring simple}. The $U$ non-zero elements are
\begin{equation}
    \begin{gathered}
        j_1 = j_0,\quad i\neq j_0,\\
        U_{i,j_0,j_1} = 1
    \end{gathered}
\end{equation}
If we want to know how many possible colorings exist for that graph with that number of colors, we only have to put a Plus Vector in all the indexes of the variables, so that all the amplitudes are added up. The resulting number is precisely the number of possible colorings.

If now the problem to be solved is a k-colouring in which we want to paint the graph with the minimum number of colors possible, we move on to an optimization problem. In this case, we only have to make the cost be of the type
\begin{equation}
    C(\vec{x}) = \sum_{i=0}^{N-1} x_i,
\end{equation}
so that we penalize using higher colors, and therefore, more colors. In this case, the $C$ tensors now apply imaginary time evolution, their new non-zero elements being
\begin{equation}
    \begin{gathered}
        j_0 = j_1 = j,\quad i\neq j,\\
        C_{j,i,j_0,j_1} = e^{-\tau j}.
    \end{gathered}
\end{equation}

If we now want to generalize the problem to one in which the cost depends on the coloring of each vertex, we will have the same network tensor, but with a cost function
\begin{equation}
    C(\vec{x}) = \sum_{i=0}^{N-1} Q_{i,x_i},
\end{equation}
so the non-zero elements of $C$ tensors now are
\begin{equation}
    \begin{gathered}
        j_0 = j_1 = j,\quad i\neq j,\\
        C_{j,i,j_0,j_1} = e^{-\tau Q_{j,x_j}}.
    \end{gathered}
\end{equation}

\subsection{Partition Problem}
Given a set of $N$ integers $S$, this problem consists in dividing it into two sets of numbers $S_1$ and $S_2$ such that the sums of their elements are the same~\cite{Partition}. This problem has been solved with quantum algorithms making use of QUBO formulation~\cite{Partition_Quantum}. To solve this problem, we have $N$ variables, in which $x_i$ indicates whether the $i$-th number $a_i$ belongs to the first set ($0$) or to the second ($1$). To solve it, we make a tensor network exactly like the Integer sum total function problem (Fig.~\ref{fig: Integer Sum TN}). Here, each variable is binary, and the signal to send is $r_m=\sum_{i=0}^{m-1} (-1)^{x_i} a_i  + c$, being $c = \sum_{i=0}^{m-1} a_i$. This means that, if the $i$-th number is in the $S_1$ set, we sum its value, but if it belongs to the $S_2$ set, we substract it. The offset is done to ensure the signal is never negative. The last tensor only allows the state on which the total sum is equal to $c$, because it means that we have the same sum in $S_1$ and $S_2$. The dimension of the signal indexes is $2c$, and the dimension of the variable indexes is $2$. The non-zero elements are

\begin{equation}
     \begin{gathered}
         \mu=i,\quad \nu=(-1)^i a_0+c,\\
         SUM(a,0)_{i,\mu,\nu} = 1,
     \end{gathered}
\end{equation}
\begin{equation}
     \begin{gathered}
         \mu=i,\quad \nu=j+(-1)^i a_k+c,\\
         SUM(a,k)_{i,j,\mu,\nu} = 1,
     \end{gathered}
\end{equation}
\begin{equation}
     \begin{gathered}
         \mu=i,\\
         (-1)^i a_{N-1} + j = c,\\
         EXPi(a,N-1)_{i,j,\mu} = 1.
     \end{gathered}
\end{equation}
By optimizing the tensors to a chain, as in the QUBO/QUDO first neighbor case, we can have a chain of matrices of dimension $(2c)\times (2c)$, which due to their sparsity actually have $O(4c)$ nonzero elements. Since we have $N$ different tensors, we need an $O(cN)$ space to store them. We can contract the tensor network from bottom to top, and reusing intermediate computations. We need $O(c)$ operations on each contraction, so the determination of the first variable requires $O(cN)$ operations. If we take into account that the following variables are determined based on a multiplication of a new first tensor that already has the previous information, the complexity of each step is $O(c)$. This makes the total computational complexity $O(cN)$, having a space complexity of $O(cN)$ due to the storage of the intermediate vectors.

\newpage
\section{Route and Graph Optimization problems}
Routing problems are the first constrained optimization problems we will solve. They consist in finding how to join vertexes/edges within a network, directed or undirected, or assign weights in a way that minimizes the cost of the configuration. They are particularly useful in real applications, such as delivery systems, server-to-server connections, board drilling, etc.

Although we are going to deal with interesting and complicated problems, there are others such as the minimal spanning tree that we have not addressed, because we have not found an efficient way to formulate them, so they are left as a task for the reader.

\subsection{Shortest Path Cost Problem}
Given a graph $G$ with $V$ vertices and $E$ edges between them, with an associated cost $E_{ij}$ to go from vertex $i$ to vertex $j$, this problem consists in obtaining the cost of the shortest path from vertex $a$ to vertex $b$. We assume all the $E_{ij}$ are natural numbers. If two vertexes $i,j$ are not connected, $E_{ij}=\infty$. The internal variables are the path vector
\begin{equation}
    \vec{x} = (x_0,x_1,x_2, \dots, x_{N-1}),
\end{equation}
where $x_t$ indicates the vertex on which we are at time step $t$. This follows $x_0=a$ and $x_{N-1}=b$. The cost function is
\begin{equation}
    C(\vec{x})=\sum_{t=0}^{N-2} E_{x_t,x_{t+1}}.
\end{equation}
If we want to generalize the problem, we can add that the network changes at every time step, so the cost function is now
\begin{equation}
    C(\vec{x})=\sum_{t=0}^{N-2} E^t_{x_t,x_{t+1}},
\end{equation}
where $E^t_{i,j}$ is the cost of going from the $i$-th vertex to the $j$-th vertex at time step $t$. In this cost can be added the cost of being at the vertex $x_t$ at time $t$.

Since we are not looking for the route, but rather the cost of the route, we will not determine the route itself. Moreover, since we do not know the number of steps of the optimal path, we make the cost of each vertex going to itself zero. In this way, unnecessary steps will be filled with the traveler at the same vertex, without increasing the cost. This problem is an optimization problem in which we want to know the cost value instead of the combination. We have studied it to show that the method can also solve these kind of problems.

\begin{figure}[h]
    \centering
    \includegraphics[width=0.9\linewidth]{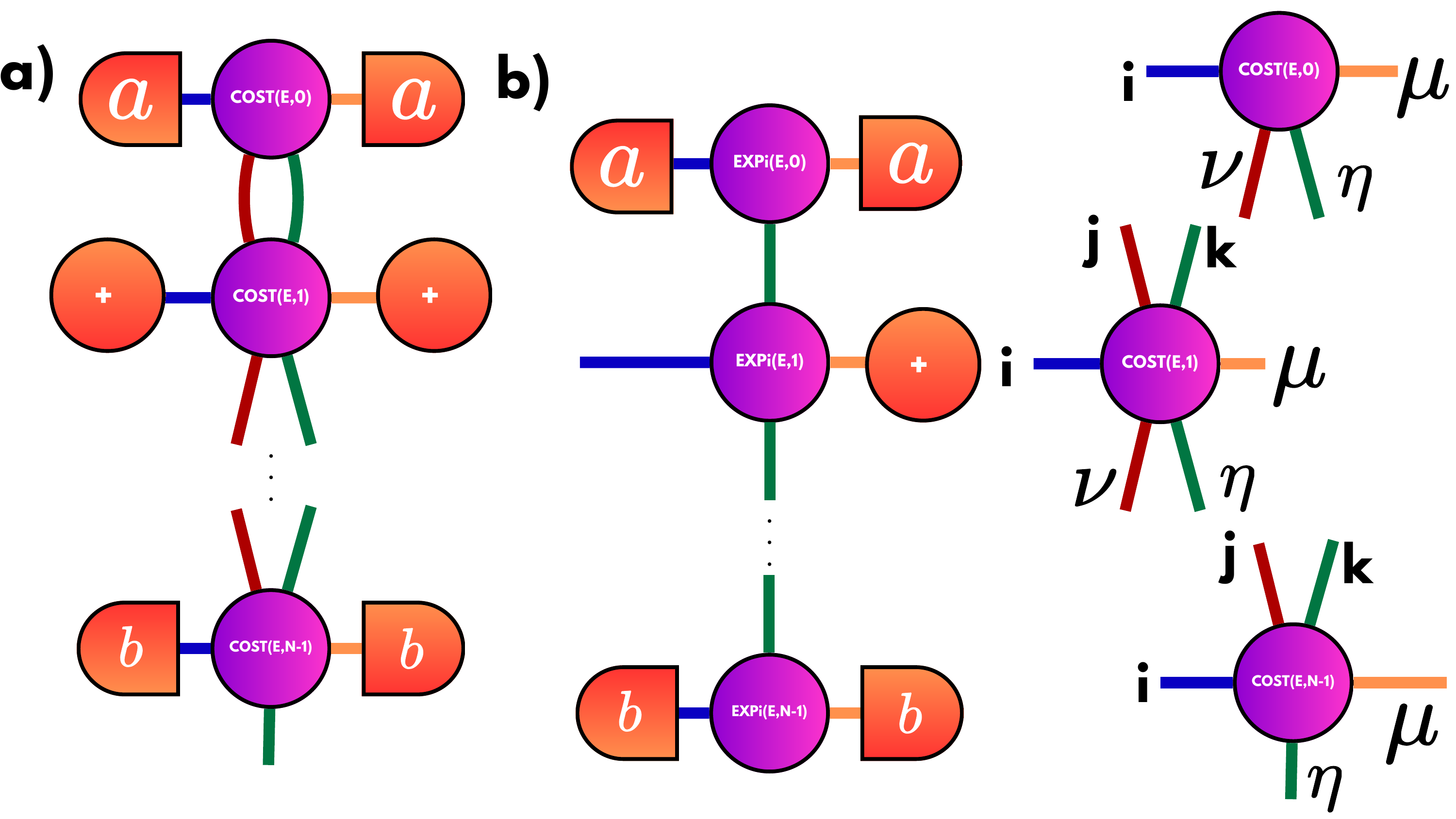}
    \caption{Tensor Network for the a) Shortest Path Cost Problem, b) Shortest Path Problem, and its notation.}
    \label{fig: Shortest Path}
\end{figure}
The tensor network to solve the problem is the one shown in Fig.~\ref{fig: Shortest Path} a, where the correct value of the first and last variable is imposed, and all the others are allowed to take any value. Each COST tensor will receive through its inputs both the value of the previous variable and the cost accumulated so far, and will return through its outputs the value of its variable and the cost after adding the one associated to its variable. The last tensor will return only the cost of the route. In this case, as we know all the costs at the end, it is not necessary to perform an imaginary time evolution, since we will only have to choose the non-zero component in lower position (which implies lower cost) as the solution. Moreover, with the resulting vector we will be able to obtain how many paths, with $N$ steps, have each of the costs.

The non zero elements of the tensors are
\begin{equation}
     \begin{gathered}
         \mu=\nu=i,\quad \eta=0,\\
         COST(E,0)_{i,\mu,\nu,\eta} = 1
     \end{gathered}
\end{equation}

\begin{equation}
     \begin{gathered}
         \mu=\nu=i,\quad \eta=k+E^t_{ji}<\infty,\\
         COST(E,t)_{i,j,k,\mu,\nu,\eta} = 1
     \end{gathered}
\end{equation}

\begin{equation}
     \begin{gathered}
         \mu=i,\quad \eta=k+E^{N-1}_{ji}<\infty,\\
         COST(E,N-1)_{i,j,k,\mu,\eta} = 1
     \end{gathered}
\end{equation}
The dimension of each of the tensors is:
\begin{itemize}
    \item $i,j,\mu,\nu$: $V,V,V,V$,
    \item $k$ for $COST(E,t)$: $\sum_{n=0}^{t-1} \max_\star(E^n)$ ,
    \item $\eta$ for $COST(E,t)$: $\sum_{n=0}^{t} \max_\star(E^n)$,
\end{itemize}
where $\max_\star(E^n)$ is the maximum finite element of $E^n$. If we want to reduce the dimensionality, we can impose that routes above a certain cost are not considered, which will be our new dimension.

With this same network we can solve the problem of the minimum number of steps to go between two vertexes $a$ and $b$, changing the matrix of edge weights to $\hat{E}^t_{i,j}=1$ for all $t,i,j$ such that $E^t_{i,j}$ is finite.

\subsection{Shortest Path Problem}
This problem is the same as the previous one, but the ultimate goal of the problem is to determine the least costly route, rather than to determine the cost itself. In this case we allow non-integer costs. This is a historically important problem, whose reference algorithm is Dijkstra's algorithm~\cite{Dijkstra1959}. As we are interested in the route and not the cost, the tensor network performs an imaginary time evolution. Each tensor receives the value of the previous variable, performs the evolution with the term of going from the previous vertex to the current one, and returns to the next tensor the value of its variable.

The tensor network is the one shown in Fig.~\ref{fig: Shortest Path} b. The non-zero elements of the tensors are 
\begin{equation}
     \begin{gathered}
         \mu=\nu=i,\\
         EXPi(E,0)_{i,\mu,\nu} = 1,\\
         EXPi(E,t)_{i,j,\mu,\nu} = e^{-\tau E^t_{ji}},
     \end{gathered}
\end{equation}

\begin{equation}
     \begin{gathered}
         \mu=i,\\
         EXPi(E,N-1)_{i,j,\mu} = e^{-\tau E^{N-1}_{ji}}.
     \end{gathered}
\end{equation}

A more in-depth study of the problem solved with this method in a more optimized version is made in {\color{red} [Pending to publish]}.

This problem can also be generalized in a similar way to the TSP we will present in the following subsection, by means of the changes in the tensor network mentioned in \cite{TSP_TN}.

\subsection{Traveling Salesman Problem (TSP)}\label{ssec: TSP}
This is a historically important problem~\cite{TSP_overview,TSP_General}, with a wide range of practical applications, and has been approached in different ways~\cite{TSP,TSP2,TSP_TN,QUBO_TSP}. Given a graph $G$ with $V$ vertices and $E$ edges between them, with an associated cost $E_{ij}$ to go from vertex $i$ to vertex $j$, the problem consists in obtaining the path with a lower associated cost which runs through all the vertexes without repeating any of them, and ends at the same initial vertex. In principle, if the network does not depend on time, we can always choose the last vertex of the path, since the problem is symmetric in time. Thus, we do not consider it as a variable, since we fix $x_{N-1}=N-1$.

To build the tensor network that solves this problem, we have two distinct and connected parts. The first part takes care of the minimization of the cost function, using the same $EXPi$ tensor layer of Fig.~\ref{fig: Shortest Path} b, slightly changing the last tensor. The second part takes care of the nonrepetition constraint, consisting of $N-2$ constraint layers. The tensor layers that apply a particular constraint to the system, without making another modification of its amplitudes, are called \textit{filter layers}.

In this case, the $k$-th constraint layer has as a signal between tensors how many times the $k$-th vertex has appeared in the path. In this way, each tensor receives the value of its variable and the signal. If vertex $k$ has not appeared, if it appears in this variable, it makes the signal 1, otherwise 0. If it has already appeared, if this variable is equal to $k$, it cancels the amplitude of the combination by multiplying it by zero. If the last tensor of the layer receives that the vertex $k$ has not yet appeared, it forces it to appear in this variable, by canceling all the amplitudes of the rest of the possibilities. With this we force the vertex $k$ to appear once and only once. If we apply this to the $V-2$ first vertexes of the graph, since there is only one remaining possibility, it will have to be that of the remaining vertex, so we do not need a layer for the $k=V-2$.
\begin{figure}
    \centering
    \includegraphics[width=0.7\linewidth]{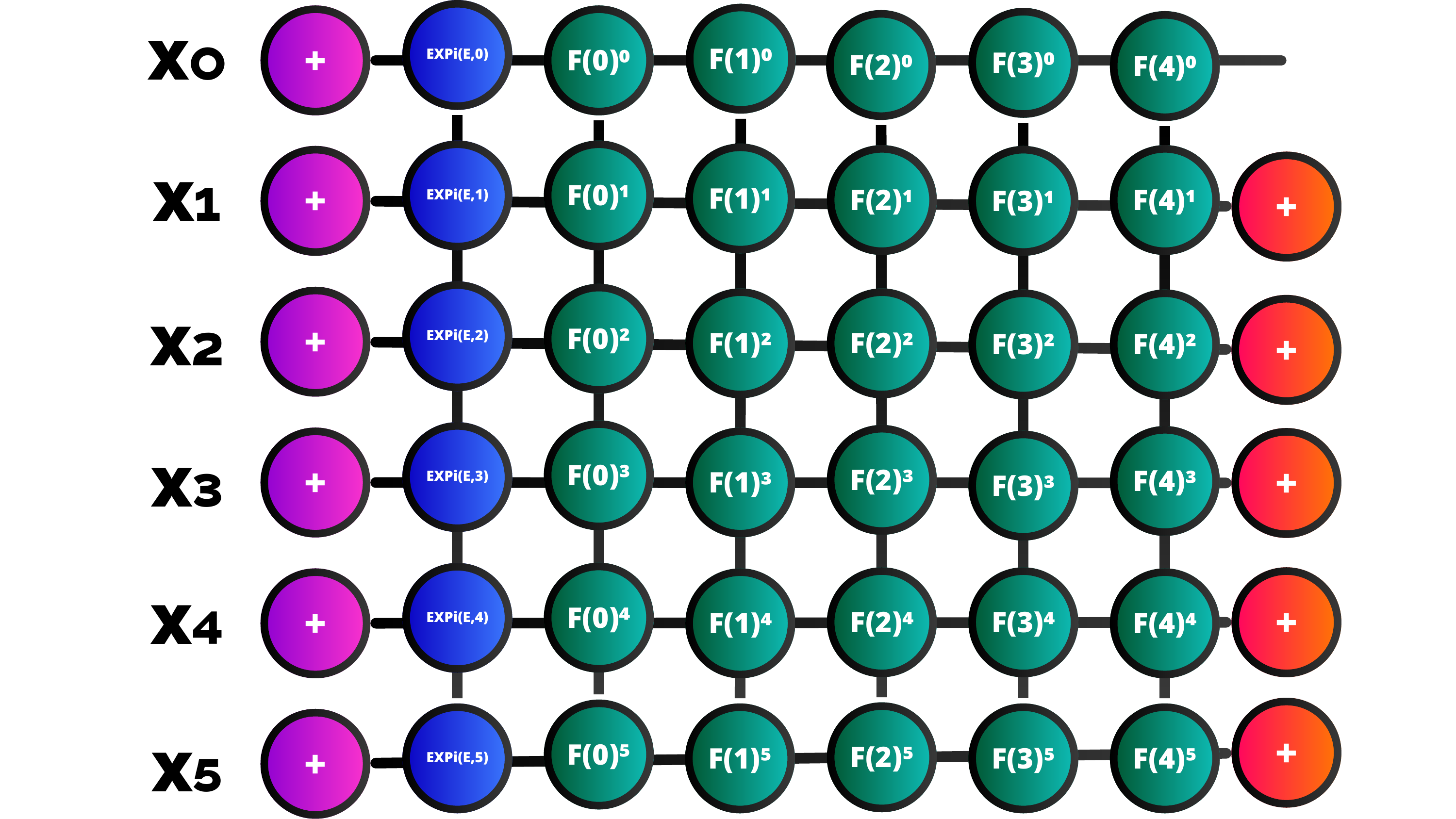}
    \caption{Tensor network for the TSP with 6 variables.}
    \label{fig: TSP TN}
\end{figure}

The tensor network of the problem is shown in Fig.~\ref{fig: TSP TN}. The non-zero elements of the tensors are
\begin{equation}
     \begin{gathered}
         \mu=\nu=i,\\
         EXPi(E,0)_{i,\mu,\nu} = e^{-\tau E_{V-1,i}},\\
         EXPi(E,t)_{i,j,\mu,\nu} = e^{-\tau E_{ji}},
     \end{gathered}
\end{equation}
\begin{equation}
     \begin{gathered}
         \mu=i,\\
         EXPi(E,V-2)_{i,j,\mu} = e^{-\tau (E_{ji}+E_{i,V-1})}.
     \end{gathered}
\end{equation}
\begin{equation}
     \begin{gathered}
         \mu=i,\\
         \text{if } i=k\Rightarrow \nu=1 ,\quad \text{ else } \nu=0,\\
         F(k)^0_{i,\mu,\nu} = 1,
     \end{gathered}
\end{equation}
\begin{equation}
     \begin{gathered}
         \mu=i,\\
         \text{if } i=k\Rightarrow j=0, \nu=1 ,\quad \text{ else } \nu=j,\\
         F(k)^t_{i,j,\mu,\nu} = 1,
     \end{gathered}
\end{equation}
\begin{equation}
     \begin{gathered}
         \mu=i,\\
         \text{if } i=k\Rightarrow j=0,\quad \text{ else } j=1,\\
         F(k)^{V-2}_{i,j,\mu} = 1,
     \end{gathered}
\end{equation}

A deeper analysis of this problem and its generalizations is made in the paper \cite{TSP_TN}, where this method is explained in more detail. It can also be generalized to a time-dependent version in the same way as in the shortest path. The longest path problem can be solved with the same tensor network, simply by changing the sign of the exponentials of the imaginary evolution, so that it penalizes the shortest paths.

\subsubsection{Grid Graph TSP}
\begin{figure}[h]
    \centering
    \includegraphics[width=\linewidth]{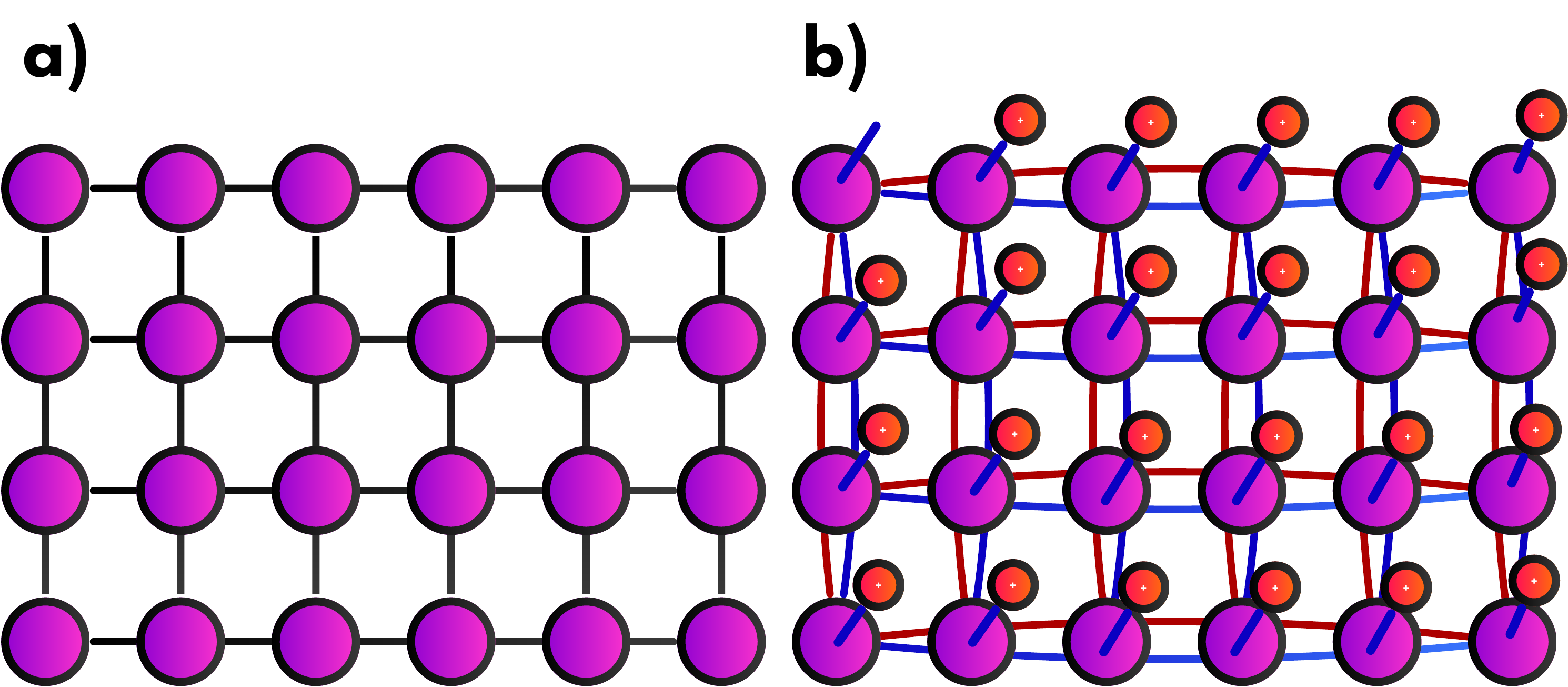}
    \caption{Graph of the TSP in square grid and its associated tensor network.}
    \label{fig: Grid_TSP_TN}
\end{figure}

An interesting particular case is the TSP with a square grid graph as in Fig.~\ref{fig: Grid_TSP_TN} a. In this case it is interesting to formulate the tensor network as a graph in which we replace each vertex by a tensor, and each edge by two indexes. Now our variables $x_i$ is the time step in which we are at vertex $i$.  With this, to perform the imaginary time evolution, each operator needs to know what is the state of its neighbors and itself. In this way, if one of the neighbors has a state one unit lower than its own, this implies that it is the previous vertex in the path, and it has to make the evolution with the cost of going to that vertex until itself. In addition, if there is no neighbor in a justly lower time, the state is destroyed, since it implies that the traveler has `teleported' to the vertex. The same if there is none just above. This applies in a circular way for the first and last time. This forces the route to be continuous and not to cut. In addition, we also eliminate, just in case, the possibilities of same time in the vertexes that are not the previous or the next. For connectivity issues of the tensor network, in this resolution we can only make the costs between vertexes change every time, but not the connectivity between them. That is, it must always be square, or remove connections with respect to it, but not introduce diagonals or higher distance connections.

In this way, each operator receives 4 inputs and its state and sends 4 equal outputs, which are its state. In tensorization, this implies that the tensors have 9 indexes of dimension $V$. This can be simplified as in the case of k-colouring, but let us treat it with the denser version. The non-zero elements of the $S^k$ tensor of the network for the $k$-variable are
\begin{equation}
     \begin{gathered}
        j_{up}=j_{down}=j_{left}=j_{right}=j,\\
        (i_{up} = j-1 \text{ or } i_{down} = j-1 \text{ or } i_{left} = j-1 \text{ or } i_{right}= j-1),\\
        (i_{up} = j+1 \text{ or } i_{down} = j+1 \text{ or } i_{left} = j+1 \text{ or } i_{right}= j+1),\\
        i_{pos} \neq i_{pos'}\ \forall pos\neq pos',\\
        S^k_{i_{up},i_{down},i_{left},i_{right},j,j_{up},j_{down},j_{left},j_{right}} = e^{-\tau E_{y,k}},
     \end{gathered}
\end{equation}
being $i_{up}$, $i_{down}$, $i_{left}$ and $i_{right}$ the inputs, $j$, $j_{up}$, $j_{down}$, $j_{left}$ and $j_{right}$ are the outputs and $y$ the vertex with the $j-1$ value. All tensors with less amount of indexes follow the same logic. It is important to note that the index $i_{pos}$ of a tensor is the $j_{pos'}$ of the tensor to which it is attached, $pos'$ being the inverse position of $pos$.

If the lattice has dimensions $N\times M$, being $M\geq N$, we have to contract it from right to left. In a simplified version, where each tensor already has only 8 indexes, it is easy to verify that the computational complexity of this contraction results in $O(V^{2N+5})$ in time, with $O(V^{2N+2})$ in space, without sparsity considerations. By reusing intermediate calculations, the total time complexity is the same, but space complexity $O(V^{2N+3})$. In the worst case, when $M=N=\sqrt{V}$, the time complexity is $O(V^{2\sqrt{V}+5})$ and the space complexity is $O(V^{2\sqrt{V}+3})$. This result shows that every TSP can be solved with a time complexity of $O(V^{2L})$, being $L$ the dimension of the longest section perpendicular to the longest dimension line of the graph.

In case there are holes in the grid such that the vertexes are connected through it, it will only be necessary to replace the tensors of the non-existent vertexes by a $Pass$ tensor whose non-zero elements are
\begin{equation}
     \begin{gathered}
        j_{up}=j_{down}, j_{left}=j_{right},\quad i_{up}=i_{down}, i_{left}=i_{right},\\
        Pass^k_{i_{up},i_{down},i_{left},i_{right},j_{up},j_{down},j_{left},j_{right}} = 1.
     \end{gathered}
\end{equation}
This tensor is responsible for passing the information from one vertex to another through the hole.

If there are holes without internal connections, simply remove both the corresponding vertex tensor and the vertex connections. The computational complexity in both cases remains the same as shown above.

\subsection{Vehicle routing problem}
This problem consists in traversing an entire graph $G$ with a number $M$ of vehicles without repeating any vertex, all leaving and returning to the same vertex, and with the minimum possible global total route cost~\cite{VRP}. This is a generalization of the TSP. This problem has been approached with quantum computing~\cite{VRP_Quantum}. In this case, instead of having a set of variables $x_t$ indicating the vertex at time step $t$, one has a set of variables $x_{v,t}$, indicating the vertex at which the vehicle $v$ is located at time step $t$.

\begin{figure}[h]
    \centering
    \includegraphics[width=0.7\linewidth]{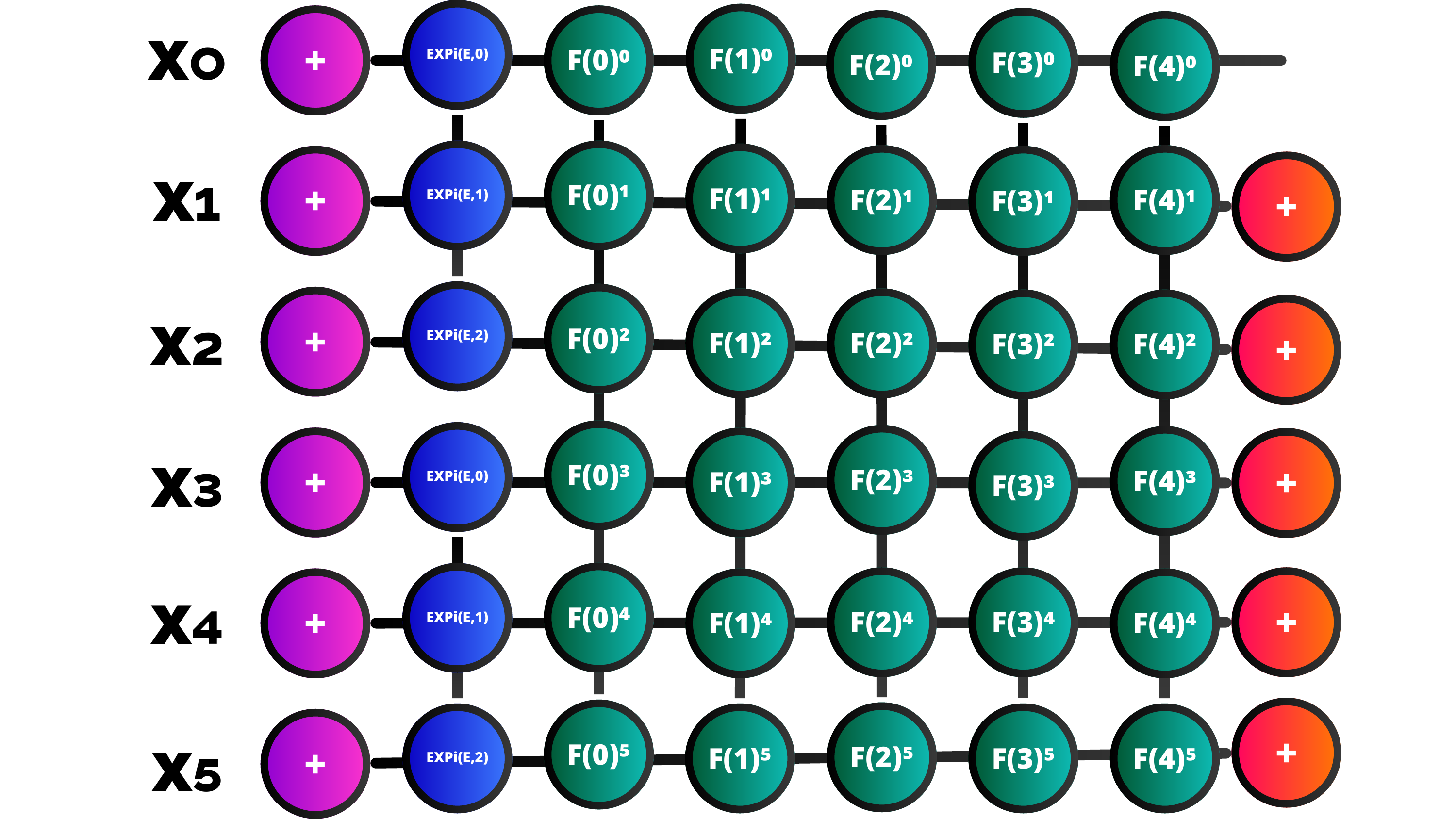}
    \caption{Tensor network for the Vehicle routing problem with two vehicles and 6 vertexes. In this case, each vehicle must visit exactly 3 vertexes.}
    \label{fig: VRP TN}
\end{figure}

To approach it with a tensor network, we only need to take the same tensor network as for the TSP (and any of its variants) for each of the vehicles. Since between them they have to cover all the vertices of the network, and pass through each one only once, the constraint layers are joined to each other, so that the constraint signal passes from the network of vehicle $n$ to that of vehicle $n+1$. The tensor network is shown in Fig.~\ref{fig: VRP TN}. It is important to note that, in case we do not want the vehicles to visit a specific number of vertices each, we must make it so that each one can visit all of them, but set a zero cost for staying at the origin vertex.

In this way, the computational complexity is exactly the same as that of the TSP presented above, and the same for all generalizations. Moreover, if we change the evolution layers between vehicles, we can generalize to the case in which the networks are different for each vehicle. We can also model the cost of using new vehicles by adding an extra cost for leaving the origin node.

\subsection{Chinese postman problem}
Given a graph $G$ of $V$ vertices and $E$ edges with costs $E_{ij}$ to go from vertex $i$ to vertex $j$, the problem consists in finding the least costly path of $T$ steps that traverses each edge at least once.

To solve this problem, the variables are a vector $\vec{x}$ where the $x_t$ component indicates on which edge the traveler is at time-step $t$. This makes the cost function dependent only on each variable
\begin{equation}
    C(\vec{x})=\sum_{t=0}^{T-1}E_{V({x_t})}, 
\end{equation}
being $V(k)$ the pair $(i,j)$ of vertexes connected by edge $k$. 

With this in mind, we have to impose the continuity constraint so that the path is a single chain. In order to apply it minimally, each edge is split into two edges, one to go from $i$ to $j$ and one to go from $j$ to $i$. The pair of edges is obtained by the function $E(i,j)$. If one of the original edges had only one possible direction, then a new one is not created for the nonexistent direction. Thus, if at time $t$ we are on edge $k$, at time $t+1$ we can only be on the edges $D(k)$ that leave from the destination vertex of $k$. This causes $x_t\in [0,2E-1]$.

Since each edge has to appear at least once, there is also the constraint that
\begin{equation}
    \exists t\ |\ x_t \in E(i,j)\ \forall i,j \in [0,V-1].
\end{equation}

\begin{figure}[h]
    \centering
    \includegraphics[width=0.7\linewidth]{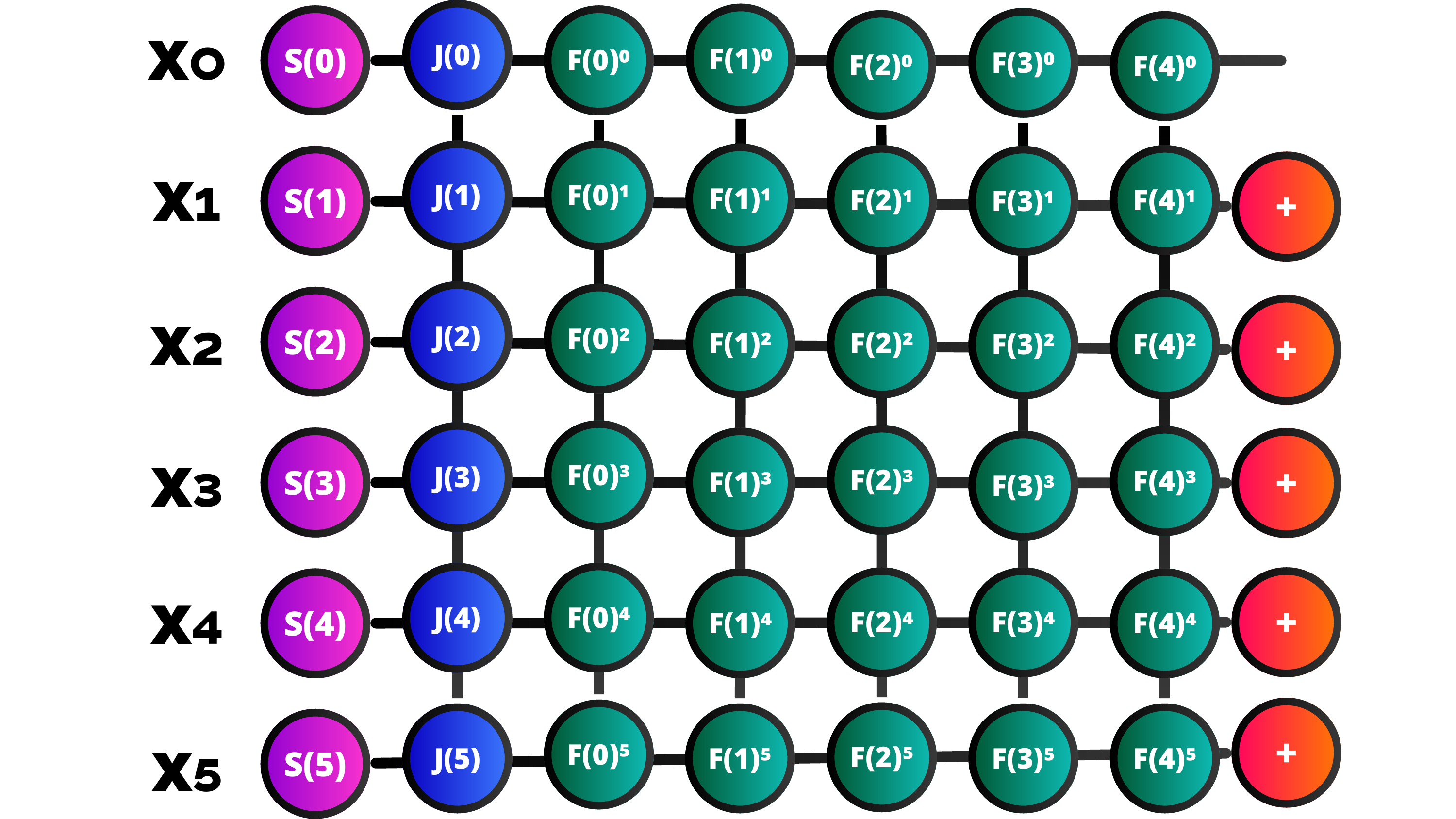}
    \caption{Tensor Network that solves the Chinese postman problem with 5 edges and 6 time steps.}
    \label{fig: Chinese TN}
\end{figure}

The tensor network we create is the same as the tensor network of the TSP in structure, but changing the tensors involved. It is shown in Fig.~\ref{fig: Chinese TN}. The first layer of tensors $S$ takes care of the evolution in imaginary time and the superposition at the same time. The second tensor layer $J$ handles the connectivity between parts of the path. The next $F$ counting layers take care of guaranteeing that each original edge is traversed at least once. The non-zero elements of each tensor are
\begin{equation}
        S(t)_i = e^{-\tau E_{V(i)}}.
\end{equation}
\begin{equation}
     \begin{gathered}
        \mu=\nu=i,\\
        J(0)_{i,\mu,\nu} = 1,
     \end{gathered}
\end{equation}
\begin{equation}
     \begin{gathered}
        i=D(j),\quad \mu=\nu=i,\\
        J(t)_{i,\mu,j,\nu} = 1,
     \end{gathered}
\end{equation}
\begin{equation}
     \begin{gathered}
        i=D(j),\quad \mu=i,\\
        J(T-1)_{i,\mu,j} = 1,
     \end{gathered}
\end{equation}
\begin{equation}
     \begin{gathered}
        \mu=i,\\
        \text{if } i=n\Rightarrow \nu=1,\quad \text{ else } \nu=0,\\
        F(n)^0_{i,\mu,\nu} = 1,
     \end{gathered}
\end{equation}
\begin{equation}
     \begin{gathered}
        \mu=i,\\
        \text{if } i=n\Rightarrow \nu=1,\quad \text{ else } \nu=j,\\
        F(n)^t_{i,\mu,j,\nu} = 1,
     \end{gathered}
\end{equation}
\begin{equation}
     \begin{gathered}
        \mu=i,\\
        \text{if } j=0\Rightarrow i=n,\\
        F(n)^{T-1}_{i,\mu,j} = 1.
     \end{gathered}
\end{equation}

If we want to address a generalization in which the weights of each edge change at each time step, we can do so by changing the cost function to
\begin{equation}
    C(\vec{x})=\sum_{t=0}^{T-1}E^t_{V({x_t})} .
\end{equation}

If we want each edge $(i,j)$ to appear between $N^0_{(i,j)}$ and $N^f_{(i,j)}$ times, we just have to change the constraint layers as shown in \cite{TSP_TN}. The new tensors are
\begin{equation}
        S(t)_i = e^{-\tau E^t_{V(i)}}.
\end{equation}
\begin{equation}
     \begin{gathered}
        \mu=i,\\
        \text{if } i=n\Rightarrow \nu=\min(j+1,N^f_n),\quad \text{ else } \nu=j,\\
        F(n)^t_{i,\mu,j,\nu} = 1.
     \end{gathered}
\end{equation}
\begin{equation}
     \begin{gathered}
        \mu=i\\
        N^0_n-1\leq j\leq N^f_n,\\
        \text{if } i=n\Rightarrow j<N^f_n,\\
        \text{if } j=N^0_n-1\Rightarrow i=n,\\
        F(n)^{T-1}_{i,\mu,j} = 1.
     \end{gathered}
\end{equation}

\subsection{Minimal Cost Closure Problem}
In this problem a closure of a directed graph is defined as a set of vertices $C$ such that none of its edges points to a vertex outside $C$. Given a weighted directed graph, the objective is to find the closure with the smallest associated weight. This problem has been solved with quantum computing algorithms~\cite{Postman_Quantum}.

\begin{figure}[h]
    \centering
    \includegraphics[width=\linewidth]{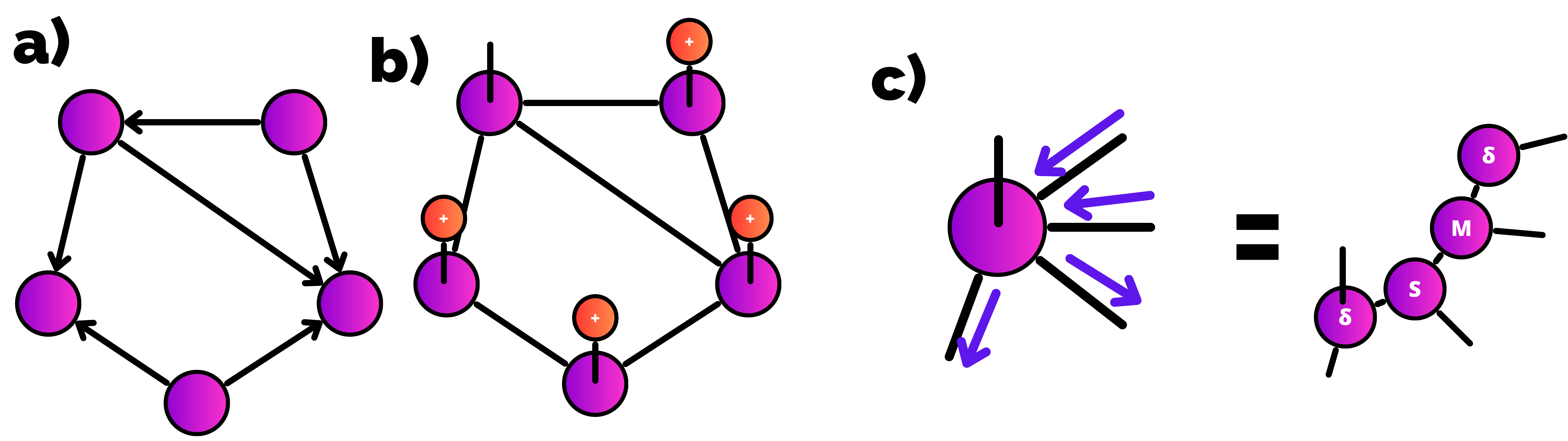}
    \caption{a) Directed graph. b) Tensor Network for its Minimal Cost Closure Problem. c) Tensor Train decomposition of the operator.}
    \label{fig: Closure 2}
\end{figure}

For this problem, we replace each vertex of the original graph by a tensor with an index for each edge of the vertex, and an index to indicate whether that vertex belongs to the closure. Each tensor will consider the superposition of whether or not it is activated (its vertex belongs to the closure). If a tensor is activated, it performs the evolution in imaginary time of the edges coming out of it. In addition, it sends by the indexes corresponding to these the signal that it is activated. If it is not activated, it does not perform the evolution, and sends the signal that it is deactivated. In case that any of the incoming edges communicates an activated signal, the tensors that receive it will also be activated and will make the evolution in imaginary time, sending its signal to the following ones. If it does not receive any activation, it may or may not be activated. The tensor network can be seen in Fig.~\ref{fig: Closure 2} b. Each tensor has $x_i$ indexes for its inputs, $y_i$ indexes for its outputs and $z$ index for its state. It can be decomposed in a tensor train of $\delta$ tensor to pass the signal of its own activation, a $S$ tensor which determines if the state has to be $1$ and performs the imaginary time evolution, and $M$ tensors, which determines if some previous vertex is active. Its non-zero elements are
\begin{equation}
    \begin{gathered}
    \mu =\nu = i,\\
    \delta_{i\mu\nu} = 1,
    \end{gathered}
\end{equation}
\begin{equation}
    \begin{gathered}
    \text{if } i=0\Rightarrow \mu\in\{0,1\},\quad \text{ else } \mu=1,\\
    \nu = \mu,\\
    S^k_{i\mu\nu} = e^{-\tau \mu\sum_{j\in D(k)} E_{k,j}},
    \end{gathered}
\end{equation}
\begin{equation}
    \begin{gathered}
    \mu =\max(i,j),\\
    M_{ij\mu} = 1,
    \end{gathered}
\end{equation}
being $D(k)$ the set of vertexes with an edge that starts in $k$-th vertex and $E_{k,j}$ the cost of the edge connecting $k$ with $j$. If we want to take into account only the number of vertexes in the closure, we can fix all the $E_{i,j}=1$.

\subsection{Maximum Flow Problem}
This problem formulated in 1954~\cite{Maximum_Flow} consists in finding the maximum flow routing for a flow network. A flow network is a directed graph of $V$ vertices and $E$ edges, with a source $S$ and a sink $T$. The capacity $E_{ab}$ of an edge going from vertex $a$ to $b$ is the maximum amount of flow it can let through the edge. The flux is a function $f$ of the edges it satisfies:
\begin{itemize}
    \item The flow through an edge from $a$ to $b$ cannot exceed its capacity. That is, $f_{ab}\leq E_{ab}$.
    \item The sum of flows entering a vertex must be equal to the sum of flows leaving it, except for the source and sink. That is, $$\sum_{a: (a,b)\in E, f_{ab}>0} f_{ab} = \sum_{a: (b,a)\in E, f_{ba}>0} f_{ba},\ \forall b\in V\backslash \{S,T\}.$$
\end{itemize}
The value of flow is the amount of flow going from the source $S$ to the sink $T$. That is,
\begin{equation}
    |f| = \sum_{b: (S,b)\in E} f_{Sb} = \sum_{a: (a,T)\in E} f_{aT}
\end{equation}
The maximum flow problem is to route as much flow as possible from the source to the sink. That is, obtain the maximum $|f|$ possible. It has been solved with quantum annealing~\cite{Max_Flow_Quantum}.

\begin{figure}
    \centering
    \includegraphics[width=0.9\linewidth]{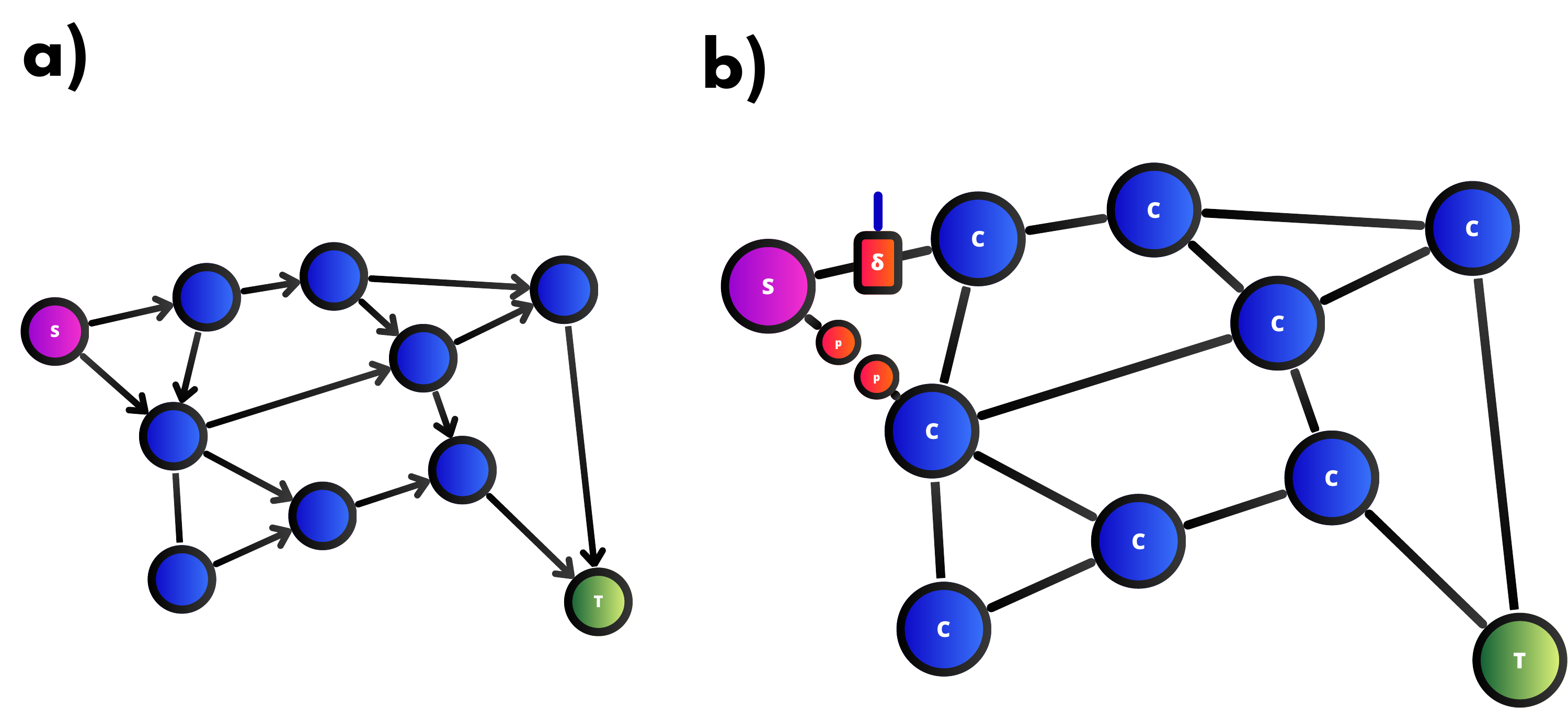}
    \caption{a) Maximum Flow graph, b) Tensor Network to determine the upper edge flow after determine the flow of the below edge.}
    \label{fig: Maximum Flow}
\end{figure}

To solve this problem with integer flows and capacities, we make a tensor network with the same structure as the network, where each vertex is a tensor and each edge an index. In addition, we add a tensor on each edge. The vertex tensors receive by their input indexes the amount of flow entering the vertex through that edge, and return by their output indexes the amount of outgoing flow through that vertex. These tensors impose the flow continuity limit constraints, while the dimension of their indexes already impose the capacity limits. The edge tensors are used to determine the amount of flow crossing that edge. To avoid an excessive use of tensors, they can be placed only on the edge to be determined in that step, as a Kronecker delta of 3 indexes, and on the edges already determined, place two Projector Vectors, one in each direction of the edge. The source tensor performs an imaginary time evolution, so that the combination with a higher flux has a higher amplitude. The indexes between vertex $a$ and vertex $b$ are of dimension $f_{a,b}+1$, to take into account the capacity limit of the edge joining them, and the possibility of having null flow. We can see an example of this tensor network in Fig.~\ref{fig: Maximum Flow}.  

\begin{figure}
    \centering
    \includegraphics[width=\linewidth]{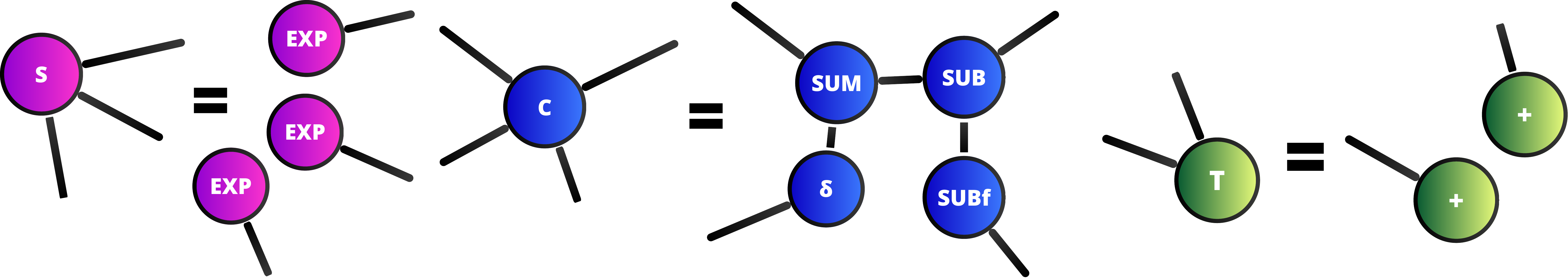}
    \caption{Decomposition of the vertex tensors.}
    \label{fig: Maximum Flow Decomp}
\end{figure}

All the vertex, source and sink tensors can be decomposed as shown in Fig.~\ref{fig: Maximum Flow Decomp}. The non-zero elements of the tensors are
\begin{equation}
    \begin{gathered}
        \mu = \nu = i,\\
        \delta_{i\mu\nu} = 1,
    \end{gathered}
\end{equation}
\begin{equation}
    \begin{gathered}
        \mu = i + j,\\
        SUM_{ij\mu} = 1,
    \end{gathered}
\end{equation}
\begin{equation}
    \begin{gathered}
        EXP_{i} = e^{\tau i},
    \end{gathered}
\end{equation}
\begin{equation}
    \begin{gathered}
        \mu = i-j,\\
        SUB_{ij\mu} = 1,
    \end{gathered}
\end{equation}
\begin{equation}
    \begin{gathered}
        \mu= i,\\
        SUBf_{i\mu} = 1.
    \end{gathered}
\end{equation}

\subsection{Maximum Independent Set Problem}
This problem consists in, given a graph $G$ of $V$ vertices and $E$ edges, finding the largest possible independent set. An independent set is a set of vertices of the graph such that no two vertices of the graph share an edge. This problem has been approached with quantum techniques~\cite{Independet_Quantum}.

To solve this problem, we only need to replace each vertex by a tensor and each edge by an index. Each vertex tensor represents a variable. If that variable is $0$, then the vertex does not belong to the independent set, and if it is $1$, then it belongs to the independent set. It is the same TLC as in Fig.~\ref{fig: kcolouring}. The operation of the tensors is exactly the same as in the case of k-colouring, with $k=2$ colors, being that the color $0$ can be repeated. Therefore, all index dimensions in the tensor network are 2. The decomposition of the tensors is the same, but now the U-tensors do not prevent the repetition of the $0$ state. In addition, now the $C$-tensors perform an imaginary time evolution for the $1$-state, to favor combinations with the largest number of vertices in the independent set. In this way, the non-zero elements of the new tensors are
\begin{equation}
    \begin{gathered}
        j_0 = j_1 = j,\\
        \text{if } i=0\Rightarrow j\in\{0,1\},\quad \text{ else } j=0,\\
        C_{j,i, j_0,j_1} = e^{\tau j},
    \end{gathered}
\end{equation}
\begin{equation}
    \begin{gathered}
        \text{if } i=0\Rightarrow j_0\in\{0,1\},\quad \text{ else } j_0=0,\\
        j_1 = j_0,\\
        U_{i,j_0,j_1} = 1.
    \end{gathered}
\end{equation}

\subsection{Minimum Vertex Cover Problem}
This problem consists in, given a graph $G$ with $V$ vertices and $E$ edges, finding the vertex cover with the smallest possible number of vertices~\cite{Vertex_Cover}. A vertex cover is the set of vertices such that every edge of the graph has at least one of its ends inside the set. This problem has been solved with quantum algorithms~\cite{Vertex_Cover_Quantum}, making use of Grover search algorithm~\cite{Grover}.

\begin{figure}[h]
    \centering
    \includegraphics[width=0.9\linewidth]{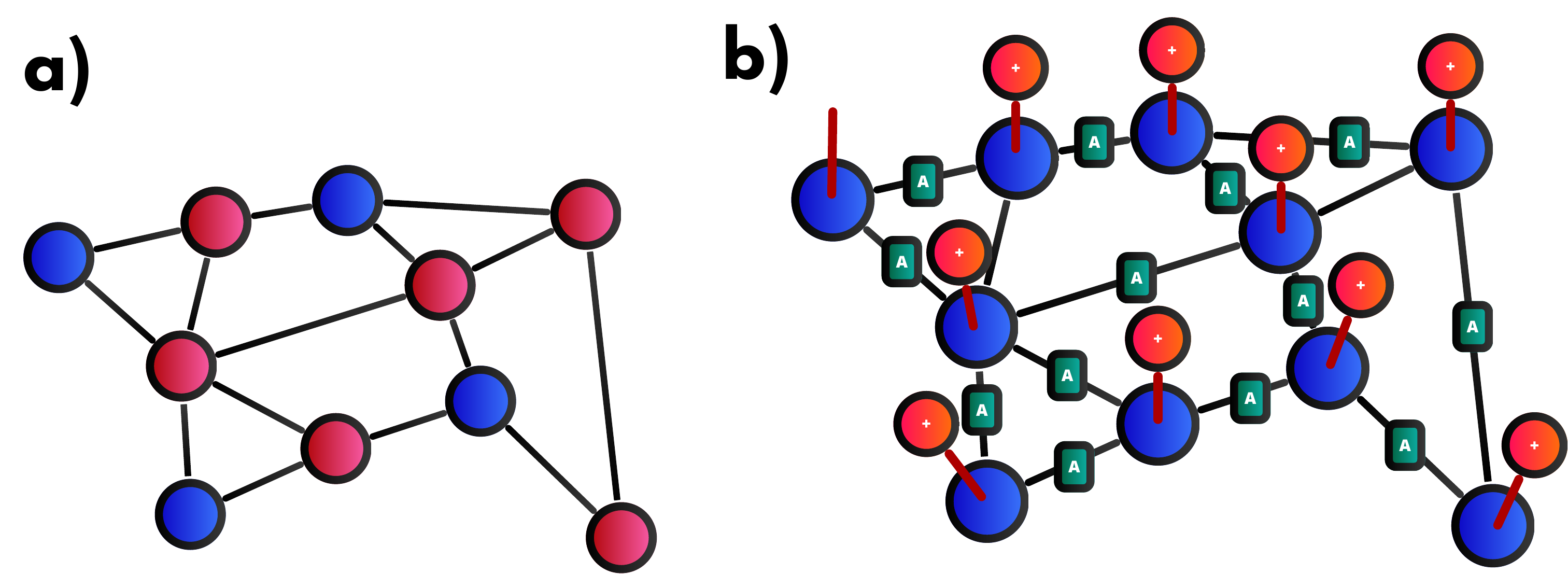}
    \caption{a) Vertex Cover Solution, b) Tensor Network for the Minimum Vertex Cover Problem.}
    \label{fig: Vertex Cover}
\end{figure}

To solve this problem, our variables $x_i$ indicate whether the $i$-th vertex belongs to the vertex cover. If $x_i=1$ it belongs to the vertex cover, otherwise it does not. Each vertex is replaced by a vertex tensor with one index per edge connected to that vertex, and an $A$-tensor is placed on each edge. The tensor network is shown in Fig.~\ref{fig: Vertex Cover}. The vertex tensors act as Kronecker deltas, sending as a signal the value of that vertex. To minimize the number of vertices in the vertex cover, they perform an imaginary time evolution that reduces the amplitude for each $1$-valued variable. This implies that this tensor is of the type $e^{-\tau i} \delta_{i,j_0,j_1,j_2,j_2,\dots}$. As always, an $n$-index delta can be decomposed into $n-2$ $3$-index deltas and two $2$-index deltas. The $A$ tensors ensure that its edge must have at least one end in the cover. Therefore, its non-zero elements are
\begin{equation}
    \begin{gathered}
        \mu \leq \max(0,1-i),\\
        A_{i\mu} = 1.
    \end{gathered}
\end{equation}

\subsection{Dominating set problem}
This problem consists in finding a dominating set for a graph $G$. A dominant set is a subset $D$ of its vertices, such that any vertex of $G$ is in or has a neighbor in $D$. The variable $x_i$ indicates whether the $i$-th vertex belongs to the dominating set or not. This problem has been solved with quantum algorithms~\cite{Dominant_Grover,Dominant_Set_Quantum}.

For solving this problem, we need a tensor network as the presented in Fig.~\ref{fig: kcolouring}. Each vertex of the graph is transformed into a vertex tensor and each edge into two indexes. Each vertex tensor sends to the adjacent ones its state through its output indexes, and receives the state of each of them through its input indexes. To guarantee that each node is or is adjacent to a vertex of the dominating set, the state is removed only if this tensor receives that all its neighbors and itself are in $0$. That is, the non-zero elements of a vertex tensor $C$ are
\begin{equation}
    \begin{gathered}
    i_0 = i_1 = \dots = i,\quad \sum_k j_k + i\geq 1,\\ 
        C_{i,i_0,i_1,\dots, j_0, j_1,\dots} = 1.
    \end{gathered}
\end{equation}
If we want to find the smallest dominating set, we can apply the cost function
\begin{equation}
    C(\vec{x}) = \sum_i c_i x_i,
\end{equation}
being $c_i$ the cost of having the $i$-vertex in the dominating set. In this case, the $C^a$ tensor for the $a$-th vertex is
\begin{equation}
    \begin{gathered}
    i_0 = i_1 = \dots = i,\quad \sum_k j_k + i\geq 1,\\ 
        C^a_{i,i_0,i_1,\dots, j_0, j_1,\dots} = e^{-\tau c_a i}.
    \end{gathered}
\end{equation}

\newpage
\section{Assignment Optimization problems}
In this section we will study some assignment problems, which consist in choosing or assigning elements of a set. They have a special industrial application and interest, especially for logistics.

\subsection{Simple Assignment problem}
This is the most basic assignment problem~\cite{Assignment_Survey}. The problem has a set of agents and a set of tasks. Any agent $i$ can perform any task $j$, with a cost $C_{ij}$. If the $i$-th agent does not perform any task, it performs the $0$-th task, so $C_{i,0}=0$. We want to perform as many tasks as possible, assigning at most one agent to each task and at most one task to each agent, minimizing the total cost of the assignment. The cost function is
\begin{equation}
    C(\Vec{x})=\sum_{i} C_{i,x_i},
\end{equation}
being $x_i$ the task assigned to $i$-th agent. This way, $x_i\neq x_{i'}$ if $i\neq i'$.

The value function is the number of tasks performed, defined as
\begin{equation}
    V(\Vec{x})=\sum_{i} V_{x_i},
\end{equation}
being $V_0=0$ and $V_j=1\ \forall j\neq 0$.

\begin{figure}[h]
    \centering
    \includegraphics[width=0.5\linewidth]{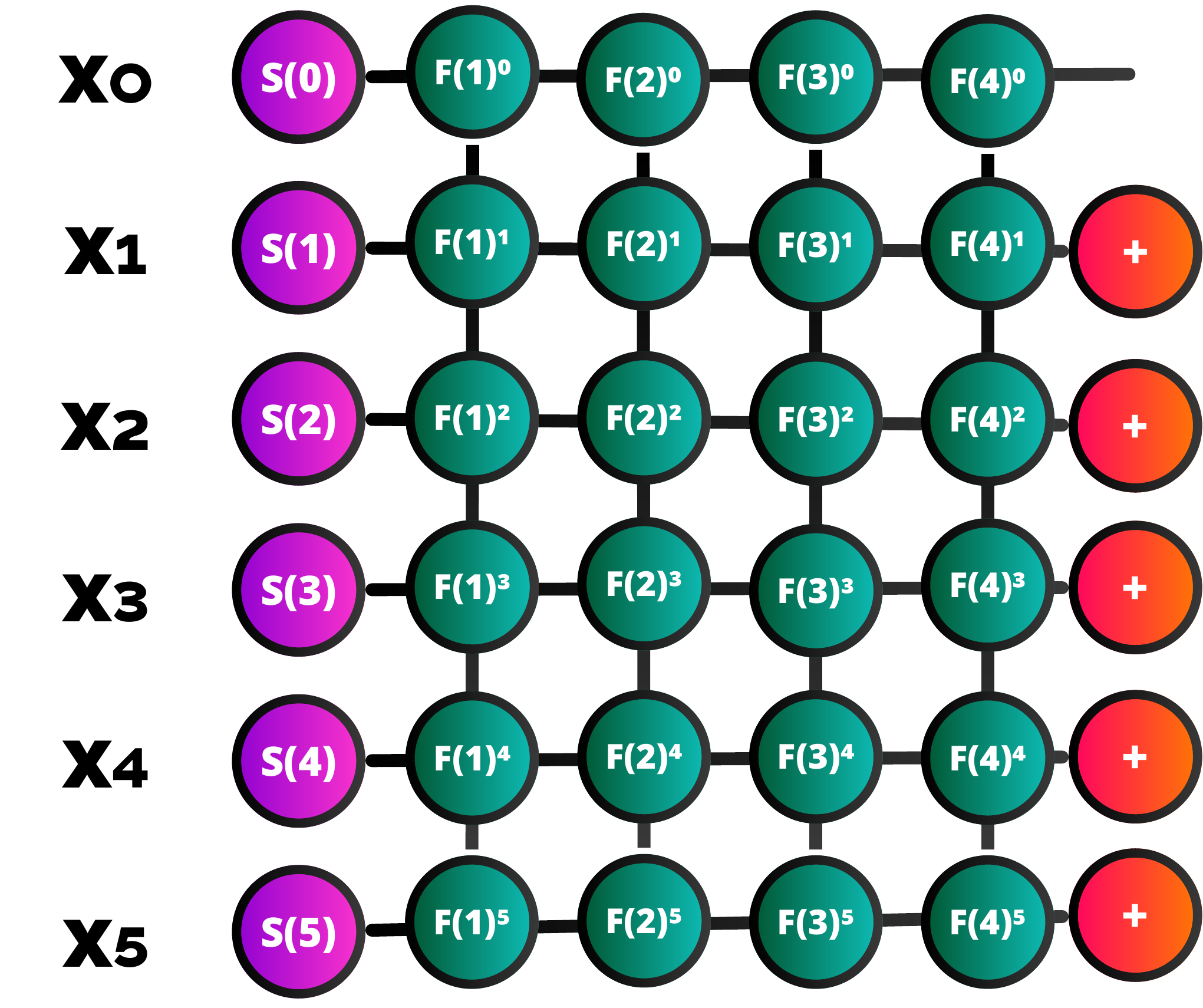}
    \caption{Tensor Network for the Assignment Problem with 6 agents and 4 tasks.}
    \label{fig: Assignment TN}
\end{figure}
To solve this problem, we only need to merge the initialization layers with evolution of the Chinese Postman problem with time dependency with the TSP repetition layers, without the layer that restricts the $0$ value. This tensor network is shown in Fig.\ref{fig: Assignment TN}. This case has also been studied in the paper~\cite{TSP_TN}. The function to minimize is
\begin{equation}
    \sum_{i} \left(C_{i,x_i} +\lambda V_{x_i}\right),
\end{equation}
being $\lambda$ a large enough factor to impose the maximum number of tasks performed.

In case the same task $t$ can be associated $c_t$ times, and each agent can have up to $d_i$ tasks, it will only be necessary to multiply each agent by $d_i$ times, making that number of variables with the same evolution. In addition, each filtering layer will have to have its upper limit of accounts to filter on $c_t$ instead of $1$.

\begin{figure}[h]
    \centering
    \includegraphics[width=0.9\linewidth]{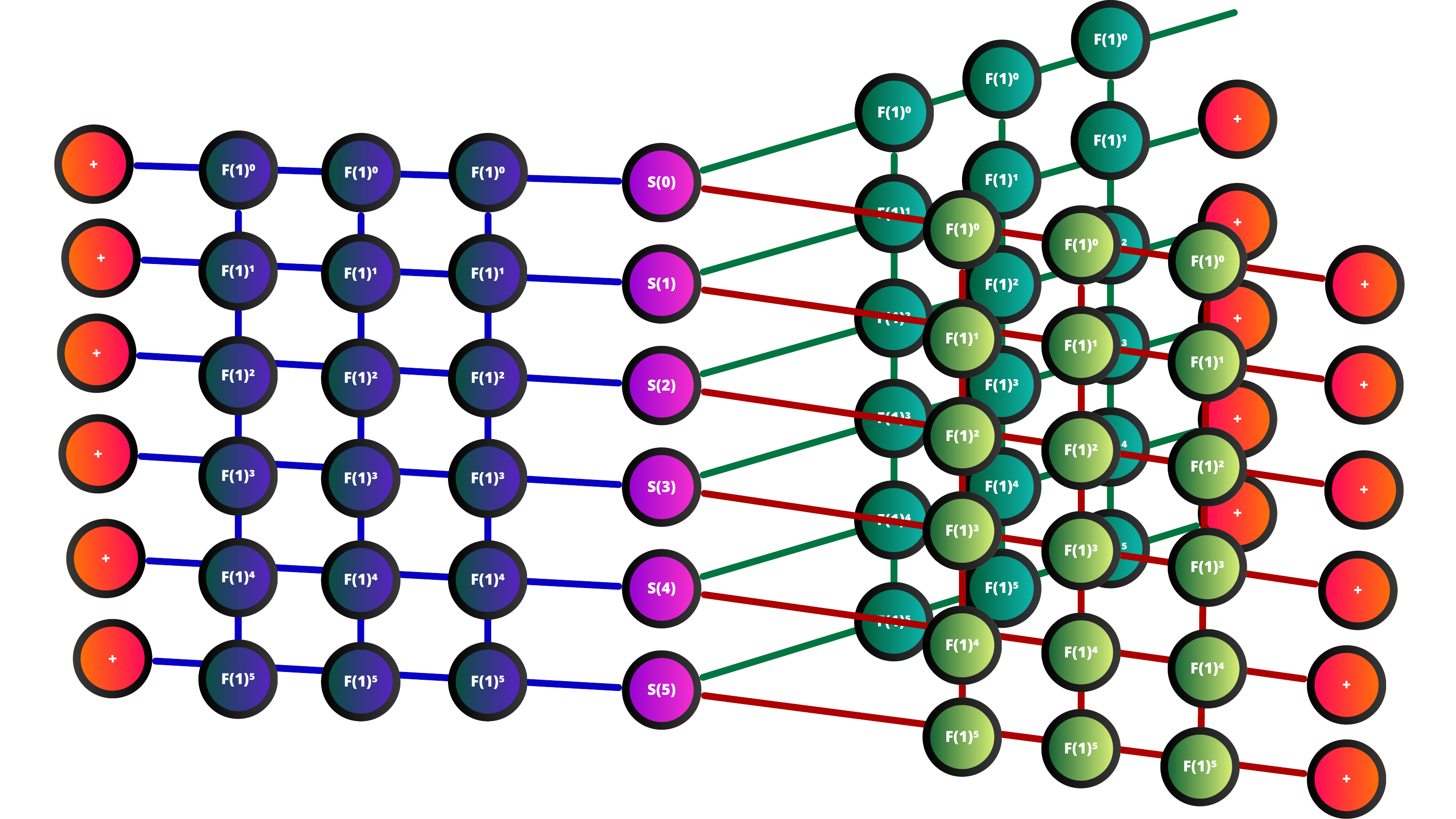}
    \caption{Tensor Network for determining the first job feature for first agent.}
    \label{fig: Multi Assignment TN}
\end{figure}

This problem can be extended to the Multidimensional assignment problem, in which each agent must do a job with a set of $M$ types of job features, which cause the job cost to change. Moreover, each job feature can only be assigned once. To address this problem, the evolution tensors $S(i)$ will have $M$ indexes, each indicating a different feature associated with the jobs, such that each of their indexes is connected to a repetition filtering network tensor equal to that of the original case. We can see the tensor network in Fig.~\ref{fig: Multi Assignment TN}.

\subsection{Knapsack Problem}

Given a set of objects, such that the $i$-th object can be selected $c_i$ times, has a weight $w_i$ and a value $v_i$, the objective is to select a combination of objects that maximizes the total value $V$ and does not exceed a maximum total weight $Q$~\cite{Knapsack_original}. The value function is calculated as
\begin{equation}
    V(\vec{x})=\sum_i v_i x_i,
\end{equation}
being $x_i$ the number of times we choose the $i$-th object, and the weight function as
\begin{equation}
    W(\vec{x})=\sum_i w_i x_i.
\end{equation}
The 0-1 version has been solved with quantum algorithms~\cite{0_1_knapsack_quantum}.

To solve this problem, the tensor network is similar to the Fig.~\ref{fig: Natural sum TN}, but instead of having the evolution in the last tensor, it takes care of eliminating all combinations that exceed the total possible weight $Q$. Each tensor in the chain performs the same task. The initialization tensors also perform the imaginary time evolution using the value of each object, such as $e^{\tau v_i x_i}$ for the $i$-th tensor, since it is a maximization problem. The optimal version of solving this problem is available in {\color{red} [pending publication]}. Based on this tensor network, two interesting generalizations can be made by changing only the tensor elements.

\subsubsection{Non Linear Knapsack Problem}
An interesting generalization of the knapsack problem is the nonlinear case. Under these circumstances, our cost and weight functions are given by the sum of nonlinear functions such that we define it as
\begin{equation}
    \begin{gathered}
        V(\vec{x})=\sum_{i=0}^{n-1}v_i(x_i) \\
        \text{subject to }W(\vec{x})=\sum_{i=0}^{n-1}w_i(x_i)\leq{Q},\\
        x_i\in [0,c_i]\quad \forall i\in [0,n-1],
    \end{gathered}
\end{equation}
where $w_i$ and $v_i$ are functions that receive natural numbers and return natural and real positive respectively.

Another simpler formulation is to convert these functions into vectors, since their inputs are natural, so that the problem is rewritten as
\begin{equation}
    \begin{gathered}
        V(\vec{x})=\sum_{i=0}^{n-1}v_{i,x_i} \\
        \text{subject to }W_{\vec{x}}=\sum_{i=0}^{n-1}w_{i,x_i}\leq{Q},\\
        x_i\in [0,c_i]\quad \forall i\in [0,n-1],
    \end{gathered}
\end{equation}
where $w$ is a natural number tensor with $w_{i,b}=\infty$ when $b>c_i$ and $v$ is a positive real number tensor with $v_{i,b}=-\infty$ when $b>c_i$ .

To address this problem, we will only need to modify the exponentials of the imaginary time evolution to add the nonlinearity and modify the output rates of the weight outputs. In this way, the tensors $M^m_{c'_m\times Q'}$ will have their non-zero elements
\begin{equation}
    \begin{gathered}
    \mu=w_{m,i} + \sum_{k=0}^{m-1}w_{k,x_k}, \\
    M^{m}_{i\mu}=e^{\tau v_{m,i}},
    \end{gathered}
\end{equation}
the tensors $K^k_{Q'\times Q'}$ will have their non-zero elements
\begin{equation}
    \begin{gathered}
    y_k \in [0,c_k], \quad \mu=i+w_{k,y_k}, \\
    K^{k}_{i\mu}=e^{\tau v_{k,y_k}},
    \end{gathered}
\end{equation}
and the tensors $K^{n-1}_{Q'}$  will have their non-zero elements
\begin{equation}
    \begin{gathered}
    d_{i}=\arg\max(\vec{\rho^{i}}),\quad \rho^{i}_{y} = \frac{1}{Q-i-w_{n-1,y}}, \\
    K^{n-1}_{i}=e^{\tau v_{n-1,d_i}}, 
    \end{gathered}
\end{equation}
where $d_i$ is the maximum number of elements of the last class that can be introduced into the knapsack without exceeding the capacity $Q$ having already a weight $i$.

In this case, the tensors involved are of the same size as in the original case, and we can also use the reuse of intermediate calculations in the same way. In addition, the optimization of the diagonals works exactly the same, only now they will not be equispaced. For all this, the computational complexity of the algorithm is the same as in the original case.

\subsubsection{Polynomial Knapsack Problem}
Another important generalization of the knapsack problem is the case where the weight function is a polynomial. To generalize as much as possible, we will take the value function to be given as a sum of nonlinear functions, and the weight function is a polynomial of a sum of nonlinear functions. With the tensorial notation, the problem is expressed as
\begin{equation}
    \begin{gathered}
        V(\vec{x})=\sum_{i=0}^{n-1}v_{i,x_i}, \quad W_{\vec{x}}=\sum_{i=0}^{n-1}w_{i,x_i},\\
        F(z)=a_0+a_1z+a_2z^2+\dots+a_pz^p\\
        \text{subject to }F(W_{\vec{x}})\leq{Q},\\
        x_i\in [0,c_i]\quad \forall i\in [0,n-1],
    \end{gathered}
\end{equation}
where $w$ is a natural number tensor with $w_{i,b}=\infty$ when $b>c_i$ and $v$ is a positive real number tensor with $v_{i,b}=-\infty$ when $b>c_i$ .

In this case, what the tensors will send to each other will be the partial result of $W$ up to that point, exactly as in the previous case. However, the change will be in the last tensor, which will be the one that will apply the $F$ function on the total accumulated $W$, eliminating the states for which it exceeds $Q$. The last tensor is
\begin{equation}
    \begin{gathered}
    d_{i}=\arg\max(\vec{\rho^{i}}),\quad \rho^{i}_{y} = \frac{1}{Q-F(i+w_{n-1,y})}, \\
    K^{n-1}_{i}=e^{\tau v_{n-1,d_i}}, 
    \end{gathered}
\end{equation}
where $d_i$ is the maximum number of elements of the last class that can be introduced into the knapsack without exceeding the capacity $Q$ having already a weight $i$.

Given the characteristics of this modification, we can also address the case in which $F$ is not a polynomial of $z$, but is a nonlinear function of $z$. In this case, the modification of the final tensor is exactly the same, taking into account the new $F$ function.

If we impose that the coefficients $a_k$ are positive integers, then $Q'\geq F(W_{\vec{x}})\geq W_{\vec{x}}$, so the matrices will have, at most, dimension $Q'$.

For the same reasons as in the previous case, the computational complexity is the same as in the original case.

\subsection{Cutting stock problem}
Given a set of $N$ types of parts, having the $i$-th part having a demand of $d_i$ and a length of $w_i$, the problem consists in determining how many parts to cut from each plate of raw material of length $W$, so that we minimize the number of plates used~\cite{Cutting_Stock}. That is, we cannot afford to exceed the total length $W$ in a single roll, and each type of piece must appear at least $d_i$ times. To formulate this problem, our variables will be $x_{i,p}$, which will indicate how many pieces of type $i$ are cut on plate $p$, and $y_p$ which is a binary variable that tells us whether that plate has been used.

\begin{figure}[h]
    \centering
    \includegraphics[width=\linewidth]{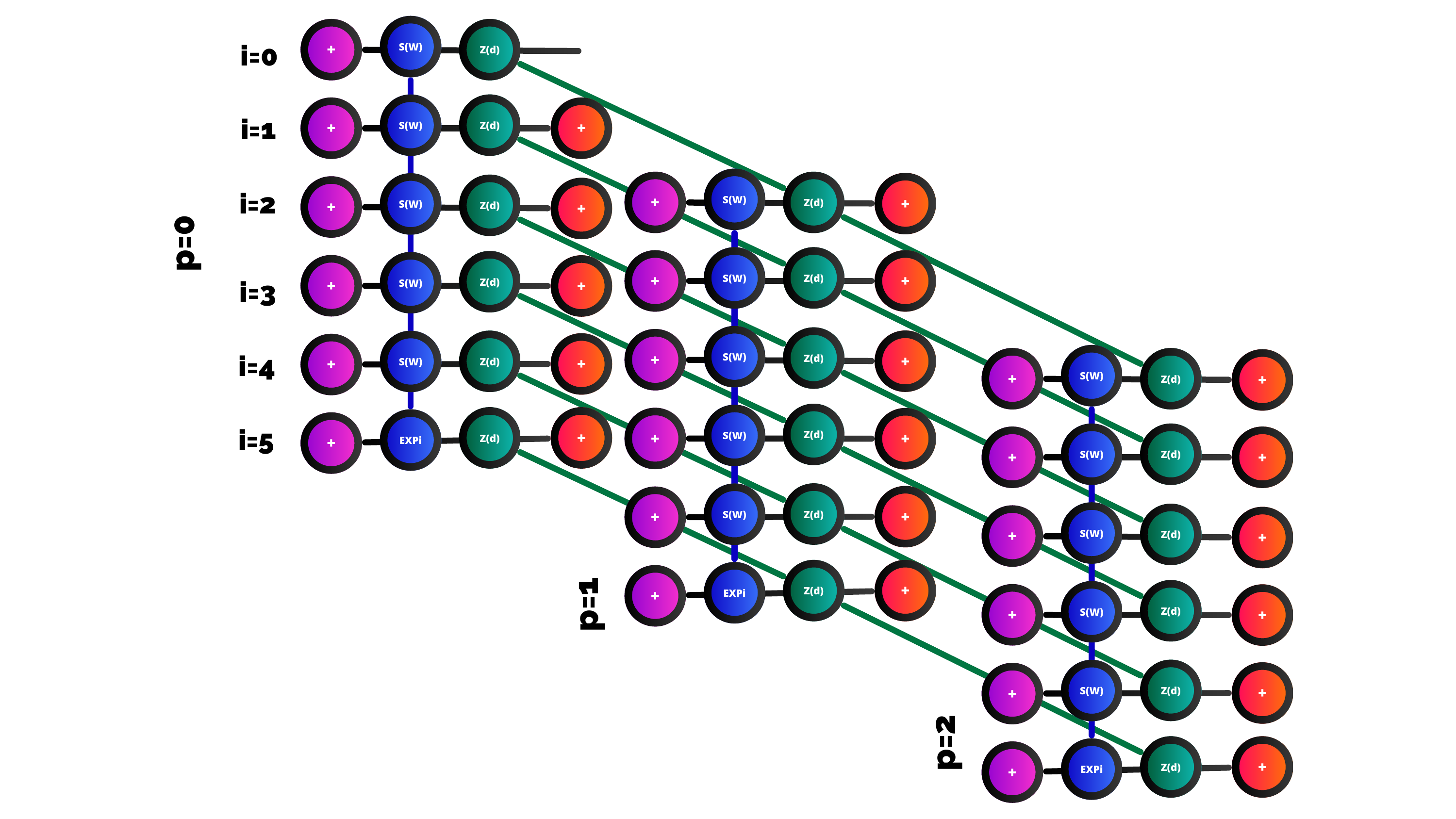}
    \caption{Cutting stock problem tensor network.}
    \label{fig: Cutting stock}
\end{figure}

To solve this problem, we will use the tensor network of Fig.~\ref{fig: Cutting stock}. In this tensor network, we have a block for each raw material plate, and a qudit line for the number of pieces $x_{ip}$ of each type. To impose the constraint of not exceeding the maximum length we use some tensors $S(W)$ which are sent what is the length up to that point, within the block of that plate. If at any time the limit is exceeded, the state is removed. To minimize the number of plates, if the length exceeds 0, we multiply the amplitude by $e^{-\tau c_i}$, being $c_i$ the cost of making use of the $i$-th plate. If all the plates have the same cost, this is the bin packing problem~\cite{Bin_Packing,Bin_packing_aprox}, which has been solved with quantum algorithms~\cite{Bin_Packing_Quantum,Bin_Packing_Quantum_Bench} and quantum annealing~\cite{Bin_Packing_Annealer,Bin_Packing_Annealer_3D}. To impose the demand condition, for each qudit, a tensor $Z(d)$ is created, and connected between blocks, keeping track of how many pieces of each type are carried. If at the end of the line, any of the types has been produced fewer times than demanded, the condition is eliminated.

\newpage
\section{Integer Programming}
In this section we will deal with two general combinatorial optimization formalisms with constraints that are well known and widely used. Integer programming is a whole branch of constrained optimization. It consists in optimization with integer variables, and is usually referred to as integer linear programming (ILP), where the cost function and constraints are linear functions.

\subsection{Integer Linear Programming}
In the case of the ILP, the problems are formulated as follows
\begin{equation}
    \begin{gathered}
        \text{maximize }C(\vec{x}) = \sum_i c_i x_i,\\
        \text{subject to } \sum_{j} A_{ij}  x_j \leq b_i \ \forall i,\\
        x_i \in \mathbb{Z}^+
    \end{gathered}
\end{equation}

In general, this kind of problems can be solved if $A$ and $\vec{b}$ consists of only positive integers. We can solve it with a tensor network similar to the presented for the systems of linear equations in Fig.~\ref{fig: Lineal Solver}. This tensor network makes the same process as in the linear solver case, but we change the post-selection nodes for $\vec{b}$. In this case, instead of using vectors with project into $b_i$ values, we use vectors with project into all values equal or less than $b_i$. This means, we change the $\delta^{b_i}$ projection vector by a step vector $s^{b_i}$ with elements
\begin{equation}
    s^{b_i}_j = 1 - H(j-b_i).
\end{equation}
This allows to take into account the restriction. To impose the maximization of the linear cost function, the superposition plus vectors are connected with a imaginary time evolution layer of matrices $S^i$ for the $i$-th variable with elements
\begin{equation}
    S^i_{j,k} = e^{\tau c_i j}\delta_{j,k}
\end{equation}
This way we can impose the restriction through the matrix multiplication tensor network, and the maximization through the imaginary time evolution layer.

\subsection{Integer Quadratic Programming}
In this case, the problem to be solved is the minimization of the cost function
\begin{equation}
    C(\vec{x}) = \sum_{i,j} Q_{ij} x_ix_j + \sum_i c_i x_i
\end{equation}
subject to the constraint
\begin{equation}\label{eq: restrict integer}
    \sum_{j} A_{ij}  x_j \leq b_i \ \forall i.
\end{equation}

\begin{figure}[h]
    \centering
    \includegraphics[width=0.9\linewidth]{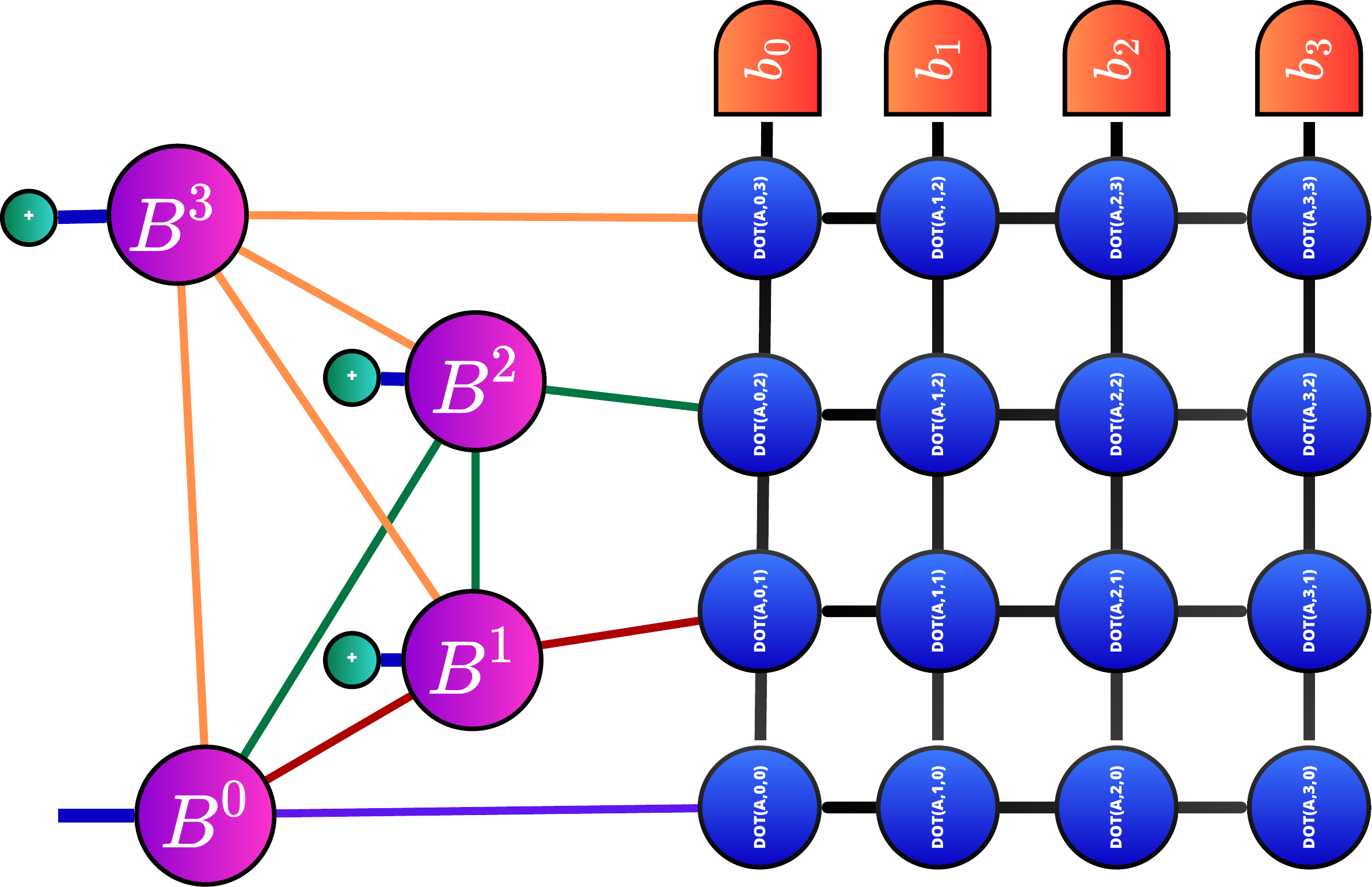}
    \caption{Tensor network to solve the Integer Quadratic Programming.}
    \label{fig: IQP}
\end{figure}

In this case, we only have to connect the tensor network that solves the QUDO problem with the tensor network that imposes the constraint, so that our resulting tensor network is the one represented in Fig.~\ref{fig: IQP}.

\subsection{Integer Polynomial Programming}
In this case, the cost function can be expressed as in \ref{eq: cost HOBO}, with the constraint \ref{eq: restrict integer}. This way, we only need to connect the TLC of the HODO problem with the TLC of the linear solver, in the same way as in Fig.~\ref{fig: IQP}.

\newpage
\section{Scheduling Optimization Problems}
In this section we will address scheduling problems, commonly related to real industrial problems, and used for performance improvement.

\subsection{Task Scheduling}
We aim to address the challenge of distributing tasks across a set of machines while adhering to specific rules regarding multiple task sets. That is, we have a set of $m$ machines and on $i$-th machine there are $P_i$ possible tasks to perform. The task $j$ of machine $i$ has an execution time $T_{ij}$. Additionally, we have a set of directed rules constraining these task combinations. An example of rules is: “If machine 1 performs task 3 and machine 2 performs task 4, machine 0 must perform task 1”. We seek a task allocation on machines that adheres to the rules while minimizing execution time. If the chosen variables are $x_i$ as the task performed by the $i$-th machine, the cost function is
\begin{equation}
    C(\vec{x}) = \sum_i T_{i,x_i}.
\end{equation}

\begin{figure}[h]
    \centering
    \includegraphics[width=0.7\linewidth]{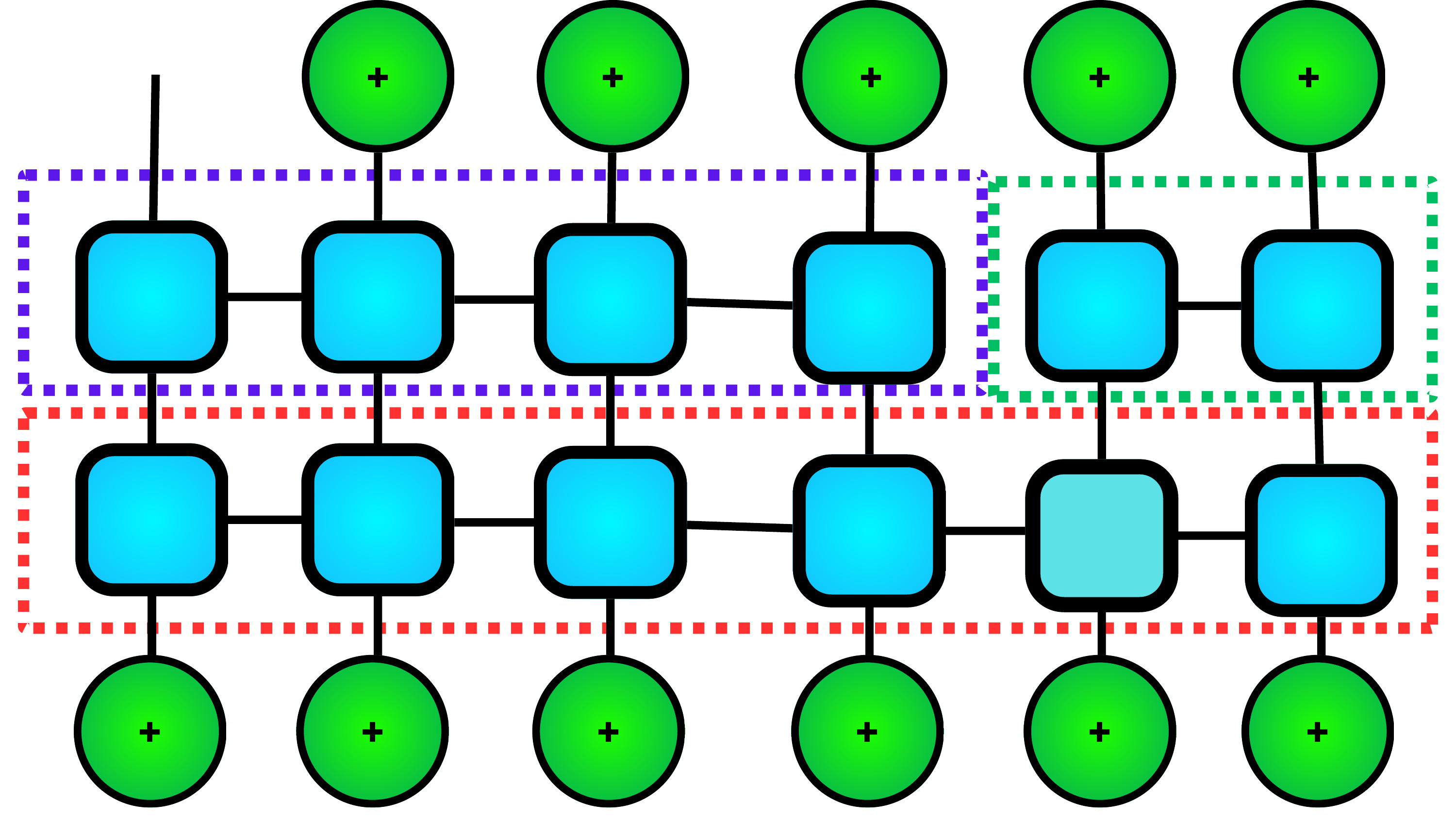}
    \caption{Task Scheduling Problem Tensor Network with 6 machines and 3 rules.}
    \label{fig: Task Scheduling}
\end{figure}

This problem is deeply analyzed in~\cite{Task_TN}, mixing this MeLoCoToN with genetic algorithms and an iterative Motion Onion. In summary, the problem is solved by a tensor network that minimizes the cost function with initialization tensors with imaginary time evolution, and a series of constraint layers that filter the states. Each layer is composed of a series of control tensors that determine whether the rule constraint is satisfied, and if it is, the projector tensor is told to project the state of the target qudit to the imposed value. In addition, such rules can be condensed under certain conditions to avoid such an exponential scaling with the number of rules. The tensor network is shown in Fig.~\ref{fig: Task Scheduling}.

In addition, the paper shows how to start the problem without constraints. It is solved until a combination is found that does not comply with a rule, in which case the constraint layer corresponding to that rule and all those that can be condensed with it are added. In this way, iteratively only the strictly necessary part of the solution space is eliminated, avoiding disproportionate scaling.

\subsection{Flow Shop Scheduling Problem}
{\color{red} Subsection not available due to paper pending publication.}

\subsection{Job Shop Scheduling Problem}
{\color{red} Subsection not available due to paper pending publication.}


\newpage
\section{Conclusions}
We have shown that every combinatorial problem that can be defined by a classical logical circuit has an associated tensor network and an exact explicit equation that solves it. Furthermore, we have shown the explicit way of obtain the associated tensor networks of a large number of different combinatorial problems. A consequence is that, if exists a physical system that allows to calculate efficiently the contraction of these tensor networks, any combinatorial problem could be solved efficiently as well. That is, the contraction of the tensor network provides an upper limit to the computational complexity of solving a problem.

Further possible lines of research are to obtain the most efficient tensor network possible with this method, a more efficient method of extracting the maximal component in optimization, or the determination of the minimum $\tau$ value to obtain the correct solution. In addition, different problems not addressed in this paper can be tackled. Another possible line is the combination of this method with other algorithms, such as genetic algorithms or heuristics, to improve their computational complexity. It is also an interesting line to analyze these problems through the tensor network that solves them, in search of new unknown properties. A last possible line is the study of how to contract such tensor networks with a quantum system.

\section*{Acknowledgment}
This work has been developed in the `When Physics Becomes Science' project of \href{https://www.youtube.com/@whenphysics}{When Physics}, an initiative to recover the original vision of science.

\bibliographystyle{unsrtnat}
\bibliography{references}

\newpage

\appendix

\section{Frequently Asked Questions}

\paragraph{Does this work imply that P=NP?}

No, this work implies that the explicit equation that solves a problem can be obtained in polynomial time, not that it is also computable in polynomial time. Moreover, polynomial time is not always with respect to the size of the problem, but is with respect to the time required to properly formulate the problem.

\paragraph{Does this work apply to all possible problems?}

This work applies to all well-formulated combinatorial problems. That is, to those in which the problem data are known and exact. For example, in the TSP, if we know exactly the edge costs, or in the CSP, if we know exactly the data imposing the constraints. Even so, the method is easily adaptable to situations in which the correct design values of the problem are not known exactly. It can even be adapted to Inverse Combinatorial Optimization~\cite{Inverse_comb}.

\paragraph{Is this formulation useful for real problems?}

Yes, provided that the problems have characteristics such that they allow efficient contraction of TLC or can be efficiently approximated by the Motion Onion.

\paragraph{In which situations is this method interesting?}

It is interesting especially in cases where there is no known algorithm to solve the problem, and when you want to perform a different mathematical analysis of the problem.

\paragraph{Which libraries are recommended to put it into practice?}

Any library that allows the use and contraction of tensors. An example in Python can be tensornetwork, tensorkrowch or Quimb.

\paragraph{Is the method approximable?}

Yes, it can be approximated both by removing logical layers, implementing them in the iteration, and by contracting the tensor network with approximate representations.

\paragraph{Is this method variational?}

No, it is a method that does not require training or derivation. It returns the solution directly.

\paragraph{I think something is missing or could be better explained.}

The reader is free to contact me directly to give feedback on the work, which may include further corrections in future versions of the document.

\newpage

\section{Glossary}

\begin{table}[h]
    \centering
    \begin{tabular}{|p{4cm}|l|p{8cm}|}
         \hline
         Term & Acronym & Meaning \\
         \hline
         Inversion Problem & - & Problem in which the objective is to obtain the input that results in a known output by means of a function.\\
         \hline
         Forward Problem & - & Problem in which the objective is to obtain the output of a known input by means of a function.\\
         \hline
         Modulated Logical Combinatorial Tensor Networks& MeLoCoToN & Algorithm that allows to obtain the tensor network equation that solves a combinatorial problem, by creating a logic circuit and its tensorization and iteration.\\
         \hline
         Signals Method & - & Construction method to build the classical logical circuits for the problems, based on the use of signals.\\
         \hline
         Logical Signal Transformation Circuits & LSTC & A logical circuit that, given an input, returns the corresponding output of a certain function. \\
         \hline
         Logical Signal Verification Circuits & LSVC & A logical circuit composed of a set of operators that send a set of internal signals to each other, each one being in charge of analyzing a specific part of the input and detecting if any constraint is violated.\\
         \hline
         Logical Signal Modulation Circuits & LSMC & Logic circuit that, given a certain input, returns the same, but alters an associated internal number according to the cost of the input. \\
         \hline
         Circuit Tensorization& CT & Conversion of a logic circuit to a tensor network by transforming its operators to constrained tensors. \\
         \hline
         Tensor Logical Circuit & TLC & Tensor network associated with a logic circuit after tensorization. \\
         \hline
         Input-Output Indexing & IOI & Transformation of the inputs and outputs of the operators to the subscripts of their associated tensors. \\
         \hline
         Half Partial Trace & - & Sum over all possible inputs or outputs for a logic circuit, leaving a single input free. Equivalent to measurement in a formalism in which the probabilities are directly the amplitudes.\\
         \hline
         Humbucker & - & Method to reduce the effect of the amplitudes of suboptimal combinations by multiplying them by a complex phase, causing them to cancel each other out when added together. \\
         \hline
         Motion Onion & -  & Group of methods to make the equation computation lighter. \\
         \hline
         Tensorial Quadratic Unconstrained D-ary Optimization & T-QUDO  & Formulation with a cost function $C(\vec{x})=\sum_{k} C_{k,x_{a_k},x_{b_k}}$ \\
         \hline
    \end{tabular}
    \caption{Glossary of new paper terms.}
    \label{tab: Glossary new}
\end{table}

\begin{table}[h]
    \centering
    \begin{tabular}{|p{4cm}|l|p{8cm}|}
         \hline
         Term & Acronym & Meaning \\
         \hline
         External Signal & - & Information of the circuit that serves as input or output, and determines the combination evaluated.\\
         \hline
         Internal Signal & - & Internal information of the circuit, which is not part of the output, comes from some operators and conditions the action of others who receive it.\\
         \hline
         Intermediate State & - & Internal signal of the circuit that is transformed along the circuit. It may ultimately be the output signal. \\
         \hline
         Amplitude & - & Internal number associated with an input in a logic circuit. Analogous to quantum amplitude. \\
         \hline
         Plus Vector (`+' tensor) & - & Vector with all its elements equal to 1.\\
         \hline
         Minus Vector (`-' tensor) & - & Vector $(-1,1)$.\\
         \hline
         Projection Vector (`$x_i$' tensor) & - & Vector that has only one non-zero element at the position corresponding to the value we want to impose. In other words, a vector of all zeros, except at position $x_i$ we impose.\\
         \hline
         Phase Vector (`P' tensor) & - & Vector with all its elements with module 1 and complex phase uniformly distributed.\\
         \hline
         Transmision Tensor Chain & TTC & A chain of tensors that share and transmit the same signal, and serve as a division of a single logical tensor with more indexes.\\
         \hline
    \end{tabular}
    \caption{Glossary of new paper objects.}
    \label{tab: Glossary new objects}
\end{table}

\begin{table}[h]
    \centering
    \begin{tabular}{|p{4cm}|l|p{8cm}|}
         \hline
         Term & Acronym & Meaning \\
         \hline
         Constraint Satisfaction Problem & CSP & Problem which consist in finding a solution
that satisfies a set of restrictions.\\
         \hline
         Optimization Problem & - & Problem in which the objective is to obtain the combination that has the lowest or highest associated value of a certain function.\\
         \hline
         Cost function & - & Function that determines how bad or good a combination is for a given problem.\\
         \hline
         Imaginary Time Evolution & ITE & Technique consisting of evolving a quantum state with its hamiltonian by making the transformation of time to an imaginary time, making the exponential real. \\
         \hline
    \end{tabular}
    \caption{Glossary of previous paper terms.}
    \label{tab: Glossary old}
\end{table}

\newpage
$ $

\newpage

\section{Tensorial Notation}
To simplify the understanding of the present and future works on tensor networks, both logical and general, we propose a new notation. The purpose of this notation is to make defining sparse tensors with logics clearer, and to make defining large tensor networks with a certain structure simpler at the graphical level.

\subsection{Notation for logical tensors}
The definition of logical tensors can be complicated and unreadable, especially in sparse cases. We have therefore taken a simpler notation to express them.

Take a 4-index $T$ tensor with dimensions $d_0$, $d_1$, $d_2$ and $d_3$, which has only non-zero elements given by a function of the indexes, at the positions given by another function of one or more indexes. This tensor can be expressed in four blocks of information: the name, the indexes, the auxiliaries and the values.

The first thing we will do is to express the dimensions of the tensor in its name, as subscript of products of dimensions. For example, in this tensor its name is
\begin{equation}
    T_{d_0\times d_1\times d_2\times d_3}.
\end{equation}

Each value of the subscripts indicates the dimension of the index of the corresponding position. For when we define the elements of the tensor, we will not add extra indexes, but replace these subscripts by the indexes of the tensor.

We differentiate the independent indexes, which run through all the values of its dimension, from the dependent indexes, whose values are obtained from the first ones. The independent ones will be expressed in roman letters, while the dependent ones will be expressed by means of greek letters. The general way of expressing them is
\begin{equation}
    \begin{gathered}
        \mu = f_\mu(i,j),\\
        \nu = f_\nu(i,j).
    \end{gathered}
\end{equation}

If we had that the tensor has an input value $n$ which creates an internal auxiliary variable $y$ to be introduced to the functions, it can be added as
\begin{equation}
    \begin{gathered}
        y = g_y(n),\\
        \mu = f_\mu(i,j),\\
        \nu = f_\nu(i,j).
    \end{gathered}
\end{equation}

The non-zero elements of the tensor will be those whose indexes satisfy the above equations, and will be obtained by a series of functions. An example is 
\begin{equation}
    T_{ij\mu\nu} =
    \begin{cases} 
    t_1(i,j,\mu,\nu)&\text{ if }  h_1(i,j,\mu,\nu)>0,\\
    t_2(i,j,\mu,\nu)&\text{ if }  h_2(i,j,\mu,\nu)>0,\\
    t_3(i,j,\mu,\nu)&\text{ else }.
   \end{cases}
\end{equation}

Thus, a tensor in general will be expressed as
\begin{align*}
    &T_{d_0\times d_1\times d_2\times d_3\times \dots\times d_N}\\
    &\begin{gathered}
        y = g_y(n,m,\dots),\\
        \vdots\\
        \mu = f_\mu(i,j,\dots),\\
        \nu = f_\nu(i,j,\dots),\\
        \vdots
    \end{gathered}\\
    T_{ij\dots \mu\nu\dots} =&
    \begin{cases} 
    t_1(i,j,\dots,\mu,\nu,\dots)&\text{ if }  h_1(i,j,\dots,\mu,\nu,\dots)>0,\\
    t_2(i,j,\dots,\mu,\nu,\dots)&\text{ if }  h_2(i,j,\dots,\mu,\nu,\dots)>0,\\
    \vdots\\
    t_M(i,j,\dots,\mu,\nu,\dots)&\text{ else }.
   \end{cases}
\end{align*}

A particular case is a tensor $T^m$ of $N$ indexes in which the non-zero elements are in the elements where the rule is
\begin{align*}
    y_j& = m^j\\
    i_n& = \sum_{j} \left(\omega_j i_j -y_j\right) 
\end{align*}
and its elements have the value
\begin{equation}
        T^m_{i_0,i_1,i_2,\dots,i_{N-1}} = 
    \begin{cases}
        \sum_{j} e^{\beta i_j+m} \text{ if } \sum_j i_j -50 > 0,\\
        \sum_{j} e^{\beta i_j-m} \text{ if } \sum_j i_j -100 > 0,\\
        1.
    \end{cases}
\end{equation}

This tensor can be expressed as
\begin{align*}
    &T^m_{d_0\times d_1\times d_2\times d_3\times \dots\times d_N}\\
    &\begin{gathered}
        y_j = m^j\\
        i_n = \sum_{j} \left(\omega_j i_j -y_j\right)
    \end{gathered}\\
     T^m_{i_0,i_1,i_2,\dots,i_{N-1}} =& 
    \begin{cases}
        \sum_{j} e^{\beta i_j+m} \text{ if } \sum_j i_j -50 > 0,\\
        \sum_{j} e^{\beta i_j-m} \text{ if } \sum_j i_j -100 > 0,\\
        1.
    \end{cases}
\end{align*}

\newpage

\subsection{Chemistry-inspired Tensor Network Notation}
This notation is inspired by the way compounds are simplified in chemistry by means of the skeletal structure. In our case, we will do the same with our tensor networks, suppressing unnecessary nodes to lighten their representation.

The motivation for this notation is that in various tensor networks, especially two-dimensional ones, representing the nodes in the figures is unnecessary and time-consuming and does not add information. Moreover, for very large tensor networks, it can blur the images.

Our notation is based on four fundamental points.
\begin{figure}[h]
    \centering
    \includegraphics[width=\linewidth]{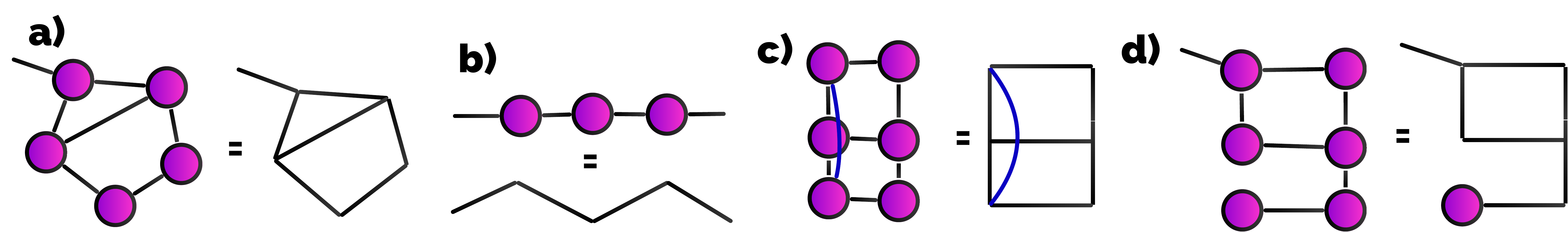}
    \caption{Representation of standard tensor network vs. its equivalent in simplified notation. a) Normal nodes. b) Chain nodes. c) Index over nodes. d) End nodes.}
    \label{fig: Notation TN}
\end{figure}

The first is the elimination of nodes in tensors with more than one index. In these cases, it will be assumed that there is a tensor whenever there are two intersecting straight lines. This is shown in Fig.~\ref{fig: Notation TN} a. 

The second is that, when a tensor has only two indexes, it will be represented as an angle, so that a linear chain remains as a chain of angles. This is shown in Fig.~\ref{fig: Notation TN} b.

The third is that if an index has to connect two nodes, so that it crosses another index, this will be represented by a curved line. In this way, we will know that the only point where there can be a node is at the ends of the curved line. This is shown in Fig.~\ref{fig: Notation TN} c.

The fourth is the non-elimination of the nodes of a single index, since if we were to eliminate them we would not be able to differentiate them from a free index. This is shown in Fig.~\ref{fig: Notation TN} d.

\newpage

\section{Common tensors and layers}
In this section we will present a series of tensors and layers of tensors that can appear in several different types of problems.

\subsection{Initialization Tensors}
These vectors are responsible for initialization, and half partial trace in some cases. We have the Plus Vector
\begin{equation}
    \begin{gathered}
        \forall i\in d,\\
        +_{i} = 1,
    \end{gathered}
\end{equation}
the Minus Vector
\begin{equation}
        - = (-1,1),
\end{equation}
the Plus Vector with local imaginary time evolution
\begin{equation}
    \begin{gathered}
        \forall i\in d,\\
        +_{i} = e^{-\tau C_i},
    \end{gathered}
\end{equation}
the Plus Vector with humbucker
\begin{equation}
    \begin{gathered}
        \forall j\in d,\\
        +_{j} = e^{2\pi i j/d},
    \end{gathered}
\end{equation}

the Plus Vector with humbucker and local imaginary time evolution
\begin{equation}
    \begin{gathered}
        \forall j\in d,\\
        +_{j} = e^{2\pi i j/d}e^{-\tau C_j},
    \end{gathered}
\end{equation}

and the projection tensor at the value $a$
\begin{equation}
    \begin{gathered}
        i=a,\\
        \delta^{a}_{i} = 1.
    \end{gathered}
\end{equation}

\subsection{Kronecker Delta}

\begin{figure}[h]
    \centering
    \includegraphics[width=\linewidth]{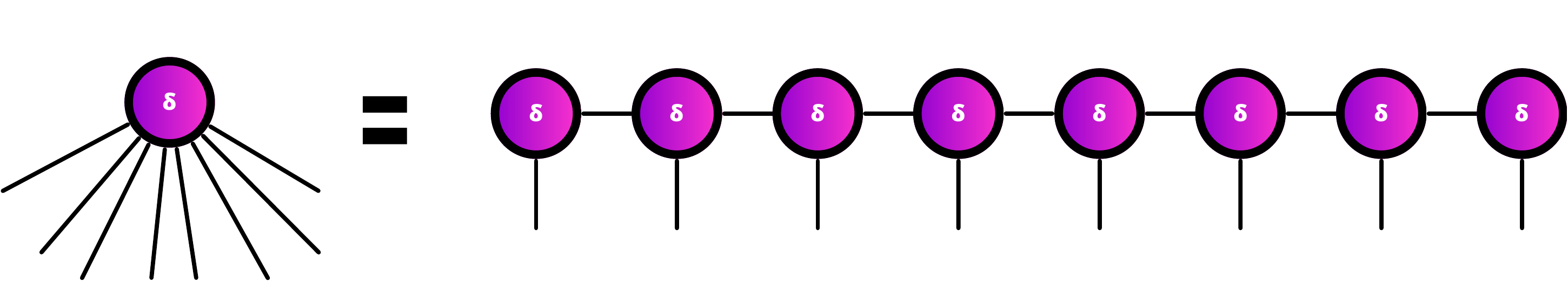}
    \caption{Kronecker delta decomposition in tensor train.}
    \label{fig: Delta}
\end{figure}

This tensor is responsible for transmitting the same signal in all directions, so that they have to be common to each other. Its non-zero elements are
\begin{equation}
    \begin{gathered}
        i_0 = i_1 = \dots = i_{N-1},\\
        \delta_{i_0, i_1, \dots, i_{N-1}} = 1. 
    \end{gathered}
\end{equation}
It can be decomposed into a train tensor of $N$ Kronecker deltas of two and three indexes as in Fig.~\ref{fig: Delta}.

\subsection{Pass Tensor}
\begin{figure}[h]
    \centering
    \includegraphics[width=0.5\linewidth]{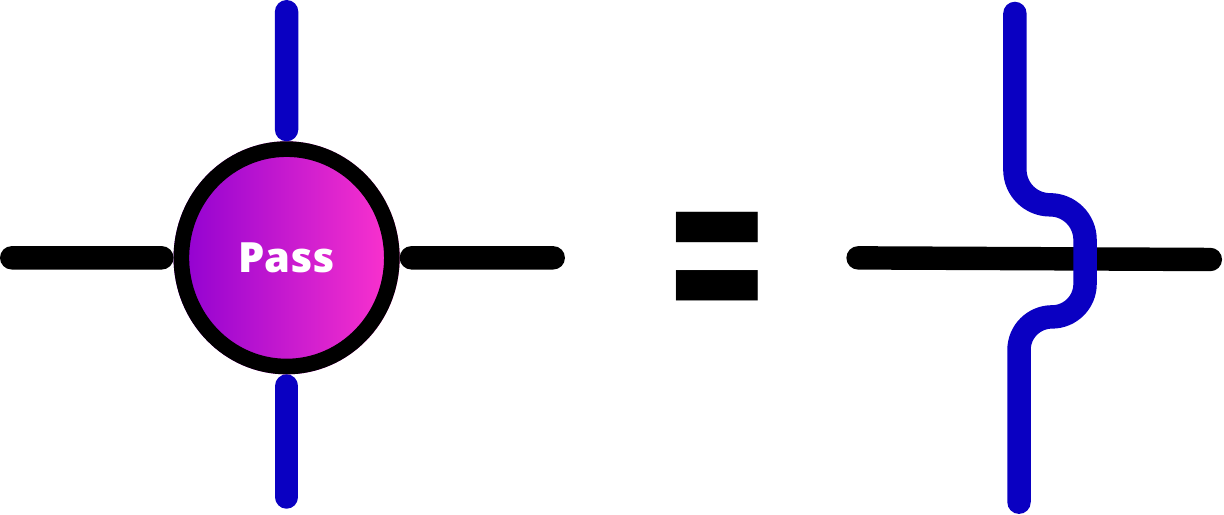}
    \caption{Pass tensor decomposition.}
    \label{fig: Pass}
\end{figure}

This 4-index tensor allows the signal to pass from top to bottom and from left to right without any interaction between them. That is, it is like having two identities acting as wires in both directions, as shown in Fig.~\ref{fig: Pass}. Its non-zero elements are
\begin{equation}
    \begin{gathered}
        \mu = i,\quad \nu = j,\\
        Pass_{i,j,\mu,\nu} = 1. 
    \end{gathered}
\end{equation}

\subsection{Counting layer}
This layer is responsible for counting how many times the value $a$ has appeared in a set of variables, and removing the state if it has appeared more than $N_a$ times. It is a Matrix Product Operator (MPO) layer, and its non-zero elements are
\begin{equation}
    \begin{gathered}
        \mu = i,\ \nu= \delta^{a}_{i},\\
        F(a)^0_{i,\mu,\nu} = 1. 
    \end{gathered}
\end{equation}
\begin{equation}
    \begin{gathered}
        \mu = i,\quad \nu= j+\delta^{a}_{i}\leq N_a,\\
        F(a)^k_{i,j,\mu,\nu} = 1. 
    \end{gathered}
\end{equation}
\begin{equation}
    \begin{gathered}
        \mu = i,\quad j+\delta^{a}_{i}\leq N_a,\\
        F(a)^{N-1}_{i,j,\mu} = 1. 
    \end{gathered}
\end{equation}

\subsection{Single Repetition Layer}
This layer ensures that a certain value $a$ appears $N_a$ and only $N_a$ times in the set of variables. Its non-zero elements are 
\begin{equation}
    \begin{gathered}
        \mu = i,\quad \nu= \delta^{a}_{i},\\
        F(a)^0_{i,\mu,\nu} = 1. 
    \end{gathered}
\end{equation}
\begin{equation}
    \begin{gathered}
        \mu = i,\quad \nu= j+\delta^{a}_{i}\leq N_a,\\
        F(a)^k_{i,j,\mu,\nu} = 1. 
    \end{gathered}
\end{equation}
\begin{equation}
    \begin{gathered}
        \mu = i,\quad j+\delta^{a}_{i}= N_a,\\
        F(a)^{N-1}_{i,j,\mu} = 1. 
    \end{gathered}
\end{equation}

\end{document}